\documentclass[8pt]{article}

\usepackage{amsmath,amsxtra,amsthm,amssymb,xr,mathtools}
\usepackage{graphicx}
\usepackage[latin1]{inputenc}
\usepackage{hyperref}
\usepackage[all]{xy}
\usepackage{xcolor}
\usepackage{stmaryrd}
\usepackage{rotating}
\usepackage{dsfont}
\usepackage{subcaption}
\usepackage{float}
\usepackage{diagbox}
\usepackage{cancel}
\usepackage{adjustbox}
\usepackage{enumitem}

\newcommand{\bit}{\begin{itemize}}
\newcommand{\eit}{\end{itemize}}
\newcommand{\bd}{\begin{description}}
\newcommand{\ed}{\end{description}}

\newcommand{\bc}{\begin{center}}
\newcommand{\ec}{\end{center}}
\renewcommand{\Ref}[1]{(\ref{#1})}

\newcommand{\C}{{\mathbb C}}

\newcommand{\R}{{\mathbb R}}
\newcommand{\Z}{{\mathbb Z}}

\newcommand{\SU}{\mathrm{SU}}

\newcommand{\SL}{\mathrm{SL}}
\newcommand{\SO}{\mathrm{SO}}

\newcommand{\be}{\begin{equation}}
\newcommand{\ee}{\end{equation}}
\newcommand{\bea}{\begin{eqnarray}}
\newcommand{\eea}{\end{eqnarray}}
\newcommand{\bs}{\begin{subequations}}
\newcommand{\es}{\end{subequations}}
\newcommand{\nn}{\nonumber}

\newcommand{\w}{\wedge}

\newcommand{\tr}{{\rm Tr}}
\newcommand{\f}{\frac}
\newcommand{\tl}{\tilde}

\newcommand{\Id}{\mathds{1}}

\newcommand{\re}{\mathrm{Re}}

\newcommand{\scr}{\scriptscriptstyle\rm}

\newcommand{\ra}{\rangle}

\newcommand{\bra}[1]{\langle {#1}|}
\newcommand{\ket}[1]{|{#1}\rangle}

\newcommand{\vet}[2]{\left( \begin{array}{cc}{#1}\\{#2}\end{array} \right)}
\newcommand{\mat}[4]{\left( \begin{array}{cc}{#1}&{#2}\\{#3}&{#4}\end{array} \right)}

\renewcommand{\a}{\alpha} 
\renewcommand{\b}{\beta} 
\newcommand{\g}{\gamma}
\renewcommand{\d}{\delta}  
\newcommand{\eps}{\epsilon} 
 
\newcommand{\z}{\zeta}
 \renewcommand{\th}{\theta}  
   
 \renewcommand{\l}{\lambda}
\let\m=\mu    \let\r=\rho \newcommand{\s}{\sigma}     \let\om=\omega
\let\G=\Gamma \let\D=\Delta  \let\Th=\Theta \let\L=\Lambda  \let\Om=\Omega

\newcommand{\norm}[1]{|\!|#1|\!|}

\usepackage[normalem]{ulem} 

\newcommand{\xg}{\hat{\xi}}
\newcommand{\hth}{\hat{\theta}}
\newcommand{\ab}{\bar{a}}
\newcommand{\J}{J} 

\newcommand{\vth}{\hat{\vartheta}}

\newcommand{\magg}{boosted orientation\ }
\newcommand{\bracing}{bracing\ }

\begin{document}

\title{\bf Asymptotics of lowest unitary \\ SL(2,C) invariants on graphs}

\author{\Large{Pietro Dona and Simone Speziale}
\smallskip \\
\small{CPT, Aix Marseille Univ., Univ. de Toulon, CNRS, Marseille, France}
}
\date{\today}

\maketitle

\begin{abstract}
\noindent 
We describe a  technique to study the asymptotics of $\SL(2,\C)$ invariant tensors associated to graphs, with unitary irreps and lowest SU(2) spins, and apply it to the Lorentzian EPRL-KKL  (Engle, Pereira, Rovelli, Livine; Kaminski, Kieselowski, Lewandowski) model of quantum gravity. 
We reproduce the known asymptotics of the 4-simplex graph with a different perspective on the geometric variables and introduce an algorithm valid for any graph.
On general grounds, we find that critical configurations are not just Regge geometries, but a larger set corresponding to conformal twisted geometries. These can be either Euclidean or Lorentzian, and can include  curved and flat 4d polytopes as subsets. For modular graphs, we show that multiple pairs of critical points exist, and there exist critical configurations of mixed signature, Euclidean and Lorentzian in different subgraphs, with no 4d embedding possible.
\end{abstract}

\begin{flushright}
\emph{To Jurek Lewandowski, for his 60th birthday}
\end{flushright}

\tableofcontents

\section{Introduction}

The spin foam formalism provides transition amplitudes for loop quantum gravity. The state of the art is the Lorentzian Engle-Pereira-Rovelli-Livine (EPRL) model \cite{EPRL,FK}, which is
based on the group $\SL(2,\C)$ and its infinite-dimensional,  unitary irreducible representations of the principal series. 
Support in favour of this model comes from the emergence of Regge geometries and the Regge action in the asymptotics of the 4-simplex vertex amplitude for large quantum numbers. This result was obtained by Barrett and collaborators \cite{BarrettLorAsymp}, building on previous work \cite{Barrett:1998gs,Barrett:2002ur,Freidel:2002mj,BarrettEPRasymp,BarrettSU2},
and has been used in a number of  applications of the model, e.g. \cite{Bianchi:2011hp,HellmannFlatness,HanZhangLor,Kaminski:2017eew,Riello:2013bzw,Engle:2011un,Engle:2015zqa,Dona:2019dkf,Han:2020fil}. 
The 4-simplex vertex amplitude is sufficient if one restricts attention to spin foams which are dual to triangulations of spacetime, and is the building block for transition amplitudes of 4-valent, simplicial spin network states. But from a canonical perspective, more general vertex amplitudes are to be included in order to provide transition amplitudes to all spin network states, and not just simplicial ones. One such generalization has been proposed by Kaminski, Kieselowski and Lewandwski (KKL) \cite{KKL}, see also \cite{Ding:2010fw}, and has been applied to cosmological models and studies of spin foam renormalization \cite{Bianchi:2010zs,Christodoulou:2016vny,Bahr:2016hwc,Bahr:2015gxa,Bahr:2018gwf,Sarno:2018ses}.
It is however not known what are the dominant geometric configurations and the asymptotic behaviour of the EPRL-KKL Lorentzian amplitude on general vertices. 
These are the open questions that we answer in this paper. 
To do so, we introduce a novel technique to study the saddle point approximation of the $\SL(2,\C)$  amplitudes. 
The technique and results presented here, while motivated by the quantum gravity applications, are of more general interest for any situation in which asymptotics of unitary $\SL(2,\C)$ Clebsch-Gordan coefficients are needed. 

The derivations of  4-simplex asymptotics  \cite{BarrettLorAsymp,Barrett:1998gs,Barrett:2002ur,Freidel:2002mj,BarrettEPRasymp,BarrettSU2} are based on 
the bivector reconstruction theorem \cite{Barrett:1997gw}: a  map from Euclidean or Lorentzian 4-simplices to a collection of bivectors satisfying certain constraints. This elegant result has played an instrumental role in driving the geometrical intuition of spin foam models, and establishing Regge calculus as a semiclassical  tool
\cite{Immirzi96NPPS,Barrett:1998gs}. But the theorem does not extend beyond 4-simplices, and the technique used in \cite{BarrettLorAsymp} cannot be directly applied to the generalized EPRL-KKL vertex.
The novel technique that we introduce 
sidesteps the bivector reconstruction theorem entirely, and allows us to by-pass its limitations. 
The idea is to focus on the 3d geometry of the boundary data, as described in terms of dihedral and twist angles, rather than on bivectors. This description follows, and is inspired by, previous work on discrete holonomy-flux geometries \cite{DittrichRyan} and on twisted geometries \cite{twigeo,IoWolfgang,Freidel:2013fia,IoFabio}. It allows us to solve the critical point equations using elementary trigonometry, without referring to bivectors, spinors, nor to algebraic maps.
We also use
a convenient choice of gauge for the boundary data of the spin foam amplitude, motivated by its clear geometric interpretation.  
This choice allows us to provide simple explicit formulas for the individual group elements at the critical point, which were necessarily left implicit in the gauge-free analysis of \cite{BarrettLorAsymp}.
 
Our first result is to show that the novel technique correctly reproduces the 4-simplex asymptotics. It offers a different perspective, with some advantages. First, solving the critical point equations and counting the number of solutions is just a matter of solving  trigonometric equations, with a prominent role played by spherical cosine and sine laws. 
Second, it makes some geometric aspects of the critical point equations more manifest, in particular it exposes the precise relation between the existence of two distinct critical points and the shape-matching conditions reducing a twisted geometry to a Regge geometry. We believe that these results simplify the understanding of the asymptotic, and can help to bring their study to a broader audience. But the main value of the novel technique lies in its general applicability beyond the 4-simplex. We present an algorithm 
that can be used to study the asymptotics of any vertex graph. 
Specifically to the purposes of this paper, we  use the algorithm to compute the asymptotics of the EPRL-KKL model for various non-simplicial vertices. 
These include the complete graph with $N$ nodes, the hypercubic graph, and the refined hypercubic graph considered in \cite{Bahr:2018vvq}. As part of our results, we extend the findings of \cite{Bahr:2015gxa,Bahr:2018gwf} and \cite{Bahr:2018vvq} to the general asymptotic properties of the Lorentzian vertex on the hypercube and refined hypercube.

From the structure of the algorithm and the specific examples treated, it is possible to draw some general conclusions about the critical asymptotics.
Let us recall from the 4-simplex asymptotics that it is the boundary data, given by spins and coherent intertwiners \cite{LS}, that determine whether the amplitudes fall-off exponentially in the large spin limit, or whether there exist one or more critical points leading to a power-law fall-off. For the 4-simplex, boundary data describing vector geometries admit one critical point, and boundary data describing 3d Regge geometries, either Euclidean or Lorentzian, admit two critical points. 
We found a similar situation for a general graph, with vector geometries having a unique critical point, and either Euclidean or Lorentzian 3d Regge geometries having more than one. 
However, there are two novelties. 
First, every 3d Regge geometry of the 4-simplex graph is also the boundary geometry of a flat 4-simplex, but this is not true for a general graph: only the subset of 3d Regge geometries that can be flat embedded, define the boundary geometry of a flat polytope.

Secondly, the configurations with more than one critical point can be more general than 3d Regge geometries, and correspond to a collection of polyhedra whose adjacent faces have partially the same shape, but not completely: the area, valence and 2d angles are uniquely defined, but not the edge lengths.
The faces can then differ by an area-preserving conformal transformation. For this reason, such angle-matched twisted geometries were called conformal twisted geometries in \cite{Dona:2017dvf}. 
The fact that non-Regge geometries admit distinct critical points was first observed in special cases of the Euclidean EPRL-KKL model \cite{Bahr:2015gxa,Bahr:2016hwc}, and then proved to be a generic feature of both SU(2) BF theory and the Euclidean EPRL-KKL model \cite{Dona:2017dvf}. Our results show that it is a generic feature of the Lorentzian EPRL-KKL model as well. 

Another new feature of general vertices is the possibility of multiple pairs of critical points. This is obvious for graph amplitudes that are reducible using  recoupling theory, since these can be written as the product of amplitudes for the smaller graphs, and data can exist leading to a pair of distinct critical points for each smaller graph. But it is also a possibility for irreducible graphs; an example for a refinement of the hypercubic graph was pointed out in \cite{Bahr:2018vvq}. 
What we find is that the increased multiplicity is a property of  $n$-modular graphs,\footnote{Graphs that can be divided in $n$ subgraphs, such that every node in each subgraph is connected to at most one node in another subgraph.} for which the angle-matched configurations admit up to $2^n$ critical points. 
Such graphs always have links that don't belong to any 3-cycle. For these links, the spherical cosine laws reconstruct 4d dihedral angles between the boundary polyhedra and the auxiliary hyperplanes identified by the larger cycles. The saddle point equations select sums and differences of these auxiliary angles, which for flat-embeddable data, can be identified as convex or concave embeddings. This explains the origin of convex and concave polytopes observed in \cite{Bahr:2018vvq}.
The same mechanism that introduces multiple solutions also unconstraints the signature of the boundary data: mixed boundary data, namely Euclidean in one subgraph and Lorentzian in another, exhibit also a critical behaviour. Such data have no flat 4d embedding.

In the conclusions we give an overview of the classification of boundary data, discuss implications for models of quantum gravity, and possible future extensions of our work.

\section{Generalized coherent EPRL-KKL amplitude}

Consider an oriented graph $\G$ with $N$ nodes connected by $L$ links. We label the nodes with $a=1,\ldots N$, and the oriented links with $(ab)$.
The generalized EPRL-KKL vertex amplitude \cite{KKL,Ding:2010fw} associates the following real function to $\G$,
\begin{equation}
\label{A1}
A_\G(j_{ab},m_{ab},n_{ab})= 
\int \prod_{a=2}^{N} dh_{a} \prod_{(ab)} D^{(\g j_{ab},j_{ab})}_{j_{ab}m_{ab}j_{ab}n_{ab}}(h_{a}^{-1}h_{b}).
\end{equation}
The integration is over ($N-1$ copies of) the non-compact manifold $\SL(2,\C)$, with $dh$ the Haar measure. The matrices $D^{(\r,k)}_{jmln}(h)$, $(j,l)\geq k$,  are the infinite-dimensional unitary irreps of the principal series in the canonical basis. Only the so-called $\g$-simple representations appear in the EPRL-KKL model, those with all $\r$'s proportional to the $k$'s, and only the lowest SU(2) spins. The common proportionality constant $\g$ is the Immirzi parameter. The integral \Ref{A1} defines an invariant tensor, and can be expressed in terms of $\SL(2,\C)$ Clebsch-Gordan coefficients \cite{Boosting}.

The graph, referred to as vertex graph, is defined surrounding the spin foam vertex with a two-dimensional sphere, and identifying each edge of the vertex puncturing the sphere as a node of the graph, and each face associated with pairs of edges as an oriented link between nodes. See \cite{KKL} for details. 
For $\G$ the 4-simplex graph, we recover the original definition of the model \cite{EPR,EPRL,FK,LS2}.
An immediate caveat in dealing with the Lorentzian theory is the non-compactness of the group integrals, which makes the convergence of \Ref{A1} non-trivial. For that to happen, it is necessary to eliminate one redundant integration (e.g. at the node 1 in the above formula), and suitably restrict the connectivity of the graph. A sufficient condition is 3-link-connectivity, namely any bi-partition of the nodes cannot be disjointed by cutting only two links, but many non-3-link-connected graphs are well defined, see \cite{Baez:2001fh,Kaminski:2010qb,Sarno:2018ses}. 
It is satisfied by all graphs explicitly considered in this paper.

The integrals \Ref{A1} can be evaluated exactly using the decomposition of $\SL(2,\C)$ Clebsch-Gordan into SU(2) ones. See \cite{Boosting} for details, and \cite{Dona:2018nev,Dona:2019dkf} for numerical applications. Here we are interested instead in analytic approximations for  large values of the spins. To study such asymptotics, it is convenient to take linear combinations of \Ref{A1} weighted by SU(2) coherent states, as suggested in \cite{LS}.
 The $\SU(2)$ coherent states are labelled by a point $(\Th,\Phi)$ on the sphere, or equivalently by unit vectors 
 \be
 \vec n:=(\sin\Th\cos\Phi,\sin\Th\sin\Phi,\cos\Th)\in\R^3.
 \ee 
In the fundamental representation $j=1/2$, any spinor $\ket\z\in\C^2$ provides a coherent state. To span an overcomplete basis, it is enough to vary the homogeneous coordinate $\z^0/\z^1\in\C P^1$, while keeping norm and the global phase fixed. 
We choose unit SU(2) norm $\norm\z^2=1$ and $\arg\z^1=0$, hence 
 \be\label{zn}
 \ket \z=  \vet{-\sin\f\Th2 e^{-i\Phi}}{\cos\f\Th2}.
 \ee
This phase convention is singled out by the relation between coherent states and holomorphic realizations of the algebra \cite{Perelomov}. We will also need the spinorial parity map,
\be\label{dualz}
|\z]:=J\triangleright\ket{\z} = \eps\ket{\bar\z} = \vet{\cos\f\Th2}{\sin\f\Th2 e^{i\Phi}}, \qquad \eps=i\s_2 = \mat{0}{1}{-1}{0}.
\ee
The states are coherent in the sense that they pick out a definite direction for the angular momentum generators, 
\be\label{sev}
\bra{\z}\vec \s \ket{\z} = -\vec n, \qquad [\z|\vec \s |\z] = \vec n,
\ee
with minimal uncertainty \cite{Perelomov}. Here $\vec\s$ are the Pauli matrices, $\bra{\z}$ and $[\z|$ the Hermitian conjugates of the two spinors $\ket\z$ and $|\z]$. This pair provides a basis of $\C^2$, orthogonal with respect to the Hermitian product, and can be put in relation with the two families of coherent states associated with the lowest and highest weights. We refer the reader to the Appendix of  \cite{Dona:2019dkf} for details and a complete list of conventions. 

To define the coherent amplitude, we take lowest weight coherent states, labeled by $\vec n_{ab}$ on the columns and by $-\vec n_{ba}$ on the rows. 
 A standard calculation exploiting the factorization property of the coherent states leads to \cite{LS,BarrettLorAsymp,Dona:2019dkf}
\be\label{Ac}
A_\G(j_{ab},\vec n_{ab}, -\vec n_{ba}) = e^{i \sum_{(ab)}j_{ab}\psi_{ab}}
\prod_{(ab)}  \f{d_{j_{ab}}}\pi
\int \prod_{a=2}^N dh_a  \int \prod_{(ab)}  \f{ d\m(z_{ab})  }{\norm {h_a^{\dagger}z_{ab}}^2 \norm {h_b^{\dagger}z_{ab}}^2} \exp S(h,z),
\ee
where the action is 
\be\label{action}
S(h,z):=\sum_{(ab)} j_{ab} \ln \frac{\bra{\z_{ab}} h_{a}^{\dagger}z_{ab}\ra^{2} \bra{h_{b}^{\dagger}z_{ab}} {\z}_{ba}]^2}{|\!| h_{a}^{\dagger}z_{ab}|\!|^2\, |\!| h_{b}^{\dagger}z_{ab}|\!|^2 } +i\g j_{ab}\ln\frac{\norm{h_{b}^{\dagger}z_{ab}}^2 }{|\!|h_{a}^{\dagger}z_{ab}|\!|^2 },
\ee
with $h_1=\Id$.
The map between the vectors on the left-hand side of \Ref{Ac} and the spinors in the right-hand side is provided by \Ref{sev}.
The detailed derivation of \Ref{Ac} with our conventions can be found in the Appendix of \cite{Dona:2019dkf}.
The continuous vectorial labels replaces the magnetic labels $m_{ab}, n_{ab}$ of \Ref{A1}, whereas the spins $j_{ab}=j_{ba}$ are untouched. 
The dummy spinors $\ket{z_{ab}}$ provide the homogeneous realization of the infinite dimensional irreps. The integrand is invariant under complex rescalings $\ket{z_{ab}}\mapsto\l_{ab}\ket{z_{ab}}$, and the spinorial integration is defined over $\C P^1$,
with measure $d\m(z):= (i/2) [z\ket{dz}\w [\bar z\ket{d\bar z}$. 
See \cite{Ruhl,BarrettLorAsymp} for more details.

The  amplitude \Ref{Ac} is complex, unlike \Ref{A1}, because of complex coefficients of the coherent states. 
The phase factors $\psi_{ab}$ depend on two choices:
the phase convention chosen for the SU(2) coherent states, and the way the parity map on the bras is introduced. We have chosen to do so taking a minus sign in the vector label. Alternatively, one can use the spinorial parity map \Ref{dualz}, or the parity map on the infinite-dimensional representations, as done in \cite{BarrettLorAsymp}. Introducing the parity map on the bras is convenient to have the closure conditions satisfied by the normals at every node, without additional signs. This makes us able to systematically interpret the normals as outgoing to the faces of the polyhedra. 
With our choices, convenient from a numerical viewpoint \cite{Dona:2017dvf,Dona:2019dkf},
 \be
\psi_{ab} = -2\Phi_{ab}.
 \ee
With the option used in \cite{BarrettLorAsymp}, $\psi_{ab} = i \arctan\gamma$. This global phase difference 
is irrelevant for the saddle point analysis and geometric interpretation of the asymptotics.
Apart from this global phase difference of the amplitude, the action \Ref{action} is related to the one used in \cite{BarrettLorAsymp} by a complex conjugation and inversion of the sign of $\g$.\footnote{And the identification $\z_{ab}=\xi_{ab}$. The two actions can equivalently be related by the map $h_a\mapsto \bar h_a$, $\z_{ab}\mapsto\bar\z_{ab}=\bar\xi_{ab}$, as stated in \cite{Dona:2019dkf}, without conjugation or changing $\g$. Notice also that we use an opposite sign for $J$, and the opposite convention for the map between spinors and normal vectors: we associate spinors with the standard lowest weights SU(2) coherent states, whereas \cite{BarrettLorAsymp} uses the highest weights. As a consequence, the signs of the vectors in \Ref{sev} would be flipped.}

The  coherent amplitude satisfies an important covariance property for the boundary data. If we rotate all the normals $\vec{n}_{ab}$ at a node $a$  
\be\label{gauge}
\vec n_{ab} \mapsto R_a \vec n_{ab} \qquad \forall b, \qquad R_a\in\SO(3),
\ee
the coherent states pick up a phase $\chi_{ab}(\vec n_{ab},R_a)$ \cite{Perelomov},
\be\label{spinorrot}
\ket{\z_{ab}} \mapsto e^{i \chi_{ab}} r_a \ket{ \z_{ab}},
\ee 
where $r_a$ is the $\SU(2)$ element corresponding to the rotation $R_a$. 
These group elements can be reabsorbed redefining  $h_a$ and using the invariance of the Haar measure,\footnote{To include rotations at the node 1 without group integration, the redefinition is $h_a\mapsto h_a r_a$ for all nodes not connected to 1, and $h_a \mapsto r_1^{\dagger}h_a r_a$ for those connected to 1.}
but the phase remains:
\be\label{Arotated}
A_\G(j_{ab},R_a \vec n_{ab}) = e^{-i\sum_{ab}j_{ab}(\chi_{ab}-\chi_{ba})} A_\G(j_{ab},\vec n_{ab}).
\ee
As a consequence, the amplitude is invariant under \Ref{gauge}, up to a phase. The
norm of the amplitude is a function of rotational-invariant quantities only, namely of the scalar products
$\vec n_{ab}\cdot \vec n_{ac}$ between normals at the same node. Their relative orientation does not matter.
The relative orientations do not affect the geometric interpretation of the asymptotic formula, and are also irrelevant to spin foam applications. Furthermore, the  global phase can always be reabsorbed changing the phase conventions for the SU(2) coherent states, if one so wishes. For these reasons, we will refer to \Ref{gauge} as gauge transformations of the boundary data. 
 It should not be confused with the invariance of the amplitude under group transformations acting on the group elements $h_a$.

\subsection{Classification of boundary data: from 3d to 4d geometries}
\label{sec:classification}

To better understand the saddle point analysis, let us first review how the boundary data assign a geometric structure to a cellular decomposition dual to the oriented graph $\G$. 
This can be done following loop quantum gravity, where the spin $j_{ab}$ gives the area of the dual face $ab$, and the vectors $\vec n_{ab}$ and $\vec n_{ba}$ are the normals to that face in two different $\R^3$ frames, one at the node $a$ and one at the node $b$. 
The triple $(j_{ab}, \vec n_{ab}, \vec n_{ba})$ on each link parametrizes the subset of the holonomy-flux phase space $T^*\SU(2)$ with vanishing twist angle $\xi_{ab}$ \cite{twigeo}.\footnote{The reason one considers boundary data with vanishing twist angles is that the amplitudes are coherent in the directions, but sharp in the areas. One can consider superpositions of vertex amplitudes also in the spins, like it is done in the propagator calculations \cite{Bianchi:2006uf}. These will be labelled also by the twist angles, and the boundary data be complete holonomies and fluxes.}

We now list a number of subsets of geometric interpretations, which are relevant to the saddle point analysis.
\begin{enumerate}
\item[$(i)$] {\bf Twisted geometries.}\footnote{Or \emph{closed} twisted geometries, if the name open twisted geometries is used for the generic parametrization of $T^*\SU(2)$ prior to imposing the closure conditions.} The data satisfying the closure condition at each node,
\begin{equation}\label{Clos}
\sum_{b\neq a}j_{ab}\vec{n}_{ab}=0, \qquad \forall a.
\end{equation} 
This condition means that the vectors describe a (typically bent, namely non-planar) polygon in $\R^3$ at each node. 
For non-coplanar vectors, the polygon also identifies a unique convex polyhedron in $\R^3$, with the spins as areas and the vectors as normal directions of the faces. 
In particular, it is the data themselves that determine the type of polyhedron, namely its adjacency matrix \cite{IoPoly}. We conventionally choose the normals to be outgoing to the polyhedra, hence the scalar products
\be\label{defphi}
\vec n_{ab}\cdot\vec n_{ac} = \cos\phi^a_{bc}
\ee
define the exterior dihedral angles $\phi^a_{bc}\in[0,\pi)$. The non-coplanar case is the most relevant one to loop quantum gravity, and our saddle point analysis will focus on this case. 
The closed, non-coplanar data describe a collection of polyhedra dual to the nodes, adjacent to one another following the connectivity of the graph. The face shared by two polyhedra has a unique area determined by the spin, but in general different shapes induced by the shape of each polyhedron. The name \emph{twisted geometries} refers to this mismatch, but also to the relation to twistors that this parametrization has led to \cite{twigeo2,WielandTwistors,IoWolfgang}. 

The normals endow each polyhedron with an orientation in $\R^3$. 
One can further reduce the description looking at the data modulo rotations at each node. This is the only information relevant to the determine the norm of the coherent amplitude, and consists of a gauge-invariant description of the twisted geometry
in terms of areas and the shape parameters of the polyhedra.

These geometries carry a notion of Euclidean and Lorentzian signature. This can be determined from the range of the functions of $\phi^a_{bc}$ that enter the spherical cosine laws (see below). In the case of a 4-simplex, it matches the signature determined by the areas and triangle inequalities.
For Lorentzian data, the polyhedra are space-like by definition of the model.\footnote{See \cite{Rennert:2016rfp,IoNull} for twisted geometries with time-like and light-like polyhedra,
and e.g. \cite{Barrett:1997gw,Conrady:2010vx,Kaminski:2017eew} for spin foam models with time-like faces.} One can also use the spherical cosine laws to define 4d dihedral angles associated to the faces. Because of the shape-mismatch, this gives as many different dihedral angles per face as its valence \cite{DittrichSpeziale,Dittrich:2012rj,Freidel:2013fia,IoFabio}. 

\item[$(ii)$] {\bf Vector geometries.} These were introduced in \cite{Barrett:2002ur} as the set satisfying closure and further the orientation conditions
\be\label{orientations}
R_a \vec n_{ab} = - R_b \vec n_{ba}, \qquad R_a \in \SO(3).
\ee
We remark that these conditions are independent of the spins, and that they can be satisfied only for twisted geometries that have a Euclidean signature everywhere, as proved in Appendix~\ref{app:vector}. 

This definition is orientation-dependent: a gauge reduction would be useful to think clearly  of the vector geometries as a subset of twisted geometries we some partial shape matching imposed, but it is not known. In the case of the 4-simplex, a partial gauge reduction of the five-dimensional space of vector geometries at fixed spins was obtained in \cite{Dona:2017dvf}, in terms of four shape parameters of the tetrahedra, and one non-gauge-invariant angle.

\item[$(iii)$] {\bf Angle-matched, or conformal twisted geometries.} These were introduced in \cite{Dona:2017dvf} and correspond to the subset of twisted geometries with a partial shape-matching, consisting of the valence and 2d angles of the faces. For triangulations, they coincide with 3d Regge geometries, but not in general.

When the faces of the polyhedra are not triangles, the angle-matching conditions don't impose the matching of the shapes: two polygons with the same area and the same 2d dihedral angles can still differ by a conformal transformation that preserves the area. 
An $n$-sided polygon in $\R^2$ has $2n-3$ degrees of freedom up to rotations, thus matching area and angles leaves $n-3$ freedoms. 

Conformal twisted geometries exist for both Euclidean and Lorentzian angles. If, and only if, they are fully Euclidean, they are a strict subset of vector geometries \cite{Dona:2017dvf}. 
Angle-matched twisted geometries have two important geometric properties, that will play a prominent role in the saddle point analysis:
\begin{itemize}
\item By having a well-defined valence for each face dual to the link, conformal twisted geometries introduce a map from cycles of the graph to edges of a cellular decomposition. 
\item Furthermore, 
they assign a unique 4d dihedral angle to each face, via the spherical cosine laws, more on this in the next Section.
\end{itemize}

\item[$(iv)$] {\bf 3d Regge geometries.} The further subset of angle-matched twisted geometries satisfying full shape-matching conditions \cite{DittrichSpeziale,IoPoly}. They split in Euclidean and Lorentzian sectors, again identified by the spherical cosine laws. The reduced data are in one-to-one correspondence with the edge lengths of the cellular decomposition,\footnote{Up to some highly symmetric configurations, e.g. a regular parallelepiped, which is only uniquely characterized by areas and angles and not by the edge lengths, } and describe a 3d Regge geometry. The standard and simplest definition of a Regge geometry uses only triangulations, but the extension to a cellular decomposition described by its edge lengths is rather straightforward. Notice that the graph $\G$ does \emph{not} fix the nature of the cellular decomposition, only the number and connectivity of the cells: whether it is a triangulation or not, it is determined by the data themselves following the Minkowski theorem \cite{IoPoly}.

\item[$(v)$] {\bf Flat-embeddable 3d Regge geometries.} 
Among the 3d Regge geometries, it is possible to single out those that can be flat embedded in 4d. This request typically  constrains the edge lengths. 
In this case, the data can be given a 4d interpretation as the boundary geometry of a flat 4d polytope,  Euclidean or Lorentzian as previously determined by the spherical cosine laws.
A special situation occurs in the case of the 4-simplex: a 3d Regge geometry on the boundary graph of the 4-simplex is always flat-embeddable, therefore all 3d Regge data admit a 4d interpretation as a flat 4-simplex. But in general this is not the case. 
When the flat embedding is not possible, one can seek a curved interpretation for instance starting from the flat polytope and adding one or more vertices in the bulk, but we are not aware a specific prescription to do so.

\end{enumerate}

Looking at the definitions, we can see that the 4-simplex case loses this fine-grained structure. In particular,  there is no distinction between ($iii$), ($iv$) and ($v$), which are all squashed in the same class of 4-simplex Regge geometries.
From the classification we also see that every Euclidean 3d Regge geometry is automatically a vector geometry. This may not be very intuitive at first, but can be given a nice geometric picture in the case of a 4-simplex, in terms of 3d objects called spike and twisted spike, we refer to \cite{Dona:2017dvf} for more details.
The 4-simplex has another special property: one can generically invert the edge lengths for the areas,\footnote{Up to isolated configurations with non-invertible Jacobian, see Appendix~\ref{AppTacchino}.} and use the latter as fundamental variables to describe the geometry. 
In this respect, the reduction from a vector geometry to a Regge geometry corresponds  to picking a configuration of normals (unique up to rotations) compatible with the areas. Furthermore, the distinction between Euclidean and Lorentzian data can be done using the areas alone, by computing the squared 4-volume and checking its sign. In the quantum model, the areas are described by the half-integer spins $j_{ab}$, and only in the large spin limit one can approach a continuum of possible shapes.

This classification of boundary data, which is common to any spin foam model with SU(2) spin network states in its boundary, is useful to classify the existence of critical points and their geometric interpretation. 

\subsection{Spherical cosine laws and edge twist angles}
\label{SecSCL}

\begin{figure}[H]
\centering
\includegraphics[width=3.5cm]{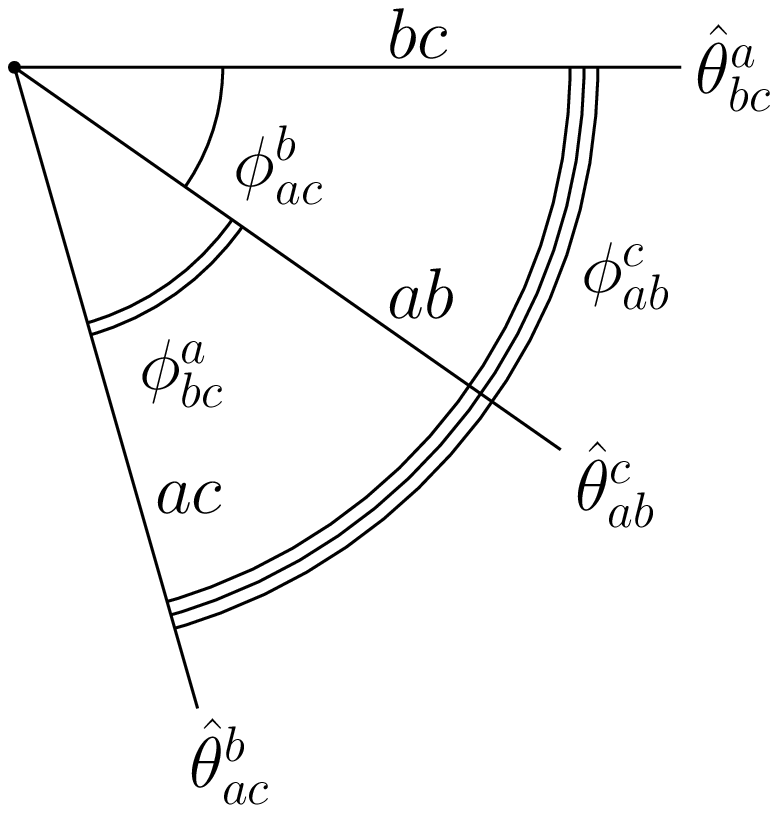}
\hspace{2cm}\includegraphics[width=4.5cm]{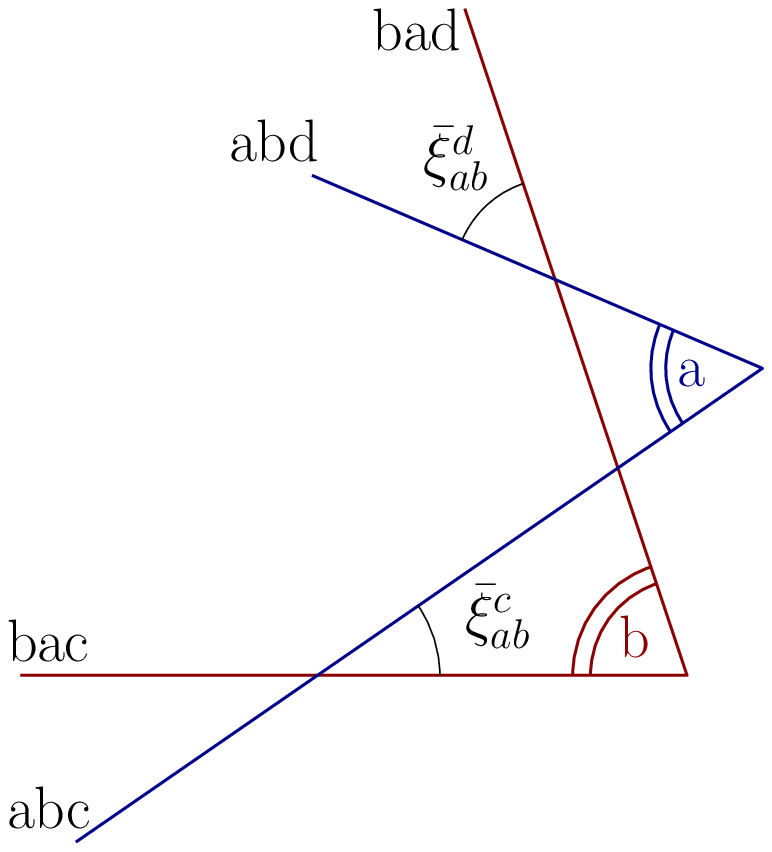}
\caption{\small{Left panel: \emph{The spherical cosine laws can be used to define edge-dependent 4d dihedral angles from  the 3d dihedral angles computed from the boundary data. } Right panel: \emph{The edge twist angles $\xg^c_{ab}$ on the face shared between the polyhedra $a$ (blue) and $b$ (red). When they match for all edges, $\xg^c_{ab}=\xg^d_{ac}$, also the 2d angles (marked by a double line) match, by Thales' theorem.}}
\label{fig:xg}
}
\end{figure}

The previous classification has made various references to the spherical cosine laws. These play a crucial role throughout our analysis, and it is useful to provide some details here. For future reference, a summary of the basic angles used is given in Table~\ref{Table1}.
\begin{table}[H]\begin{tabular}{|c|l|}
\hline
$\phi^a_{bc}$ & 3d angle of polyhedron $a$ between the faces identified by its adjacency to the polyhedra $b$ and $c$ \\
$\hth^a_{bc}$ & 4d angle between polyhedra $b$ and $c$ defined using the edge identified by their adjacency to $a$ \\
$\a^{ab}_{cd}$ & 2d angle of the polyhedron $a$, in the face identified by its adjacency to $b$, \\ & between the edges identified by its adjacency to $c$ and $d$ \\
$\xg^a_{bc}$ & angle between the vectors associated to the same edge \\ & (the one in the face shared by the polyhedra $b$ and $c$, identified by their adjacency to $a$) \\& but computed using either the geometry of $b$ or that of $c$\\
\hline
\end{tabular}\caption{\label{Table1}\emph{\small{Various geometric angles and their notation. The first three are defined in $[0,\pi)$, and taken to be external by convention, while the twist angle is defined in $[0,2\pi)$.}}}\end{table}
Consider first three hyperplanes in $\R^4$, call them $a$, $b$, and $c$ for later convenience. Generically, they intersect at an edge, and pairwise at planes.
The spherical cosine laws give a relation between the  3d dihedral angles among the planes, denote them $\phi$, and the 4d dihedral angles among the hyperplanes, denote them $\hth$. 
With reference to Fig.~\ref{fig:xg}, left panel, and using conventions with exterior dihedral angles, we have
\begin{equation}
\label{defhth}
\cos\hth_{ab}^c =\frac{\cos \phi^{c}_{ab}+ \cos \phi^{b}_{ac} \cos \phi^{a}_{bc} }{\sin \phi^{b}_{ac} \sin \phi^{a}_{bc}}\in\R, \qquad \re(\hth_{ab}^c)\in[0,\pi].
\end{equation}
This is the angle between $a$ and $b$.
Since the hyperplanes are given, the left-hand side does not depend on $c$: as we move the third hyperplane in $\R^4$, all three $\phi$'s change, while keeping the result unchanged.
In other words, the 4d angle depends on two hyperplanes only, the third is introduced only to compute it from 3d angles.
The reality of the angles $\hth^c_{ab}$ depends on the values of the 3d angles, as follows:
\begin{itemize}
\item If the RHS takes values in $(-1,1)$ then $\hth_{ab}^c=\th_{ab}^c\in[0,\pi)$. These are `Euclidean' configurations, with $\hth$ the angle between two Euclidean vectors, or space-like Minkowski vectors.
\item If the RHS takes values in $(1,\infty)$ then $\hth_{ab}^c=i\th_{ab}^c$ is purely imaginary, and we fix by convention $\th_{ab}^c>0$. These are `Lorentzian co-chronal' configurations, with $\hth$ the angle between two time-like vectors pointing both to the future or the past (called `thick wedge' configurations  in \cite{BarrettLorAsymp}).
\item If the RHS takes values in $(-\infty,-1)$ then $\hth_{ab}^c=i\th_{ab}^c+\pi$ is complex with real part $\pi$, and we fix by convention $\th_{ab}^c<0$. These are `Lorentzian anti-chronal' configurations, with $\hth$ the angle between one future-pointing and one past-pointing time-like vector (or `thin wedge' configurations).
\end{itemize} 
We use the notation $\hth$ as bookkeeping for both real and complex possibilities, and $\th$ for the Euclidean or Lorentzian real angles.

Consider now a different setting, where we are not given the hyperplanes, but rather just a triple of 3d angles $\phi^c_{ab}$. This is the situation described by the boundary data of the amplitude for three nodes  $a$, $b$ and $c$ mutually connected, namely belonging to a 3-cycle 
${\cal C}$ of the graph. Assuming non-degenerate data, each node describes a polyhedron in $\R^3$, and the 3-cycle identifies an edge shared by the three polyhedra. 
With reference to Fig.~\ref{fig:xg}, left panel, the edge is the vertex where the three lines meet.
We can now use \Ref{defhth} as a \emph{definition} of 4d angles from the boundary data. This definition provides a local flat embedding of the polyhedra in $\R^4$,
or equivalently or their hyperplanes, like a discrete version of the Gauss-Codazzi equation. This embedding works independently of the shapes of the polyhedra.
If we now swap the triple for a new triple $\phi^d_{ab}$, where $d\neq c$ is a fourth mutually connected node, we obtain in general different 4d angles and a different embedding of the polyhedra. Since each choice of triple is associated with an edge shared by two of the polyhedra, we refer to \Ref{defhth} defined in this way as \emph{edge-dependent} 4d dihedral angles. 
Such angles allows us to locally classify the boundary data into Euclidean or Lorentzian, according to the range of the right-hand side of \Ref{defhth}. 

This is the situation for generic boundary data.
We are now interested in characterizing a special subset of the data, for which the reconstructed 4d angle does \emph{not} depend on the choice of edge.
To provide an answer to this question, we consider the spherical cosine laws
relating 
the 3d dihedral angles to 2d dihedral angles:
\be\label{2dSCL}
\cos\phi_{bc}^{a}=\frac{\cos\alpha_{bc}^{ad}+\cos\alpha_{bd}^{ac}\cos\alpha_{cd}^{ab}}{\sin\alpha_{bd}^{ac}\sin\alpha_{cd}^{ab}}, \qquad 
\cos \a^{ab}_{cd} = \frac{\cos \phi^{a}_{cd}- \cos \phi^{a}_{bc} \cos \phi^{a}_{bd} }{\sin \phi^{a}_{bc} \sin \phi^{a}_{bd} }.
\ee
Here $\a^{ab}_{cd}$ is the external 2d angle in the face $ab$, between the edge shared by $a$ and $c$ and the edge shared by $a$ and $d$. Looking at the right-hand side, we see that the notation for the angle is symmetric in the lower indices, $\a^{ab}_{cd}=\a^{ab}_{dc}$. This is a constant property of the notation for all angles used in this paper. It is a priori not symmetric on the upper indices, $\a^{ab}_{cd}\neq \a^{ba}_{cd}$. This amounts to using the 3d angles of $a$ or the 3d angles of $b$, and this will in general produce different answers, since adjacent polyhedra have different shapes. The equality $\a^{ab}_{cd}= \a^{ba}_{cd}$ defines the special boundary data that we call angle-matched, or conformal, twisted geometries.
It is easy to show that angle-matched data define also edge-independent 4d dihedral angles. To see that, it suffices to express $\hth^c_{ab}$ and $\hth^d_{ab}$ in terms of the 2d angles using \Ref{2dSCL}. It follows by inspection that
\be\label{eureka}
(\a^{ab}_{cd}=\a^{ba}_{cd}, \ \a^{ac}_{bd}=\a^{ca}_{bd}, \ \a^{ad}_{bc}=\a^{da}_{bc}, \ \a^{bc}_{ad}=\a^{cb}_{ad}, \ \a^{bd}_{ac}=\a^{db}_{ac})
\quad\Rightarrow\quad \hth^c_{ab}=\hth^d_{ab}.
\ee
The 4d dihedral angle computed at $c$ matches the one at $d$ if the 2d angles at the vertex $abcd$ for the faces $ab$, $ac$, $ad$, $bc$, and $bd$ match. To have edge independence of $\hth$ at a face $ab$, we need angle matchings not only at $ab$, but also on adjacent faces.
Reversing the procedure, we also obtain 
\be\label{eurekainv}
(\hth^c_{ab}=\hth^d_{ab}, \ \hth^b_{ac}=\hth^d_{ac},\ \hth^b_{ad}=\hth^c_{ad}, \ \hth^a_{bc}=\hth^d_{bc}, \ \hth^a_{bd}=\hth^c_{bd},
\  \hth^a_{cd}=\hth^b_{cd})
\quad\Rightarrow\quad \a^{ab}_{cd}=\a^{ba}_{cd}.
\ee
A local double arrow corresponding to a necessary and sufficient condition is clearly not possible, because the 2d and 4d angles have common $\phi$'s but also uncommon ones: they differently on the structure of the graph. Nonetheless, 
if all (independent) cycles of the graph are 3-valent, it is possible from these relations to conclude that all 4d dihedral angles are edge-independent if and only if all 2d angles match:
\be\label{19}
 \{\a_{ab}^{cd}\} =\{ \a^{dc}_{ab} \} \qquad \Leftrightarrow \qquad \{\hth^c_{ab}\}=\{\hth_{ab} \}.
\ee

This is for instance the case of the 4-simplex: all nodes are mutually connected, and all edges of the tetrahedra are dual to 3-cycles. In this case, it is easy to see that angle-matching conditions for all 2d angles imply the edge-independence of all 4d angles.
For a general graph we need additional formulas, to take into account the edges of the polyhedra which are not dual to 3-cycles but to larger cycles of the graph.
In this case, it turns out that one has to consider first the (edge-dependent) 4d angles between one polyhedron and the auxiliary hyperplanes identified by the higher cycles, and then sum them up to signs. The relative signs determines locally convex or concave embedding of the polyhedra sharing the edge dual to the higher cycle. Let us illustrate this more general situation in the case of a 4-cycle.

\begin{figure}[H]
\centering
\includegraphics[width=5cm]{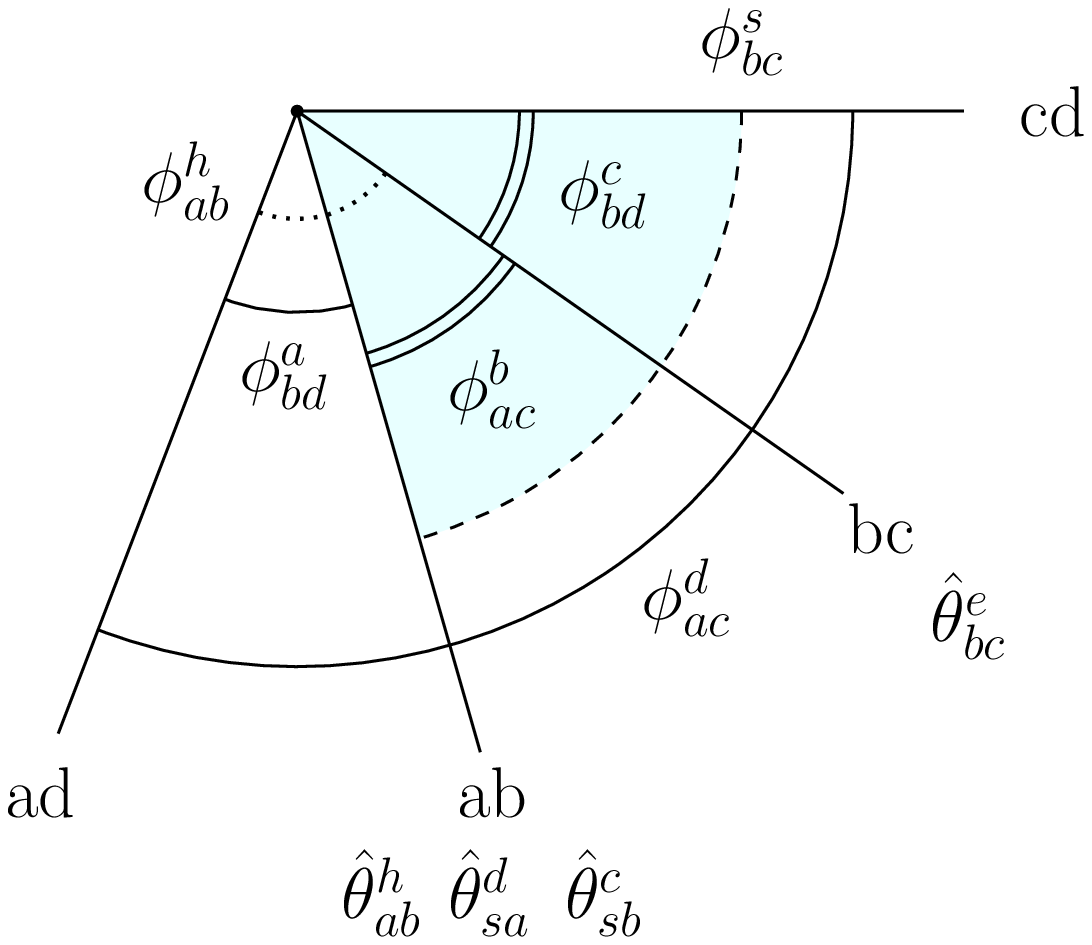}\hspace{2cm}\includegraphics[width=4.5cm]{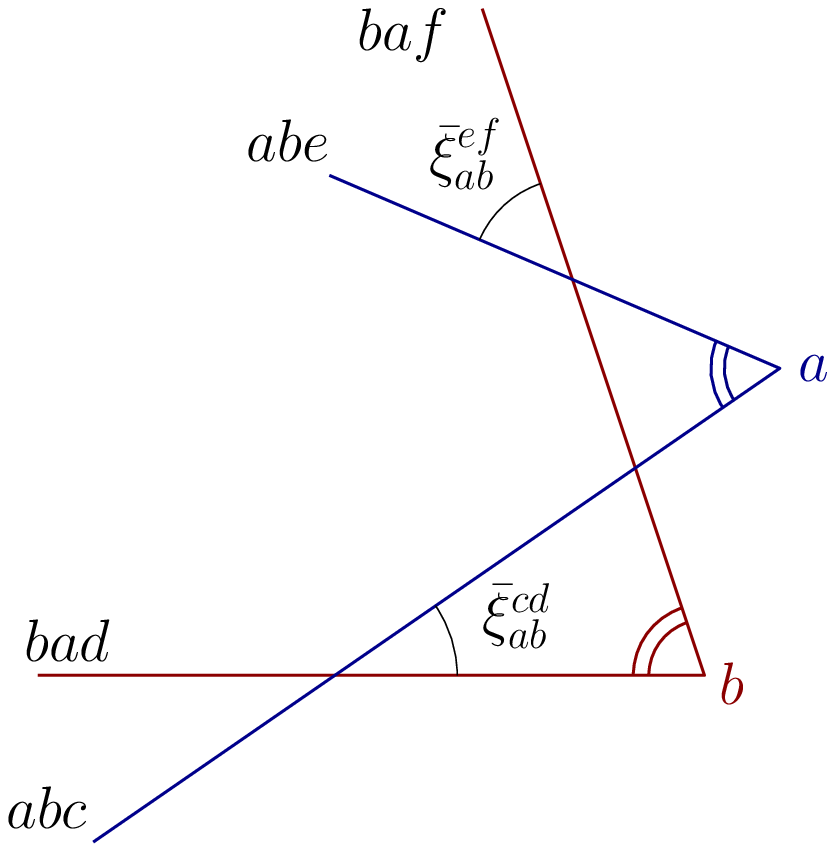}
\caption{\small{Left panel: \emph{Use of spherical cosine laws for a 4-cycle, here labelled by $(abcd)$. The 4-cycle identifies an auxiliary hyperplane $s=(ab,cd)$ colored in blue. If $bc$ belongs also to a 3-cycle, say with a node $e$, the dashed 3d dihedral angle can be determined by the boundary data and the 4d angle computed using the 3-cycle data. The result is the identification of two 4d angles between $a$ or $b$ and $s$. } \\
Right panel: \emph{Visual proof that edge-independence of the twist angles \Ref{twist4} implies matching of the 2d angles:
when $\xg_{ab}^{cd}=\xg_{ab}^{ef}$, the 2d angles in red and blue coincide, by Thales' theorem. Unlike in the 3-cycle case of Fig.~\ref{fig:xg}, the third nodes $c$ and $e$ need not coincide with $d$ and $f$. As a consequence, the lines drawn in red and blue are not necessarily edges of the polytope.} }
\label{fig:genSCL}
}
\end{figure}
Consider a face $ab$ in a 4-cycle ${\cal C}=abcd$. With the help of Fig.~\ref{fig:genSCL}, left panel, we see that 
if we know all six 3d angles, we can use the spherical cosine laws to define three different 4d angles at $ab$:
\be
\cos\hth_{ab}^{h}:=\frac{\cos\phi_{ab}^{h}+\cos\phi^{a}_{bd}\cos\phi_{ac}^{b}}{\sin\phi^{a}_{bd}\sin\phi_{ac}^{b}},
\ee
\begin{equation}\label{4daux}
\cos\hth_{sb}^{c}:=\frac{\cos\phi_{bd}^{c}+\cos\phi_{ac}^{b}\cos\phi^{s}_{bc}}{\sin\phi_{ac}^{b}\sin\phi^{s}_{bc}},
\qquad 
\cos \hth_{sa}^{d}:=\frac{\cos\phi_{ac}^{d}+\cos\phi_{bd}^{a} \cos\phi^{s}_{bc}}{\sin\phi_{bd}^{a}\sin\phi^{s}_{bc}}.
\end{equation}
The first option uses the left-most triple of 3d angles, and defines an angle between $a$ and $b$ constructed in reference to an auxiliary hyperplane $h$ spanned by the faces $ad$ and $bc$. 
The second option uses the dashed 3d angle and the two marked by a single line. It defines a 4d angle between $b$ and the auxiliary hyperplane $s$ spanned by the faces $ab$ and $cd$.
The third uses the right-most triple of 3d angles, the dashed one and the two marked by a double line. It defines a 4d angle between $a$ and the same hyperplane $s$.
These three definitions provide a local flat embedding of three hyperplanes $a$, $b$ and $s$.
By construction, the three hyperplanes share a common face $ab$, and the three angles are related by
\be\label{localemb}
\cos(\hth^h_{ab}-\hth^c_{sb}) = -\cos\hth^d_{sa}.
\ee
It follows that
\be\label{matchingconvex}
\hth^h_{ab} - \hth^c_{sb} = \eta_s\hth^d_{sa} +\pi, \qquad \eta_s=\pm 1.
\ee
The sign of $\eta_s$ can be understood as a convex or concave relative embedding of the hyperplanes. 
The  $\pi$ here is to take into account our convention of always using external dihedral angles.
We stress that this formula holds always, be the configuration Euclidean or Lorentzian, or even mixed in same cases.

There is a catch, however: the boundary data do \emph{not} determine all 3d angles needed. The dotted and dashed ones in the figure are not given,
because their relative faces don't belong to the same polyhedron. Therefore we do not have direct access to the 4d angles just defined.
It is however possible to obtain them using the boundary data and 4d angles previously determined using 3-cycles. For instance if $ab$ belongs to a 3-cycle with a node $f$, we can use that 3-cycle to determine the dihedral angle $\hth^f_{ab}$. If $ab$ does not belong to a 3-cycle, but say $bc$ does, with a node $e$, then we can use $\hth^e_{bc}$ to determine $\phi^{s}_{bc}$ via
\begin{equation}\label{faisola}
\cos\phi^{s}_{bc} = \cos\hth_{bc}^{e}\sin\phi_{ac}^{b}\sin\phi_{bd}^{c}-\cos\phi_{ac}^{b}\cos\phi_{bd}^{c}.
\end{equation}
This quantity allows us to know the two angles \Ref{4daux}. At this point, we can use \Ref{matchingconvex} as a definition of $\hth^h_{ab}$. This procedure leaves $\eta_s$ undetermined: we have no way to fix from the boundary data the local convex or concave embedding for a 4d angle associated to a face that does not belong to any 3-cycle. There is also freedom left in the sector of the data: looking at the 3d angles entering the definitions, we see that  $\hth^c_{sb}$ is in the same sector (Euclidean or Lorentzian) of $\hth^e_{bc}$, but the sector of $\hth^d_{sa}$ is independent.

Once again, the 4d angles and hyperplanes defined are all edge-dependent: choosing different cycles containing $ab$, we get different values. The subset of data that we want to characterise are those that identify a unique 4d angle on $ab$, namely those that give the same value 
\be\label{sind}
\hth_{ab}:=\hth^c_{sb} + \eta_s\hth^d_{sa}=\hth^{c'}_{s'b} + \eta_{s'}\hth^{d'}_{s'a},
\ee 
independently of the 4-cycle considered. Furthermore, if there is also a 3-cycle containing $ab$, say with node $f$, this can also be used to compute a 4d angle $\hth^f_{ab}$ associated with the same face, and we must also require
\be\label{3=4+4}
\hth^f_{ab} \equiv \hth^c_{sb} + \eta_s\hth^d_{sa} +\pi = \hth_{ab}.
\ee
In this case, the relative embedding described by $\eta_s$ is fixed. Notice that the edge-independence we are looking for, \Ref{sind} and/or \Ref{3=4+4}, does not extend to the auxiliary angles between polyhedra and $s$. These are free to vary with $s$, only their sum and/or difference must not.
Proceeding as before, one can derive this edge independence from 2d angle matching conditions. Notice in fact that \Ref{2dSCL} only require a 3-valent node $a$ of the graph, irrespective of the size of the cycles passing through it.
The general angle matching conditions take the form
\be
\a^{ab}_{cd} = \a^{ba}_{c'd'},
\ee
since for general cycles $b$ has no face in common with $c$ and $d$, but rather with $c'$ and $d'$, and vice versa for $a$.\footnote{
Notice that with this procedure one can associate to a face also a spurious 4d angle that corresponds to a dependent 4-cycle, namely a 4-cycle that can be obtained composing 3-cycles. In this case, the angle-matching conditions guarantee that this spurious angle matches the edge-independent dihedral angle.
}
A similar reasoning allows to define edge-dependent 4d dihedral angles for faces whose independent cycles are arbitrarily large, in terms of boundary data and previously determined 4d dihedral angles. Iterating use of \Ref{2dSCL} one identifies
the relevant angle-matching conditions in each case. Listing all the required angle matching conditions for the edge-independence of a given face, with its edges belonging to arbitrary cycles, may be cumbersome. Fortunately, it is also unnecessary, since all we need is a global characterization. Our investigations lead us to conclude that \Ref{19} holds in general, with the edges corresponding to arbitrary cycles:
\be\label{at}
\{\a^{ab}_{cd}\} = \{\a^{ba}_{c'd'}\} \qquad \Leftrightarrow\qquad \{\hth^{\cal C}_{ab}\}=\{\hth_{ab}\}.
\ee
We have however no formal proof to offer. The main geometric interpretation pointed out in this paper, that conformal twisted geometries are the general type of data admitting multiple critical points, relies on the validity of \Ref{at}.
Notice that the 2d angles are defined in a purely combinatorial way: they can belong to a vertex which is not part of the boundary of the polyhedra. Therefore the left-hand side is clearly a redundant requirement, as it would be enough to restrict attention to those that do belong to the boundary of a polyhedra. While redundant, this characterization is simpler. On the right-hand side, it is sufficient to focus on independent cycles.

To get some intuition about these auxiliary hyperplanes, it is useful to consider an example in one less dimension, given by a polyhedron with the combinatorial shape of a house (a tetragonal-based pyramid stacked on top of a cuboid). The dihedral angle between one wall and its adjacent part of the roof cannot be computed using the standard spherical cosine laws, because there is no other face touching both. The solution is to introduce an auxiliary plane, the (tetragonal) attic of the house. The spherical cosine laws permit to compute the angles between this auxiliary plane and the wall or the roof. The desired angle is then the sum of the two (or the difference in case the polyhedron was concave). If the polyhedron is modified in such a way that there is a face touching both one wall and its adjacent part of the roof, then it is possible to compute the dihedral angle using the spherical cosine laws with this face. If the angle-matching conditions are satisfied, the two computations must give the same result, and this in particular fixes the relative convex or concave embedding. We will see this example in full details in Section~\ref{sec:example3} below.

\medskip

There is an alternative, equivalent way to characterize the angle-matched data:
instead of the matching of the 2d angles, we can look at the relative angles between the edge vectors, and see whether they are 
constant all along the face. 
We refer to these as \emph{edge twist angles}. To define them in the way most relevant to the saddle point analysis, we proceed as follows.
For two adjacent polyhedra $a$ and $b$  we can always consider the gauge in which $\vec n_{ab}=-\vec n_{ba}$. In the case of a 3-cycle with $c$, the triple $abc$ defines an edge, to which we can associate two distinct vectors, one computed in the frame of $a$, and one in the frame of $b$. In the chosen gauge, we define the angle between these two vectors as
\be\label{defxg}
\cos\xg^c_{ab}:= \frac{\vec{n}_{bc}\times\vec{n}_{ba}\cdot\vec{n}_{ab}\times\vec{n}_{ac}}{\norm{\vec{n}_{bc}\times\vec{n}_{ba}} \, \norm{\vec{n}_{ab}\times\vec{n}_{ac}}}, \qquad 
\sin\xg^c_{ab}:= - \frac{\vec{n}_{bc}\cdot\vec{n}_{ab}\times\vec{n}_{ac}}{\norm{\vec{n}_{bc}\times\vec{n}_{ba}} \, \norm{\vec{n}_{ab}\times\vec{n}_{ac}}}, \qquad
\xg^c_{ab}\in [0,2\pi). 
\ee
This edge twist angle is always real, however it is not gauge-invariant. It is invariant under the remaining gauge freedom of rotations at the node $b$ but not at the node $a$.\footnote{It should not be confused with the gauge-invariant twist angle used in the twisted geometry literature, 
see e.g. \cite{Freidel:2013fia,IoMiklos},
$$
\cos\bar{\xi}^c_{ab}:= \frac{\vec{n}_{bc}\times\vec{n}_{ba}\cdot g_{ab}\triangleright (\vec{n}_{ab}\times\vec{n}_{ac})}{\norm{\vec{n}_{bc}\times\vec{n}_{ba}} \, \norm{\vec{n}_{ab}\times\vec{n}_{ac}}}, \qquad
\sin\bar{\xi}^c_{ab}:= - \frac{\vec{n}_{bc}\cdot\vec{n}_{ba}\times g_{ab}\triangleright \vec{n}_{ac}}{\norm{\vec{n}_{bc}\times\vec{n}_{ba}} \, \norm{\vec{n}_{ab}\times\vec{n}_{ac}}}, \qquad  
\xg^c_{ab}\in [0,2\pi). 
$$
This is independent of the orientation thanks to the $\SU(2)$ parallel transport $g_{ab}$ along the link connecting the two tetrahedra.
To distinguish the two but hint at their common structure, we used $\xg$ as opposed to $\bar{\xi}$. The 2d angle-matching conditions also follows from  the edge-independence of the gauge-invariant twist angles, the only reason to work with \Ref{defxg} is that those are the ones that appear in the saddle point analysis below.
} 
If  $\xg^c_{ab}=\xg^d_{ab}$, the 2d angle between the two edges $abc$ and $abd$ computed in the frame of $a$ matches the one computed in the frame of $b$, see the right panel of Fig.~\ref{fig:xg} for a visual proof. 
Notice that for a polygonal face with $n$ sides, we need $n$ edge independent conditions for all 2d angles to match.
Edge-independence of  twist angles thus implies matching of  2d angles, and in turn, edge-independence of  4d angles. When this happens, there exist a relative orientation of two adjacent polyhedra with all  twist angles vanishing.
For the edges dual to 4-cycles, 
we look at the scalar products between the two vectors associated to the edge shared by $a$ and $b$, identified by the connected pair $(c,d)$,
\begin{equation}\label{twist4}
\cos\xg_{ab}^{cd}:=\frac{
\left(\vec{n}_{ac}\times\vec{n}_{ab}\right)\cdot\left(\vec{n}_{ba}\times\vec{n}_{bd}\right)}
    {|\vec{n}_{ab}\times\vec{n}_{ac}|  |\vec{n}_{ba}\times\vec{n}_{bd}|},
\qquad \sin\xg_{ab}^{cd}:= - \frac{\vec{n}_{ac}\cdot\vec{n}_{ba}\times\vec{n}_{bd}}
    {|\vec{n}_{ab}\times\vec{n}_{ac}|  |\vec{n}_{ba}\times\vec{n}_{bd}|},
    \qquad \xg_{ab}^{cd}\in[0,2\pi).
\end{equation} 
Proceeding in the same way, one can define edge-dependent 4d dihedral and twist angles for larger cycle, and deduce their edge-independence from  2d angle-matching conditions. 
When all edge twist angles match, the 2d angles at the face also match:
\be\label{axi}
\a^{ab}_{cd} = \a^{ba}_{c'd'} \qquad \Leftrightarrow\qquad \xg^{\cal C}_{ab} = \xg_{ab} \qquad \forall \ {\rm independent \ cycles}.
\ee
This equivalence is completely local, valid face by face. It may involve redundant quantities: for instance some of the edges or the vertices may not be in the boundary of the polyhedra, but lie outside of the polyhedra, where non-adjacent faces intersect. We refer to edges and vertices defined at the generic intersection of faces in $\R^3$, not in the boundary of the polyhedra, as `virtual ones'.

The resulting geometric picture is the following. For every (closed) twisted geometry, we can identify a unique convex polyhedron at each node, and define edge-dependent 4d dihedral angles among the polyhedra, and edge twist angles. 
The twist angle varies in general from one edge to the next, forbidding the reconstruction of a unique shape for the face $ab$.
But, if the data are such that the twist angles at a face are all the same, the valence and 2d angles of the face $ab$ are uniquely determined. The only possible mismatch left is that of (area-preserving, since the spins are uniquely assigned) conformal transformations. 
In this case, also the 4d angles defined by the spherical cosine laws are edge independent.

\section{Saddle point analysis}

We are interested in the homogeneous large spin behaviour of \Ref{Ac}, namely 
\be
j_{ab}\mapsto \l j_{ab}, \qquad \l\rightarrow \infty.
\ee
The action \Ref{action} is linear in the spins, and therefore the asymptotics of the vertex amplitude can be studied with saddle point techniques. The critical points, or saddles, are those where the gradient vanishes. Among them, those maximizing the real part of the action give the dominant contributions to the asymptotic behaviour of the integral, and we restrict attention to these. With a little abuse of language, we will still call them critical points, even though strictly speaking they should be called absolute critical points.
Accordingly, the analysis of \cite{BarrettLorAsymp} obtains three sets of equations. Requiring the absolute maximum of the real part of the action, one finds
\begin{align}
\label{eq:maxRe}
\ket{\z_{ab}}= e^{i\upsilon_{ab}} \f{h_a^\dagger \ket{z_{ab}}}{\norm{h_a^\dagger z_{ab}}}  &&& |\z_{ba} ] = e^{i\upsilon_{ba}} \f{h_b^\dagger \ket{z_{ab}}}{\norm{h_b^\dagger z_{ab}}} .
\end{align}
Here $\upsilon_{ab}$ are arbitrary phases, and we used the fact that the boundary spinors have unit norm.
These two equations can be combined to give
\begin{subequations}\label{eq:crit1e2}
\begin{equation}
\label{eq:crit1}
(h_{a}^{\dagger})^{-1}\ket{\z_{ab}} = \frac{\norm{h_{b}^{\dagger}z_{ab}}}{\norm{h_{a}^{\dagger}z_{ab}}} e^{i(\upsilon_{ab}-\upsilon_{ba})} (h_{b}^{\dagger})^{-1}|\z_{ba}].
\end{equation}
From the vanishing of the (imaginary part of the) spinorial gradient, on-shell of \Ref{eq:maxRe}, one finds
\begin{equation}
\label{eq:crit2}
h_{a}\ket{\z_{ab}}= \frac{\norm{h_{a}^{\dagger}z_{ab}}}{\norm{h_{b}^{\dagger}z_{ab}}} e^{i(\upsilon_{ab}-\upsilon_{ba})}h_{b}|{\z_{ba}}].
\end{equation}
\end{subequations}
From the vanishing of the gradient in the group variables, on-shell of \Ref{eq:crit1e2}, one finds the closure conditions \Ref{Clos}. These equations were derived in \cite{BarrettLorAsymp} for the 4-simplex graph, but it is immediate to see that the same equations arise from \Ref{action} for any vertex graph.
To proceed,  we can further combine (the hermitian conjugate of) \Ref{eq:crit1} with \Ref{eq:crit2} to derive
\begin{subequations}\label{C2}\begin{align}\label{Cvec}
& \bra{\z_{ab}} h^{-1}_a \vec\s h_a\ket{\z_{ab}} = [\z_{ba}| h_b^{-1}\vec \s h_b |\z_{ba}], \\
& [\z_{ba}| h^{-1}_b h_a\ket{\z_{ab}} = \frac{\norm{h_{a}^{\dagger}z_{ab}}}{\norm{h_{b}^{\dagger}z_{ab}}} e^{i(\upsilon_{ab}-\upsilon_{ba})}.\label{Cphases}
\end{align}\end{subequations}
Equations \Ref{Cvec} turn out to be enough to fully determine the group elements at the critical point. The equations \Ref{Cphases} can then be used to determine the norm ratios and the phase differences $\upsilon_{ab}-\upsilon_{ba}$. 
As for the sums $\upsilon_{ab}+\upsilon_{ba}$, these are part of the irrelevant choice of section for the dummy spinors $z_{ab}$ and therefore are undetermined.
Finally \Ref{eq:maxRe} determine the dummy spinors $z_{ab}$, up to their arbitrariness under complex rescalings.

The system of non-linear critical point equations for the integration variables $z_{ab}$ and  $h_a$, made of  \Ref{Clos}, \Ref{eq:maxRe} and \Ref{C2}, has no solution for general boundary data. 
This is manifest from the closure conditions \Ref{Clos}, which don't involve the integration variables at all, but are directly a restriction on the boundary data. Satisfying the closure conditions \Ref{Clos} is not enough: the equations \Ref{C2}, or equivalently \Ref{eq:crit1e2}, impose additional restrictions on the boundary data, known from \cite{BarrettLorAsymp} for a 4-simplex and to be revisited in this Section. Critical points exist only for special boundary data, corresponding to a strict subset of twisted geometries.

If a critical point {\small (c)} is found, and the Hessian $H^{\scr (c)}$ is not degenerate, the saddle point approximation of \Ref{Ac} gives $6(N-1)+2L$ Gaussian integrals to evaluate. 
From the general formula of the saddle point approximation we have the leading order asymptotics  
\begin{equation}
\label{eq:asymp}
A_\G(\l j_{ab},\vec n_{ab},-\vec n_{ba}) =
\f{(2\pi)^{3(N-1)} 2^{2L+N-1} \J}{\l^{3(N-1)}} 
\sum_{\scr c} \f{\Om^{\scr (c)}}{\sqrt{\det{-H^{\scr (c)}}}} \exp S^{\scr (c)} +O(\l^{-3N+2}).
\end{equation}
The sum is over distinct critical points, and there is a factor $2^{N-1}$ because the action is even in all group elements, so if $h^{\scr (c)}_a$ is a critical point, also $-h^{\scr (c)}_a$ is. 
We have used the short-hand notation \be
\label{bigJ}
\J:= \prod_{(ab)}j_{ab},
\ee
and $\Om^{\scr (c)}$ is Barrett's notation for the value of the spinorial measure at the critical point.

The value of the action at the critical points is 
\begin{equation}
S^{\scr (c)}=
-2i \l \sum_{(ab)} j_{ab} \left(\upsilon_{ab}^{\scr (c)}-\upsilon_{ba}^{\scr (c)}\right) 
+ \gamma j_{ab}\log\frac{\norm{h_{a}^{\scr (c)\dagger}z_{ab}^{\scr (c)}}}{\norm{h_{b}^{\scr (c)\dagger}z_{ab}^{\scr (c)}}}. \label{eq:actioncp}
\end{equation}
Both terms of this on-shell action can be determined from the LHS of \Ref{Cphases}. We notice also that the explicit solution for the dummy spinors from \Ref{eq:maxRe} is not needed. They are needed however to compute the Hessian determinant. This can be done choosing explicitly a section of the tautological bundle, following the calculations presented for the 4-simplex case in \cite{Dona:2019dkf}.
This dependence on the choice of section is exactly cancelled by the measure factor
$\Om^{\scr (c)}$.  

\subsection{From spinors to 3d vectors}
\label{secSpinToVec}
The key equations to solve for the saddle point analysis are \Ref{Cvec}. These can be rewritten in terms of the vectors only, without reference to the spinors, as
\begin{equation}
\label{maggica}
H_a \vec n_{ab} = - H_b \vec n_{ba},
\end{equation}
where $H$ is the 3-dimensional, non-unitary representation of $h\in\SL(2,\C)$, defined as
\be
H\vec n:=\bra{\z} h^{-1} \vec\s h\ket{\z}, \qquad \vec n = -\bra{\z}\vec \s \ket{\z}.
\ee 
This mapping plays an important role in the analysis, and it is useful to give more details about it.
We use the polar decomposition $h=bu$, where $u\in\SU(2)$ and  $b$ is a boost, with rapidity $r$:
\be
b := e^{-r\vec v\cdot\f{\vec \s}2},
\qquad r\in\R, \qquad \vec v^2=1.
\ee
From elementary properties of the Pauli matrices (see Appendix~\ref{app:pauli} for details) we derive
\be\label{3dHirrep}
\bra{\z} h^{-1} \vec\s h\ket{\z} = \cosh r \, U \vec n -i\sinh r\, \vec v\times U \vec n + (1-\cosh r)(\vec v\cdot U \vec n) \vec v,
\ee
where $U$ is the vectorial representation of $u$. This is the formula defining \Ref{maggica} explicitly.
It can be recognized as the $\SL(2,\C)$ transformation of the self-dual part of a bivector with vanishing magnetic part. The latter property follows from the reality of $\vec n$. This special class of bivectors 
is singled out by the use of the canonical basis in the vertex amplitude, which selects the SU(2) subgroup stabilizing the  time direction $t^I=(1,0,0,0)$. Introducing the notation $n^I:=(0,\vec n)$, the  bivector can be written $B^t_{IJ}:=\eps_{IJKL}t^K n^L$. 

Notice that $\pm h_a$ give the same $H_a$. 
The action \Ref{action} is even in the group elements $h_a$, therefore we can replace in the saddle point analysis \Ref{Cvec} with \Ref{maggica} without loss of generality. Each solution obtained for $H_a$ will induce two solutions for $h_a$, distinguished by the sign. 

Written in form \Ref{maggica}, one class of solutions to the equations becomes manifest. If the boundary data satisfy
\be
\vec n_{ab}=-\vec n_{ba},
\ee
then \Ref{maggica} is solved with $H_a=\Id$ for all nodes. This leads to two solutions for $h_a=\pm\Id$, but we still refer to it as a single critical point, discounting the $\Z_2$ symmetry. More generally, if the boundary data satisfy \Ref{orientations} for some rotations $R_a$, then \Ref{maggica} is solved with $H_a=R_a\in\SO(3)$. This shows that vector geometries can be immediately identified as boundary data with a critical point, for any graph.
Finding if there are more solutions, or solutions with $H_a\notin \SO(3)$, requires all the work.

\section{4-simplex asymptotics revisited}
\label{sec:4simplex}
The reader familiar with the results of  \cite{BarrettLorAsymp} can recognize from \Ref{maggica} the main classes of critical configurations for the 4-simplex: vector and Euclidean Regge geometries, for which $H_a\in\SO(3)$, and Lorentzian Regge geometries for which  $H_a$ have non-vanishing boosts.
The novelty of our procedure lies in the way we solve these equations. 

The 4-simplex amplitude is associated to $K_5$, the complete graph with 5 nodes. We pick the node 1 as the one without group integration. Then, we perform a partial gauge fixing at the 4 remaining nodes: we require the normal at the triangles shared with the tetrahedron 1 to be anti-parallel to the corresponding normal of the tetrahedron 1, namely 
\be\label{partialgauge}
\vec n_{1a} = -\vec n_{a1}.
\ee
This condition involves only one normal per tetrahedron, and can then be realized for \emph{any} set of boundary data (even those not satisfying the closure conditions). It fixes 2 gauge freedoms per node, leaving one gauge freedom to perform rotations in the plane of the triangle $1a$. 
If desired, the latter can be fixed  for instance aligning one edge of the triangle $1a$ in the tetrahedron $a$ with the corresponding edge in the tetrahedron $1$; the other two edges will be unaligned in general, because of the shape mismatch of generic data. 

The advantage of the partial gauge \Ref{partialgauge} is to make the direction of the critical group elements straightforward. 
Taking $b= 1$ in  \Ref{maggica}, we have
\be\label{crit1a}
\vec n_{1a} = -H_a\vec n_{a1} = H_a\vec n_{1a}.
\ee
This equation implies that the critical group elements contain a rotation and a boost along the same direction determined by the boundary data between $a$ and the root 1. $\SL(2,\C)$ transformations of this type are called four-screws, and can be conveniently parametrized in the fundamental representation as 
\be\label{polifemo}
h_a = \pm\exp \Big(\f i2\om_a \vec n_{a1}\cdot\vec\s\Big),
\qquad 
\om_a:=\a_a + i \b_a, \qquad (\a_a,\b_a)\in[-\pi,\pi)\times\R.
\ee
To find the complex angle $\om_a$, we insert the special form \Ref{polifemo} in \Ref{maggica}, using equations with
both $a$ and $b$ different from $1$. After some simple algebra we find
\begin{align}
\label{critDexpl}
&\cos\om_a\ \vec n_{ab} +\sin\om_a\ \vec{n}_{a1} \times \vec n_{ab} + (1-\cos\om_a) (\vec n_{a1}\cdot\vec n_{ab})\ \vec n_{a1} \nn\\
&\quad =-\cos\om_b\ \vec n_{ba} -\sin\om_b\ \vec{n}_{b1} \times \vec n_{ba} - (1-\cos\om_b) (\vec n_{b1}\cdot\vec n_{ba})\ \vec n_{b1}.
\end{align}

It is at this point that we exclude from our analysis configurations with coplanar normals at the same node.  For the 4-simplex, these have tetrahedra with zero volume, and such degenerate configurations are excluded also from the standard analysis based on the bivector reconstruction theorem.
This vectorial equation can be decomposed projecting along the basis provided by $\vec{n}_{b1}$, $\vec{n}_{a1}$
and $\vec n_{b1}\times \vec n_{a1}$. After simple trigonometry, we obtain the following scalar equations,
\begin{subequations}\label{capitani2}
\begin{align}
\cos(\omega_{a}+\xg^b_{a1})&=\cos\hth_{a1}^b, \label{Totti1}\\
\cos(\omega_{b}+\xg^a_{b1})&=\cos\hth_{b1}^a, \label{Totti2} \\
\sin\phi^a_{b1} \sin(\omega_{a}+\xg^b_{a1})&= \sin\phi^b_{a1} \sin(\omega_{b}+\xg^a_{1b}),
\label{DeRossi2}
\end{align}\end{subequations}
where $\hth_{a1}^b$, {$\hth_{b1}^a$} and $\xg_{a1}^b$ are the functions of the boundary normals defined in \Ref{defhth} and \Ref{defxg}.
Thanks to a natural choice of basis to project the vectorial equations, the (edge-dependent) twist and 4d angles appear naturally from the critical point equations! All expressions are well defined since $\sin\phi^a_{bc}\neq 0$ for all combinations of nodes, having excluded the degenerate configurations. 

As we vary $a$ and $b$, the cosine equations determine the complex angles $\om_a$ in terms of the 3d normals, giving two solutions at most. The sine equations involve $\om_a$ at different nodes and can introduce global conditions restricting the space of solutions.
To proceed, we split \Ref{capitani2} into real and imaginary parts, in terms of the angles $\a_a$ and boosts $\b_a$:
\begin{subequations}
\label{scudetto}
\begin{align}
\cosh\beta_{a}\cos(\alpha_{a}+\xg^b_{a1})&=\cos\hth_{a1}^b, \label{Cafu}\\
\sinh\beta_{a}\sin(\alpha_{a}+\xg^b_{a1})&=0, \label{Zanetti}\\
\sin\phi^a_{b1}\sin(\alpha_{a}+\xg^b_{a1})\cosh \b_a &= \sin\phi^b_{a1} \sin(\alpha_{b}+\xg^a_{b1})\cosh \b_b, \label{Delvecchio}\\
\sin\phi^a_{b1} \cos(\alpha_{a}+\xg^b_{a1})\sinh \b_a&= \sin\phi^b_{a1} \cos(\alpha_{b}+\xg^a_{b1})\sinh \b_b, \label{Montella}
\end{align}\end{subequations}
to be valid for all $a$ and $b$.

The solutions to \Ref{scudetto} can  be classified according to whether \Ref{Zanetti} is solved by $\b_a=0$, or by $\a_{a}+\xg^b_{a1}=0$ modulo $\pi$. 
It follows from \Ref{Delvecchio} and \Ref{Montella} that whichever case is chosen for the first $a$, it has to be chosen for all remaining nodes as well, since they are all connected to 1 and $a$. In the first case, all critical group elements are pure rotations, that is $H_a\equiv R_a\in\SO(3)$. From \Ref{maggica}, we see that such critical points exist only for boundary data satisfying the orientation conditions
\be\label{vector}
R_a \vec{n}_{ab}=-R_b\vec{n}_{ba}, \qquad R_a\in\SO(3).
\ee
This implies that the 3d dihedral angles satisfy the Euclidean condition $\hth_{a1}^b\equiv \th_{a1}^b\in[0,\pi)$ (see Appendix~\ref{app:vector} for an explicit proof). Accordingly, we will refer to these solutions as Euclidean. In the second case, $\b_a\neq 0$ and no critical group element is a pure rotation, that is $H_a\notin\SO(3)$ for $a\neq 1$. 
The presence of a boost in $H_a$ means that the 3d angles $\phi^b_{a1}$ violate the Euclidean conditions, and 
$\hth^b_{a1}=i\th_{a1}^b$ up to a possible real part $\pi$. We will refer to these solutions as Lorentzian.

\subsection{Euclidean solutions}
Taking $\b_a=0$ $\forall a$ solves \Ref{Zanetti} and \Ref{Montella}. This leaves us with \Ref{Cafu} and \Ref{Delvecchio}, which read
\begin{align}
\label{Antonioli}
& \cos(\alpha_{a}+\xg^b_{a1})=\cos\th_{a1}^b,\\
\label{DDR3}
& \sin\phi^a_{b1} \sin(\a_{a}+\xg^b_{a1})= \sin\phi^b_{a1} \sin(\a_{b}+\xg^a_{b1}).
\end{align}
The first equation requires
\be\label{a1}
\alpha_{a} = \eps_a^b \th_{a1}^b - \xg^b_{a1}, \qquad \eps_a^b=\pm 1
\ee
to hold for any choice of $b$ different from 1 and $a$, and to be solved modulo $2\pi$ within the interval $[-\pi,\pi)$. 
These gives two candidate solutions per node different from 1, namely $2^4$. But the value of the RHS depends a priori on the choice of $b$, hence it is not obvious that the solutions are admissable. It turns out that \Ref{vector} guarantees that the plus sign in \Ref{a1} is \emph{always} a solution. The simplest way to prove this is to notice that for data satisfying \Ref{vector}, the partial gauge-fixing \Ref{partialgauge} used so far can be completed to have all normals pairwise-opposite, and not just those at the faces of the first tetrahedron. Namely, there exist a complete gauge-fixing such that
\be\label{twistedspike}
\vec{n}_{ab}=-\vec{n}_{ba} \quad \forall \, a,b.
\ee
In this gauge we have $\xg^b_{a1} \equiv  \th_{a1}^b$ for all $a$ and $b$. This can be immediately seen expanding 
the numerator of \Ref{defxg}, then using \Ref{twistedspike} and the definition of the 3d angles in terms of the normals \Ref{defphi}, to recover \Ref{defhth}.
Since $\xg^b_{a1}$ is affected only by rotations at $a$, it follows that  the most general configuration satisfying \Ref{vector} has
\be\label{varphi}
\th_{a1}^b = \xg^b_{a1} + \nu_a,
\ee
for some angles $\nu_a$ determined by the orientation chosen. Then, $\a_a=\th_{a1}^b - \xg^b_{a1}\equiv\nu_a$ is manifestly $b$-invariant and solves \Ref{Antonioli}. It also solves identically the remaining equation \Ref{DDR3}, which become the spherical sine laws. 
We have thus found a solution to all critical point equations, and this is in general the only solution, since  the minus sign in \Ref{a1} is ruled out by the  $b$-dependent RHS. The corresponding critical group elements $h^{\scr (c)}_a$ are given by \Ref{polifemo} with $\om_a^{\scr (c)}=\nu_a$. In other words, we have a unique critical configuration of $R^{\scr (c)}_a=D^{(1)}(h^{\scr (c)}_a)\in\SO(3)$, to which there correspond $2^4$ critical configurations $h_a^{\scr (c)}\in\SU(2)$. Notice that in the gauge \Ref{twistedspike}, $h_a^{\scr (c)}=\pm\Id$ $\forall a$.

The boundary data satisfying the closure and orientation conditions, given respectively by \Ref{Clos} and \Ref{vector}, are the vector geometries, and we have just recovered that such data admit a unique critical configuration  (up to spin lifts). Here we used equations that assumed non-coplanar normals, but vector geometries can be shown to be always solutions regardless of this assumption, as discussed in Section~\ref{secSpinToVec}. A direct counting of conditions shows that the space of vector geometries for this graph has 15 dimensions, up to rotations.

Taking the minus sign in \Ref{a1} becomes possible only if the right-hand side is independent of $b$. 
As discussed in Section~\ref{SecSCL}, we know that this happens for the special subset of data with matching 2d angles, for which
\be\label{edgeindep1}
\th_{a1}^b=\th_{a1}, \qquad \xg^b_{a1}=\xg_{a1},
\ee
both independent of the choice of edge, provided we also fix $\eps_a^b=\eps_a$. 
The conditions \Ref{edgeindep1} involve only the faces shared between the tetrahedron $1$ and $a$. 
This is a consequence of our choice of solving the equations starting from the node 1.
But we could have repeated the analysis taking any other node as reference, therefore the boundary data that admit additional critical points must satisfy  angle-matching on all faces. 
Since the faces are triangles with a well-defined area, this condition imposes full shape-matching of the faces. 

Plugging these solutions in the sine equations \Ref{DDR3}, we have
\begin{equation}
\sin\phi^a_{b1} \sin(\eps_a \theta_{a1})= \sin\phi^b_{a1} \sin(\eps_b \theta_{b1}).  
\end{equation}
These are satisfied identically as the spherical sine laws, and 
from the parity of the functions and positivity of all quantities involved, we conclude that $\eps_a\equiv \eps_b$. Therefore we only have two distinct solutions 
$\om^{\scr (c)}_a=\eps\, \th_{a1} - \xg_{a1}\in\R$ which differ by a global sign $\eps=\pm1$. 

We have found that the subset of boundary data describing a Euclidean 4-simplex admits two distinct critical points, determined up to a global sign by the  dihedral angles $\th_{ab}$ of the 4-simplex. This is the same set obtained in \cite{BarrettLorAsymp} using the bivector reconstruction theorem. Our construction can not exclude per se the presence of special configurations for which the minus sign is an acceptable solution also if the shapes don't match, but we know from the analysis of \cite{BarrettLorAsymp} that these don't exist. Therefore we can either restrict the validity of our result to generic configurations, or invoke the bivector reconstruction theorem to exclude the existence of special configurations bypassing the angle-matching conditions. In the case of general graphs below, we will only be able to choose the first option.

The critical points for the group elements in the partial gauge used are
\be
h^{\scr (c)}_a = \pm h^{\eps}_a, \qquad h^{\eps}_a:=\exp\Big(\f i2 (\eps\, \th_{a1} - \xg_{a1})  \vec{n}_{a1}\cdot \vec{\s}\Big).
\ee
These values are manifestly orientation-dependent, including  the norm of the rotation $\om^{\scr (c)}_a$: if we rotate a critical configuration of the normals by $R_a$, we obtain again a critical point, but with a different $\om_a^{\scr (c)}$.

Among the geometrically equivalent configurations with angle-matching data, there are two convenient choices to fix the residual gauge freedom. We refer to \cite{Dona:2017dvf} for a more accurate construction and to Fig.~8 of the same paper for a graphical representation. The first gauge choice is the \emph{spike} gauge, characterized by a rotation of the boundary normals so that $\xg^b_{a1}$ vanish for all $a\neq 1$. 
The second choice is the \emph{twisted spike} gauge, where the boundary normals are rotated so that $\xg^b_{a1}=\hth^b_{a1}$, as in  \Ref{twistedspike}. Namely, the edge angles match precisely the 4d dihedral angles. 
This choice allows the extrinsic curvature (the 4d dihedral angle) to be encoded among the intrinsic data (the holonomies and fluxes) in a simple way, similarly to the analysis done in \cite{IoFabio}.
These two gauges provides a useful visualization of the geometry of the problem, and are very convenient in the explicit reconstruction of the data needed in calculations of the amplitudes. We summarize their properties in the following list: 
\smallskip\begin{center}
\begin{tabular}{lllll}
\emph{spike gauge:} & $\xg_{a1}=0$, & $\vec{n}_{a1}=-\vec{n}_{1a}$, 
& $\om^{\scr (c)}_a = \pm\th_{a1}$, & $h_a^{\scr +}=h_a^{{\scr -}\dagger}$ \\ 
&& $\vec{n}_{ab}\times \vec n_{ac}=-\vec{n}_{ba}\times\vec n_{bc}$ &&\\ &&&&\\
\emph{twisted spike gauge:} & $\xg_{a1}=\th_{a1}$, & $\vec{n}_{ab}=-\vec{n}_{ba}$, 
& $\om^{\scr (c)}_a=0,-2\th_{a1}$, & $h_a^{\scr +}=\Id$.
\end{tabular}\end{center}

\medskip

From the boundary data corresponding to a 4-simplex one can also reconstruct its `spacetime' orientation in 4d Euclidean space, namely the individual 4-normals to each tetrahedron, up to a global rotation. 
This reconstruction is described for instance in \cite{Dona:2017dvf}, and produces $\SO(4)$ holonomies that rotate the twisted spike into a 4d object. The critical group element provide also these holonomies, but with a catch: a single set of Euclidean critical group elements is only $\SU(2)$, therefore it would not suffice.\footnote{And for this reason, Euclidean critical points were considered `degenerate' in \cite{BarrettLorAsymp}.} The solution to recover the correct $\SO(4)$ group element is to use both sets of critical points. Explicitly, we first assign a reference 4d normal $t^I=(1,0,0,0)$ to the tetrahedron 1, the node without the integration. 
From the two sets of critical group elements labelled by $h_a^{\scr\pm}$ we define the $\SO(4)$ transformations
\begin{equation}\label{Leuc}
\Lambda_a{}^I{}_J := \f{1}{2} \tr( \s^I h^{\scr+}_a \s_J h_a^{{\scr-}\dagger} )\in\SO(4), 
\end{equation}
where $\sigma_I := (\mathds{1},\vec{\sigma})$, and indices are contracted with the Euclidean metric $\d_{IJ}$. We construct the 4d normal to the tetrahedron $a$ acting with $\Lambda_a$ on the reference vector. In any (partial) gauge of our analysis, 
\be
N^I_a:=\L_a{}^I{}_J t^J = (\cos\th_{a1},\sin\th_{a1}\vec{n}_{a1}).
\ee
If we exchange the two critical points in \Ref{Leuc} we get $\L^{\rm T}=\L^{-1}$, and thus parity-transformed normals and 4-simplex.
It is necessary on the other hand to choose the same sign of the spin lift, to have $\det \L=1$.

Notice that if we use the group elements from a single critical point, we obtain instead
\be
R_a{}^I{}_J := \f{1}{2} \tr( \s^I h^{\eps}_a \s_J h_a^{\eps\dagger} )\in\SO(3),
\ee
which are the rotations preserving the canonical time direction, that is $R_a{}^I{}_J t^J\equiv t^I$. Hence, for vector geometries with a single critical point, the reconstructed 4d normals are all parallel. 
Similarly, for Euclidean 4-simplex data, if one uses a single critical group element and not both as in \Ref{Leuc}, the reconstructed 4d normals are all parallel. 
For this reason, Euclidean 4-simplex as well as vector data are considered `degenerate' in \cite{BarrettEPRasymp,HanZhangLor}. We find this name misleading, since there is a well-defined Euclidean 4-simplex with non-zero 4d volume associated with these data, and we will not use it. We can then keep `degenerate' in reference to boundary data with vanishing 3d volumes.

\subsection{Lorentzian solutions}

In the Lorentzian sector we solve \Ref{Zanetti} taking $\beta_{a}\neq 0$ and $\sin(\alpha_{a}+\xg^b_{a1})=0$, $\forall a$. The latter condition  gives 
\be\label{Zebina}
\alpha_{a} = - \xg^b_{a1} + \Pi_{a1}, \qquad \text{with } \Pi_{a1}=0, \pi.
\ee
These are $2^4$ tentative solutions, but they are admissible only if $\xg^b_{a1}=\xg_{a1}$ is independent of the choice of $b$. As explained earlier, this happens if the boundary data satisfy the angle-matching conditions, implying in turn that also $\hth^b_{a1}=\hth_{a1}$ is independent of the choice of $b$.  In addition, the arbitrariness of choosing the node $1$ as reference imply that $\hth^c_{ab}$ is edge independent. 
Since \Ref{maggica} is now satisfied with $H_a\notin\SO(3)$, the RHS of \Ref{defhth} takes values outside $(-1,1)$.
For such Lorentzian boundary data, we have followed \cite{Barrett:1993db} and defined the 4d dihedral angles $\th_{a1}$ as
\begin{equation} 
\label{boostconv}
\begin{split}
\text{co-chronal (`thick wedge'):}\qquad & \cos\hth^b_{a1} = \cosh\th_{a1}, \phantom{-} \qquad \th_{a1}>0, \\
\text{anti-chronal (`thin wedge'):}\qquad &  \cos\hth^b_{a1} = -\cosh\th_{a1},       \qquad     \th_{a1}<0.
\end{split}
\end{equation}
The choice of sign for $\th_{a1}$ is purely conventional, and it is made to simplify the parametrization of the solutions. Taking \Ref{Zebina}  solves \Ref{Zanetti} and also \Ref{Delvecchio}. This leaves us with \Ref{Cafu} and \Ref{Montella}, which read
\begin{align}\label{Candela}
& \cos\Pi_{a1}\cosh\b_a = \cos\hth_{a1}, \\ 
&\cos\Pi_{a1} \sin\phi^a_{b1} \sinh\b_{a}= \cos\Pi_{b1} \sin\phi^b_{a1} \sinh\b_{b}.
\end{align}
The first equation fixes $\Pi_{a1}=0$ for a co-chronal configuration (`thick'),  $\Pi_{a1}=\pi$ for an anti-chronal (`thin') configuration, 
and $\b_a = \eps_a\th_{a1}$, where $\eps_a=\pm1$. 
In other words, on-shell
$\hth_{a1}=i\th_{a1}+\Pi_{a1}$ for both co- and anti-chronal data. Thanks to the convention \Ref{boostconv}, 
we can write the solutions as $\b_a=\eps_a \cos\Pi_{a1} |\th_{a1}|$.
Inserting these values 
in the second equation we find 
\be\label{sphsine}
\eps_a \sin\phi^a_{b1} \sinh|\th_{a1}|= \eps_b \sin\phi^b_{a1} \sinh|\th_{b1}|.
\ee
To solve this equation we must fix $\eps_a=\eps_b$,  since by  definition all 3d angles are in $(0,\pi)$. This leaves a single, global sign freedom,  $\b_a=\eps\th_{a1}$, $\eps=\pm1$. Having done so, \Ref{sphsine} can be recognised as the spherical sine laws, which are identically satisfied since we are already satisfying the spherical cosine laws. 

We have found that in the Lorentzian branch, there is no subset of data with a single critical point: either there are no critical points, or there are two distinct ones, labelled by $\eps$. The critical values of the group elements contain a boost and a rotation, and are given by
\be\label{hcritLor}
h_a^{\scr (c)} = \pm h_a^{\eps}, \qquad  h_a^{\eps} = \exp\Big(-\f\eps 2 \th_{a1} \vec{n}_{a1}\cdot \vec{\s}\Big) 
\exp \Big(-\f i2(\xg_{a1} - \Pi_{a1} ) \vec{n}_{a1}\cdot \vec{\s}\Big).
\ee
The boundary data admitting critical points satisfy closure and angle-matching conditions at all faces. From the same argument used in the Euclidean case, it follows that these data describe Lorentzian 4-simplices. Since the boundary data are 3d Euclidean, the 4-simplices must have all tetrahedra space-like, namely the 4d normals to the tetrahedra are time-like. This is confirmed looking at the critical group elements, which are boosts representing the extrinsic curvature between time-like normals.
The space-like 4-simplices are of two types: 1-4, with one 4d normal anti-chronal to the other four, and $2-3$, with two 4d normals anti-chronal to the remaining three. The types are distinguished by the values of $\Pi_{a1}$, which are in turn determined by the boundary data.

In the Lorentzian case, a single set of critical points captures the holonomies mapping the 3d data into a 4d object, and changing the set used has the effect of a parity map on the 4d object.
To reconstruct the 4d normals explicitly, we assign the 4d normal $t^I=(-1,0,0,0)$ to the reference tetrahedron 1. From the critical group element we define
\begin{equation}\label{Llor}
\Lambda_a{}^I{}_J := \f{1}{2} \tr( \hat \s^I h^{\scr +}_a \s_J h_a^{{\scr +}\dagger} )\in \SO^+(1,3)\simeq \SL(2,\C), 
\end{equation}
with $\sigma_I = (\mathds{1},\vec{\sigma})$, $\hat \s_I=(-\Id, \vec\s)$, and the indices are now contracted with the Minkowski metric $\eta_{IJ}$. 
This can be recognized as the mapping between $\SL(2,\C)$ matrices and the future-pointing, identity-connected part of the Lorentz group.
The 4d-normals are obtained from these transformations and the reference vector. In our partial gauge,
\be
N^I_a:=\L_a{}^I{}_J t^J = \cos\Pi_{a1} (\cosh\th_{a1}, \sinh\th_{a1}\vec n_{a1}).
\ee
Changing to the second critical point achieves $\L_a\mapsto \L^{\rm T}_a=\L_a^{-1}$, and one reconstructs the parity transformed Lorentzian 4-simplex.

The twisted spike gauge \Ref{twistedspike} is not accessible for Lorentzian data, because the presence of a boost in \Ref{maggica} prevents them from satisfying \Ref{vector}.
Among the geometrically equivalent gauge configuration for a Lorentzian 4-simplex, there is one convenient choice: the spike configuration with the 3d normals rotated so to have $\xg_{a1}=0$.  
However, the geometric meaning of this gauge is slightly different than in the Euclidean case: it still represents the 4-simplex as the 3d object obtained gluing 4 tetrahedra to the exterior of the reference tetrahedron 1. But unlike in the Euclidean case, this configuration cannot be obtained with the group elements \Ref{Llor} acting on the 4-simplex embedded in Minkowski space: these are $\SL(2,\C)$ transformations, and will
rotate the anti-chronal tetrahedra \emph{inside} the reference one. To obtain the spike configuration one needs a parity or time-reversal transformations on these tetrahedra.

\subsection{Action at the critical points}
\label{sec:Scrit}
To evaluate the action at the critical point \Ref{eq:actioncp}, we need the matrix elements \Ref{Cphases}. We start with the partial gauge \Ref{partialgauge}, and use the general expression \Ref{polifemo} for the critical group elements. For the links connected to 1, we obtain
\begin{equation}
\label{hrocket1}
[\z_{b1}|h_{b}^{-1}\ket{\z_{1b}}=e^{-i\om_b}[{\z_{b1}}\ket{\z_{1b}}=e^{-i\om_b+ i \arg [\z_{b1}\ket{\z_{1b}}}.
\end{equation}
For the other links, after some algebra based on the properties of the coherent states (see Appendix~\ref{app:razzetto}), we obtain
\begin{align}
\label{hrocket}
& [{\z_{ba}}|h_{b}^{-1}h_{a}\ket{\z_{ab}}= 
{[\z_{ba}\ket{\z_{ab}}}\left(\cos{\frac{\om_{a}}{2}}\cos{\frac{\om_{b}}{2}}-\sin{\frac{\om_{a}}{2}}\sin{\frac{\om_{b}}{2}}\vec{n}_{a1}\cdot\vec{n}_{b1}\right)\\
&\ - i[\z_{ba}\ket{\z_{ab}} \f{ 
\vec{n}_{ab}-\vec{n}_{ba}+ i \vec{n}_{ab} \times \vec{n}_{ba}
} {1-\vec{n}_{ab}\cdot\vec{n}_{ba}} \cdot
\left(\cos{\frac{\om_{a}}{2}} \sin{\frac{\om_{b}}{2}} \,\vec{n}_{a1}-\sin{\frac{\om_{a}}{2}}\cos{\frac{\om_{b}}{2}}\,\vec{n}_{b1}-\sin{\frac{\om_{a}}{2}}\sin{\frac{\om_{b}}{2}}\,\vec{n}_{a1}\times\vec{n}_{b1}\right). \nn
\end{align}

\medskip

For vector geometries, we have  $\om_a=\th_{a1} - \xg_{a1}$. 
In the twisted spike gauge \Ref{twistedspike}, $\om_a=0$ and  
\be
S^{\scr (c)}=-2i\l \sum_{(ab)} j_{ab}\arg [\z_{ba}\ket{\z_{ab}}.
\ee
To write the on-shell action in an arbitrary gauge, recall that upon a rotation of the 3d normals, the spinors transform as in \Ref{spinorrot}. 
Hence the general form of the action is 
\be\label{Svector}
S^{\scr (c)}= i\Psi, \qquad \Psi:=-\l\sum_{(ab)} j_{ab} (\chi_{ab}-\chi_{ba} + 2  \arg  [\z_{ba}\ket{\z_{ab}} ),
\ee
where $\chi_{ab}$ is the phase of the spinor $\z_{ab}$ under the change of orientation of the normal $\vec{n}_{ab}$ \eqref{spinorrot}.
\medskip

For Regge data, we can choose the spike gauge, which is accessible for both Euclidean and Lorentzian data. In this gauge, $\xg^b_{a1}=0$ and
the vectors $\vec{n}_{ab}\times\vec{n}_{ba}$ are aligned with $\vec{n}_{1a}\times\vec{n}_{1b}$. Additional simple relations hold between the scalar products at different tetrahedra and the 3d dihedral angles, such as $\vec{n}_{ab}\cdot\vec{n}_{ba}=\cos\left(\phi^{1}_{ab} + \phi^{b}_{a1} + \phi^{a}_{b1} \right)$. Using this formula and similar ones proved in Appendix~\ref{app:razzetto}, \Ref{hrocket} reduces to
\begin{align}
\label{hrocketspike}
& [\z_{ba}|h_{b}^{-1}h_{a}\ket{\z_{ab}}
\\\nn & \quad = 
\left[ \cos{\frac{\om_{a}}{2}}\cos{\frac{\om_{b}}{2}}\sin\left(\frac{\phi_{ab}^{1}+\phi_{b1}^{a}+\phi_{a1}^{b}}{2}\right)+
\sin{\frac{\om_{a}}{2}}\sin{\frac{\om_{b}}{2}}\sin\left(\frac{-\phi_{ab}^{1}+\phi_{b1}^{a}+\phi_{a1}^{b}}{2}\right) \right.\\\nn
&\qquad\quad -i  
\left. \cos{\frac{\om_{a}}{2}}\sin{\frac{\om_{b}}{2}} \sin\left(\frac{\phi_{ab}^{1}-\phi_{b1}^{a}+\phi_{a1}^{b}}{2}\right) 
-i \sin{\frac{\om_{a}}{2}} \cos{\frac{\om_{b}}{2}} \sin\left(\frac{\phi_{ab}^{1}+\phi_{b1}^{a}-\phi_{a1}^{b}}{2}\right) \right]
e^{i \arg  [\z_{ba}\ket{\z_{ab}} }.
\end{align}

At this point we can specialize to Euclidean or Lorentzian solutions. In the first case we have $\om_a= \eps \th_{a1}$.
Spherical trigonometry, see \Ref{eq:halfcompo},  gives
\begin{align}
&\cos{\frac{\th_{a1}}{2}}\cos{\frac{\th_{b1}}{2}}\sin\left(\frac{\phi_{ab}^{1}+\phi_{b1}^{a}+\phi_{a1}^{b}}{2}\right)+
\sin{\frac{\th_{a1}}{2}}\sin{\frac{\th_{b1}}{2}}\sin\left(\frac{-\phi_{ab}^{1}+\phi_{b1}^{a}+\phi_{a1}^{b}}{2}\right)
=\cos \frac{\th_{ab}}{2},\\
&\cos{\frac{\eps\th_{a1}}{2}}\sin{\frac{\eps\th_{b1}}{2}}\sin\left(\frac{\phi_{ab}^{1}-\phi_{b1}^{a}+\phi_{a1}^{b}}{2}\right)+
\sin{\frac{\eps\th_{a1}}{2}}\cos{\frac{\eps\th_{b1}}{2}}\sin\left(\frac{\phi_{ab}^{1}+\phi_{b1}^{a}-\phi_{a1}^{b}}{2}\right)
=\eps \sin \frac{\th_{ab}}{2}.
\end{align}
It follows that
\begin{equation}
\label{eq:euclrazz}
[\z_{ba}|h_{b}^{-1}h_{a}\ket{\z_{ab}}= \exp \Big(- \f i2 \eps\, \theta_{ab} + i \arg [\z_{ba}\ket{\z_{ab}} \Big).
\end{equation}
Comparing with \Ref{Cphases} we read $\norm{h_{a}^{\dagger}z_{ab}}/\norm{h_{b}^{\dagger}z_{ab}}=1$ and $\upsilon_{ab}-\upsilon_{ba}= -\eps\, \theta_{ab}/2 + \arg [\z_{ba}\ket{\z_{ab}}$.  The action at the two critical points in the spike gauge is thus

\begin{equation}
S^{\scr (c)}= i \eps\l \sum_{(ab)} j_{ab} \th_{ab} - 2i\l \sum_{(ab)} j_{ab} \arg  [\z_{ba}\ket{\z_{ab}}. 
\end{equation}
Since $\th_{ab}$ are the 4d dihedral angles functions of the 10 areas $j_{ab}$ uniquely determining the 4-simplex, the first term above 
contains the Regge action for a 4-simplex,
\be\label{Regge}
S_{\rm R}(j):=\l \sum_{(ab)} j_{ab} \th_{ab}(j).
\ee
The general form of the  action at the critical points reads
\be\label{SER}
S^{\scr (c)}= i \eps S_{\rm R} + i\Psi.
\ee
The first term is independent of the orientation of the boundary data, as well as of the phase conventions for the coherent states. The second depends on both, and can always be put to zero choosing a convenient orientation or adapting the phases of the coherent states.
In the twisted spike gauge $\chi_{ab}-\chi_{ba}\equiv -\th_{ab}(j)$ \cite{Dona:2017dvf}, and $S^{\scr (c)}=i(\eps-1)S_{\rm R} - 2i\l \sum j_{ab}\arg  [\z_{ba}\ket{\z_{ab}}$.

For Lorentzian data $\om_a= i\eps\, \th_{a1} + \Pi_{a1}$ in the spike gauge. Using this time hyperbolic trigonometry, see \ref{eq:halfcompohyp}, we arrive at
\begin{align}
\label{eq:lorerazz}
[\z_{ba}| h_{b}^{-1}h_{a}\ket{\z_{ab}}= \exp \Big(-\f12 \eps\, \theta_{ab} +  \f i2{\Pi_{ab}} + i\arg [\z_{ba}\ket{\z_{ab}}  \Big).
\end{align}
From this matrix element we read $\norm{h_{b}^{\dagger}z_{ab}}/\norm{h_{a}^{\dagger}z_{ab}}=\exp(\eps\, \th_{ab}/2)$ and  
$\upsilon_{ab}-\upsilon_{ba}= \Pi_{ab}/2 + \arg [\z_{ba}\ket{\z_{ab}}$.  
The action at the two critical point in the spike gauge is thus
\be
S^{\scr (c)} = i \eps \g \l \sum_{(ab)} j_{ab} \th_{ab} - i \l\sum_{(ab)} j_{ab}(\Pi_{ab} + 2 \arg  [\z_{ba}\ket{\z_{ab}} ).
\ee
The first term contains the Regge action \Ref{Regge}, now for a Lorentzian 4-simplex. The critical action for an arbitrary orientation of the 3d normals is 
\be\label{SLR}
S^{\scr (c)}= i \eps \g S_{\rm R} +i\Psi + i\l \sum_{(ab)} j_{ab} \Pi_{ab},
\ee
where $\Psi$ is the same as before, and we have kept separated the phase induced by the number of anti-chronal pairs (thin wedges) since it is independent of orientations and phase conventions.

\subsection{Asymptotic formulae}\label{4sLO}
To write the asymptotic formulas we use the general expression \Ref{eq:asymp}. The Hessian for the 4-simplex amplitude was partially studied in \cite{BarrettLorAsymp} and explicitly computed in \cite{Dona:2019dkf}.
Numerical investigations in \cite{Dona:2019dkf} showed that it is non-degenerate, and that  $H^{+}=\overline{H^{-}}$, for various critical configurations considered. An analytic proof that the Hessian in non-degenerate on a dense set of critical points was then given in \cite{Kaminski:2019dld}.\footnote{The Hessian determinant vanishes for instance for degenerate data with coplanar normals, or when the 4-volume reconstructed from the areas  vanishes.} 
With this information and the results of the previous analysis we obtain the following asymptotic formulas. 
\begin{itemize}
\item For vector data:
\begin{equation}
A_\s(j_{ab},\vec n_{ab},-\vec n_{ba})= \f{2^{36}\pi^{12}\J}{\l^{12}}  \f{\Om^{\scr (+)} e^{i\Psi} }{\sqrt{\det (-H^{\scr (+)} ) }}  +O(\l^{-13}),
\end{equation}
with $\Psi$ given by \Ref{Svector}, the factor $\J$ is the product of all the spins in the vertex as in \eqref{bigJ} and $\Om^{\scr (c)}$ is the value of the spinorial measure at the critical point ${\scr (c)}$ as in \cite{BarrettEPRasymp}. 

\item For Euclidean Regge data:
\begin{equation}
A_\s(j_{ab},\vec n_{ab},-\vec n_{ba})= \f{2^{36}\pi^{12}\J}{\l^{12}}
\f{\Om^{\scr +} e^{i\Psi } }{\sqrt{|\det{(-H^{\scr +})}|}} \cos \Big( \lambda S_{\rm R} 
- \f12\arg \det(-H^{\scr +})\Big) +O(\l^{-13}),
\end{equation} 
with $S_{\rm R}$ given by \Ref{Regge}.
\item For Lorentzian Regge data:
\begin{equation}
A_\s(j_{ab},\vec n_{ab},-\vec n_{ba})= \f{2^{36}\pi^{12}\J}{\l^{12}}
\f{\Om^{\scr +}  (-1)^\chi e^{i\Psi }  }{\sqrt{|\det{(-H^{\scr +})}|}}  \cos \Big( \l \g S_{\rm R}
- \f12\arg \det(-H^{\scr +})\Big)
 +O(\l^{-13}),
\end{equation}
with $S_{\rm R}$ given by \Ref{Regge}, this time with Lorentzian angles, and 
where $\chi=\l \sum_{(ab)} j_{ab} \Pi_{ab}$ depends on the number of  anti-chronal pairs of tetrahedra (thin wedges).
\end{itemize}
The main qualitative difference between Euclidean and Lorentzian data concerns the dependence on $\g$, only in the Hessian in the first case, also in the frequency of $\l$-oscillations in the second. We also recall that with the phase conventions used here, the amplitude for the Euclidean 4-simplex is real in the twisted spike gauge, and $\arg H^{\scr {(c)}}=0$.

These formulas complete the reproduction of the results of \cite{BarrettLorAsymp}, amended with the reality properties of the Hessian found in \cite{Dona:2019dkf}. In the next section, we describe the novel application of our technique to general graphs.

\section{The algorithm for a general vertex asymptotics}

The critical point equations for a general graph are the closure conditions of the normals at the same node \Ref{Clos}, and the equations \Ref{eq:maxRe} and \Ref{C2} obtained maximizing the real part of the action and from the gradient respect to the dummy spinors $z_{ab}$. The first one is solved restricting the boundary data to describe bent polygons in $\R^3$, or convex polyhedra in the non-coplanar case. The crucial one is \Ref{Cvec}, which we replace with \Ref{maggica}, as explained earlier. Once this is solved, 
\Ref{Cphases}  and \Ref{eq:maxRe} are straightforward. 
The focus is then on finding group elements solving the \magg equations \Ref{maggica}, which we report here for convenience of the reader,
\be\nn
H_a \vec n_{ab}=-H_b \vec n_{ba}.
\ee
As discussed in Section~\ref{secSpinToVec}, vector geometries always admit one critical point, with $h_a\in\SU(2)$ for all nodes. The non-trivial quest is to find the most general set of boundary data that admits solutions, and determine the solutions.

\begin{figure}[H]
\centering
\includegraphics[width=6cm]{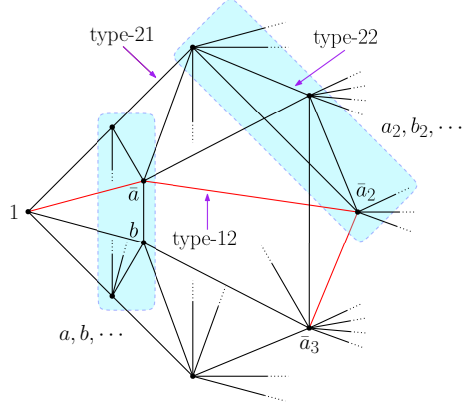}
\caption{\small{\emph{Schematic representation of the algorithm for a generic graph. We choose a starting node, denoted 1 in the figure. We determine the critical group elements on the nodes first neighbours to it (vertical shaded area), by solving the \magg equations for all links $(1a)$ and $(ab)$. We then choose a new seed, $\ab$, among the first neighbours, and determine the critical group elements on the nodes first neighbours to $\ab$ not already known (diagonal shaded area), by solving the \magg equations for all links $(\ab a_2)$ called type-12, $(a_2b_2)$ called type-22,
and $(a a_2)$ called type-21. We distinguish the equations as type-12, type-22, and type-21, respectively. 
If all unfinished links (the dashed ones) are removed from the picture to get an example of a closed graph, the 
dominating set can be chosen to be
$1\ab \ab_2 \ab_3$. Notice that it is not minimal: a minimal dominating set such as $1\ab b$ cannot be used to run the algorithm, and can be excluded because it defines a cycle. }}  \label{fig:general} }
\end{figure}

Our algorithm requires picking a
set of connected nodes that don't define a cycle, and such that all remaining nodes of $\G$ are first neighbours to it. 
A set of nodes such that all remaining nodes are first neighbours to it is known as `dominating set' in graph theory. The path that connects them is the `dominating path', and provides the trunk of a rooted spanning tree of $\G$. For the examples considered explicitly below, all graphs are such that a minimal dominating set does not define cycles, and can be thus used for the algorithm. But in more general cases, one may need a non-minimal set of nodes to avoid cycles, see Fig.~\ref{fig:general}.

We denote the root as node $1$. To keep a uniform notation with the 4-simplex case, we label the nodes first neighbours to $1$ with $a,b,$ etc. Among them  is the second node of the dominating set,
which we label $\ab$. The first neighbours to $\ab$ \emph{distinct} from the first neighbours to 1 are labelled $a_2,b_2,$ etc. We refer to them as second neighbours (they are indeed second neighbours to 1 along the dominant path chosen). Among them is the third node of the dominant set, which we label $\bar a_2$. The first neighbours to $\bar a_2$ \emph{distinct} from the first neighbours to 1 and $\ab$ are labelled $a_3,b_3,$ etc, and called third neighbours. We proceed in this way until all nodes have been reached.
This dominating set is used to order the \magg equations \Ref{maggica} and solve them hierarchically. 

We now describe the explicit steps of the algorithm. 
Each step is associated with a node in the dominant set chosen, and solves all \magg equations associated to its first neighbours.
We assume that the closure conditions have been already solved restricting the boundary data, and we focus on non-degenerate configurations, with non-coplanar normals and non-zero 3-volumes of the polyhedra.\footnote{For general polyhedra, flat 3d dihedral angles can also occur for non-degenerate configurations with parallel faces. These cases can be treated as limits of slightly non regular polyhedra.}
 We use the partial gauge-fixing introduced in the previous Section, anti-aligning the normals along all links in the dominating path.

\subsubsection*{Step 1: First neighbours}

The first step of the algorithm involves only the first neighbours to 1, and has many similarities with  the analysis done for the 4-simplex.
We use the equations linking 1 to $a$ to determine the directions in the anti-aligned partial gauge, and those between first neighbours to determine the complex angles.

\begin{itemize}[leftmargin=*]
\item $\mathbf{Eqs. \ [1a]}$. We take the root as the node without group integration. The equations linking 1 to $a$ are
\be
\label{eq:direction}
\vec n_{1a} = -H_a\vec n_{a1}.
\ee
We partially fix the gauge at the nodes $a$ requiring 
$\vec n_{1a} = -\vec n_{a1}.$ The equations are then solved by four-screws with directions  $\vec n_{a1}$, 
\be
\label{4screw}
h_a = \pm\exp \Big(\f i2\om_a \vec n_{a1}\cdot\vec\s\Big),
\ee
and $\om_a\in\C$ still to be determined.
\item $\mathbf{Eqs. \ [ab]}$. For all first neighbours belonging to at least one 3-cycle with the root, we have the $ab$ equations
\be
\label{eq:type11}
 H_a \vec n_{ab} = -H_b\vec n_{ba}.
\ee
We pick a basis $n_{a1}, n_{b1}$ and $n_{a1}\times n_{b1}$ to project them into scalar equations, obtaining 
\begin{subequations}\label{Eqstep1}\begin{align}
& \cos(\om_{a}+ \xg^b_{a1})=\cos \hth_{a1}^b, \\
& \cos(\om_{b}+ \xg^a_{b1})=\cos \hth_{b1}^a, \\
&\sin\phi^a_{b1} \sin(\om_{a}+ \xg^b_{a1})= \sin\phi^b_{a1} \sin(\om_{b}+ \xg^a_{b1}).\label{DDR}
\end{align}\end{subequations}
for all $a$ and $b$ belonging to 3-cycles. Here $\hth^b_{a1}$ are the edge-dependent dihedral angles defined in \Ref{defhth}, and $\xg^b_{a1}$ the edge twist angles \Ref{defxg}. 

Finding solutions for these equations follows the analysis detailed in the 4-simplex Section. The cosine equations require $\om_{a}=\eps_a^b\hth_{a1}^b - \xg^b_{a1},$ with $\eps_a^b=\pm 1$.
Writing the solution in this form allows us to treat at once all possibilities. Whichever possibility is realized will depend on the boundary data at the nodes involved in this step.
The plus signs  $\eps_a^b=1$ are admissible if the data satisfy the orientation conditions \Ref{vector} at the links $ab$.
This can be proved as we did for the 4-simplex: 
use the twisted spike gauge \Ref{twistedspike} and the linear transformation of $\xg^b_{a1}$ under rotations at $a$.
This is the vector geometry solution. 
Solutions with $\eps_a^b=-1$ are  possible only when the right-hand side of \Ref{sol1} is independent of $b$. This requires 
generically\footnote{\label{caveat} We put a word of caution here because we cannot rule out the possibility of individual configurations for which the equations can be solved by numerical coincidence, without requiring $b$-independence of the right-hand side. We are merely providing a constructive algorithm here, not an existence theorem. It may also happen that for a certain $a$, there is a single choice of $b$. In this case there are conditions for $1a$ arising at this step, but they will arise at a later step.}
edge-independent angles 
\be\label{ei1}\hth^b_{a1}\equiv \hth_{a1}, \qquad \xg^b_{a1}\equiv \xg_{a1},\ee and also fixes 
$\eps_a^b\equiv \eps_a$. These are angle-matching data, as explained in Section~\ref{SecSCL}, but notice that not all 2d angles of the faces $1a$ are already required to match:
there may be additional edges of this face that have not yet entered the algorithm at this step.

Inserting \Ref{sol1} in \Ref{DDR}, we find the spherical sine laws between $\hth$'s and $\phi$'s. These are identically satisfied, provided all nodes are in the same Euclidean or Lorentzian sector, and $\eps_1:=\eps_a=\eps_b$.
For Euclidean angle-matched data, $\hth_{a1}=\th_{a1}$ is real, and for Lorentzian data $\hth_{a1}=i \th_{a1} + \Pi_{a1}$,
with $\Pi_{a1}=0$ or $\pi$ if the boundary data are respectively co-chronal or anti-chronal. 
There is thus a unique sign freedom, and the solution for any critical data can be written in the generic form
\be
\label{sol1}
\om_{a}=\eps_1\hth_{a1}^b - \xg^b_{a1}.
\ee

\item {\bf Lonely neighbour.} If one first neighbour does not belong to any 3-cycle with the root, there are no equations to fix its complex angle at this step.

\end{itemize}

\noindent If this step has determined all group elements, the algorithm stops here. 

\subsubsection*{Step 2: Second neighbours}

In the second step, we choose a new seed among the first neighbours of 1, labelled $\ab$, and determine the group elements at the nodes $a_2,b_2,$ etc.. namely those first neighbours to $\ab$ that were not determined in the first step. We refer to these nodes as second neighbours of the root along the trunk.
 
\begin{itemize}[leftmargin=*]
\item $\mathbf{Eqs. \ [\ab a_2]}$. We consider the `type-12' equations from the new seed to the second neighbours, 
\be
\label{type12}
\vec n_{\ab a_2} = -H_{\ab}^{-1}H_{a_2}\vec n_{a_2\ab}.
\ee
We define $\tl H_{a_2}:= H_{\ab}^{-1}H_{a_2}$, and partially fix the gauge at the nodes $a_2$ requiring $\vec n_{\bar a a_2} = -\vec n_{a_2\bar a}$. The equations are then solved by  a four-screw
\be\label{4screw2}
\tl h_{a_2} = \pm\exp \Big(\f i2\om_{a_2} \vec n_{a_2\ab}\cdot\vec\s\Big),
\ee
with $\om_{a_2}$ to be determined. 

\item $\mathbf{Eqs. \ [a_2 b_2]}$. For all second neighbours belonging to at least one 3-cycle with the new seed $\ab$, we have the `type-22' equations.

Multiplying them on both sides by $H_{\ab}^{-1}$, we can write them as
\be
 \tl H_{a_2} \vec n_{a_2b_2} = -\tl H_{b_2}\vec n_{b_2a_2},
\ee
so to involve 4-screws with known directions.
We pick a basis $\vec{n}_{a_2 \ab}, \vec{n}_{b_2 \ab}$ and $\vec{n}_{a_2 \ab}\times \vec{n}_{b_2 \ab}$ to project the equations into scalar components. These have the same structure of \Ref{Eqstep1},
\begin{subequations}\label{Capitani2}\begin{align}
& \cos(\om_{a_2}+ \xg^{b_2}_{a_2\ab})=\cos \hth_{a_2\ab}^{b_2}, \\
& \cos(\om_{b_2}+ \xg^{a_2}_{b_2\ab})=\cos \hth_{b_2\ab}^{a_2}, \\
&\sin\phi^{a_2}_{b_2\ab} \sin(\om_{a_2}+ \xg^{b_2}_{a_2\ab})= \sin\phi^{b_2}_{a_2\ab} \sin(\om_{b_2}+ \xg^{a_2}_{b_2\ab}),
\end{align}\end{subequations}
and lead to 
\be\label{Eqstep2}
\om_{a_2}=\eps_2\hth_{a_2\ab}^{b_2} - \xg_{a_2\ab}^{b_2}, 
\qquad \eps_2=\pm 1,
\ee 
for all $a_2$ and $b_2$ belonging to a 3-cycle with $\ab$.
These group elements must thus belong to the same Euclidean or Lorentzian sector, there is single sign freedom $\eps_2$. The plus solution requires orientation conditions \Ref{vector} for the links $a_2\ab$, $a_2b_2$ and $\ab b_2$, and the existence of a second solution requires edge-independent conditions
\be\label{ei2}
\hth_{a_2\ab}^{b_2}=\hth_{a_2\ab},\qquad \xg_{a_2\ab}^{b_2}=\xg_{a_2\ab},
\ee

\item $\mathbf{Eqs. \ [a_2 b]}$. The `type-21' equations involve a second and a first neighbours. The first neighbour need not be linked to the new seed $\ab$.
These equations have a new structure with respect to those appearing at step 1, because they involve a 4-cycle, $1\ab a_2 b$. They can be rewritten as
\be\label{3H}
\tl H_{a_2} \vec n_{a_2b} = - H_{\ab}^{-1} H_{b}\vec n_{ba_2},
\ee
so to have all directions already determined. But unlike the previous equations, they involve three group elements. 
We project them on the basis $\vec n_{\ab b}$, $\vec n_{a_2\ab}$ and $\vec n_{\ab b} \times \vec n_{a_2\ab}$.\footnote{Alternative versions of \Ref{capitani3} follow choosing different triples of vectors as basis. The one chosen here is the most convenient because of the simplicity of \Ref{Totti3}. Other options lead to equations of the type \Ref{Totti3b}.
}
The first projection has the usual structure encountered so far. The other two are slightly more complicated, but it is possible to simplify them using not just the directions, but also the complex angles found in the previous steps. We then obtain
\begin{subequations}\label{capitani3}
\begin{align}
& \cos(\omega_{a_2}+\xg^b_{a_2 \ab})=\cos\hth^{b}_{a_2 \ab}, \label{Totti3} \\
& \cos(\eps_1 \hth^c_{\ab 1} - \xg^c_{\ab 1} +\xg^{a_2b}_{\ab 1} - \eps_1 \vth^{b}_{s1}-\pi)=\cos\vth^{a_2}_{s\ab}, \label{Totti3b} \\
& \sin\phi^{a_2}_{b \ab} \sin(\omega_{a_2}+\xg^b_{a_2 \ab})= \sin\phi^{b}_{a_2 \ab} \sin(\eps_1 \hth_{b\ab}^{a_2}).
\label{DeRossi3}
\end{align}\end{subequations}

The first cosine equations require 
$\om_{a_2}=\eps_{a_2}^{b}\hth_{a_2\ab}^{b} - \xg_{a_2\ab}^{b},$ with $\eps_{a_2}^{b}=\pm 1.$
The plus sign is valid if the boundary data satisfying the orientation conditions at the links $a_2 \ab$, $a_2 b$ and $b \ab$, and the minus sign requires edge-independence conditions at $\ab a_2$, \be\label{ei3}
\hth_{a_2\ab}^{b} = \hth_{a_2\ab}, \qquad \xg_{a_2\ab}^{b}=\xg_{a_2\ab},
\ee
and $\eps_{a_2}^b=\eps_{a_2}$.
If there are no type-22 equations in the graph, \Ref{Totti3} replaces \Ref{Eqstep2} in determining $\om_{a_2}$. Otherwise, it must be consistent with the previous solution. This requires $\eps_{a_2}=\eps_2$ and the additional edge-independence conditions that \Ref{ei3} must match \Ref{ei2} previously determined.
The sine equations require $\eps_{a_2} = \eps_{\ab}=\eps_1,$
as well as a matching of sectors of the boundary data:
mixed Euclidean and Lorentzian configurations would fail to satisfy the spherical sine laws. 
The type-21 equations thus force the matching of signs between first and second neighbours,
\be \eps_1=\eps_2. \ee
The second cosine equations involve explicitly the 4-cycles, but all complex angles have been already determined. They thus give only restrictions on the boundary data. These will be examined in details below in the examples of Sections~\ref{sec:examples}, 
where the new angles $\vth$ entering this formula will also be defined. They are closely related to 
the dihedral angles with the auxiliary hyperplane $s=(\ab 1, b a_2)$ described in Section~\ref{SecSCL}. 
  The required edge-independence conditions turn out to be
\be\label{ei4} \xg^{a_2b}_{\ab 1} =\xg_{\ab 1}, \qquad \xg^{a_2\ab}_{b 1} =\xg_{b 1},\ee 
involving the edges of the face $1a$ dual to 4-cycles. When these are satisfied, $\vth^b_{s1}=\hth^b_{s1}$ and $\vth^{a_2}_{s\ab}=\hth^{a_2}_{s\ab}$,
and \Ref{Totti3b} becomes 
\be\label{ei5}
\hth^c_{\ab 1} = \hth^b_{s1} + \eps_1\eps_s^{a_2}\hth^{a_2}_{s\ab} +\pi, \qquad \eps_s^{a_2}=\pm1.
\ee
This equation fixes the sign $\eps_s^{a_2}$ in terms of the boundary data and $\eps_1$, but to be valid requires also new edge-independence conditions, which we see are of the type  \Ref{3=4+4} relating the 4d angles of the face $1\ab$ computed using 3-cycles and 4-cycles.
The important point here is that additional orientation or edge-independence conditions appear for the faces $1a$ already involved in the first step. It is this type of additional restrictions that in the end guarantees that there are as many angle-matching conditions as possible valence of the faces connected to the dominating set.

\item {\bf Lonely neighbour.} If a second neighbour does not belong to any 3-cycle with the new seed, both type-22 and type-21 equations are missing, and there are no equations to fix its complex angle at this step.

\end{itemize}

\noindent If this step has determined all group elements, the algorithm stops here.

\subsubsection*{Step 3}

We iterate the procedure moving to the next node of the chosen dominating set, $\ab_2$, and use it as a new seed to solve the \magg equations around it. There can be type-23 equations involving the nodes $\ab_2a_3$, type-33 equations involving $a_3b_3$, and type-32 involving $a_3b_2$.
These are treated as before, leading to solutions 
\be\label{sol3}
\om_{a_3}=\eps_3\hth_{a_3\ab_2}^{b_3} - \xg_{a_3\ab_2}^{b_3}, \qquad \eps_3 =\pm 1,
\ee
and matching signs $\eps_3=\eps_2$ and signature of the boundary data from the type-32 equations.
The novelty appearing at step 3 is the possibility of type-31 equations, connecting a first neighbour $b$ to a third neighbour $b_3$. These can be rewritten as
\be\label{type31}
\tl H_{\ab_2} \tl H_{b_3} \vec n_{b_3b} =  H_{\ab}^{-1} H_{b}\vec n_{bb_3},
\ee
so to have all directions known. Four group elements appear now, and their scalar projections involve a 4-cycle $(\ab b b_3 \ab_2)$ and a 5-cycle 
$(\ab 1 b b_3 \ab_2)$.
Using  solutions obtained from the previous steps, we can write the scalar projections as follows,
\begin{subequations}\label{step31}\begin{align}
& \cos(\om_{\ab_{2}}+\xg_{\ab_{2} \ab}^{b_3 b} - \eps_3\vth_{s\ab_{2}}^{b_3} -\pi)=\cos \vth_{s\ab }^{b},\\
& \cos(\eps_1\hth^{c}_{\ab 1}-\xg^c_{\ab 1}+\xg_{\ab 1}^{\ab_2 b} - \eps_1 \vth_{s' 1}^{b} -\pi)=\cos \vth_{s'\ab }^{s},\\
& \sin\varphi^s_{ b_3 \ab_2}\sin(\om_{\ab_{2}}+\xg_{\ab_{2} \ab}^{b_3 b} - \eps_3\vth_{s\ab_{2}}^{b_3}-\pi)=
\sin\phi^b_{\ab_2 \ab}\sin(\eps_1 \vth_{b\ab}^{s}), 
\end{align}
\end{subequations}
where $s=(\ab_2\ab,bb_3)$ and $s'=(\ab 1,bb_3)$ are auxiliary hyperplanes associated respectively with the 4-cycle and 5-cycle, and $\vth$'s mismatched versions of the auxiliary 4d angles, defined below in Section~\ref{sec:example3}.
If $\om_{\ab_2}$ was a lonely neighbour of a previous step, the first cosine equation can be used to determine it. 
So the lonely neighbours require equations associated with other nodes of the dominant set to be determined.
This shows that while the (complex angles of the) non-lonely neighbours can be determined always using the simple 3-cycle expressions like \Ref{sol1}, the lonely neighbours require more complicated expressions with higher cycles. Otherwise, $\om_{\ab_2}$ goes also on-shell using a previous step, and then both cosine equations are only restrictions on the boundary data. These turn out to be additional orientation conditions or angle-matching conditions on boundary data that had not yet entered the analysis, and refer to edges associated to the higher cycles.

An important feature is that the type-31 sine equations don't provide restrictions between the signs $\eps_1$ and $\eps_3$, but between $\eps_1$ and the new sign in front of $\vth_{s\ab }^{b}$. Therefore, only the equations of type-21, 32, etc, rigidify the structure.

\subsubsection*{Iteration}

At step 4 one encounters equations with five group elements involving at least 5-cycles, and so on. 
One proceeds in this way until all group elements have been determined, and all unused  \magg equations have been verified to see if they are identities or require additional constraints.

\subsubsection*{Modular graphs}

The equations of type-21, 32, etc, play a key role in rigidifying the geometric structure defined by the boundary data: their sine equations require matching signs  among different parts of the graph, thus reducing the overall number of solutions, and matching sectors, thus selecting globally Euclidean or Lorentzian data. For this reason, we refer to them as \bracing equations.
A special situation occurs if for any choice of dominating set, one or more consecutive pairs of \bracing equations are missing.
This `unbraced' situation occurs for modular graphs, namely graphs that can be divided in subgraphs whose nodes are connected to at most one node outside the subgraph. 
Such graphs appear naturally for instance if one refines an existing graph replacing one of its nodes with a new subgraph. In this case, there is nothing relating the signs of the subgraphs, nor the sector of the boundary data.
Furthermore, there will always be a lonely neighbour for any choice of dominating set, given by the first seed of the subgraph. Its complex angle is then  necessarily determined through a higher cycle.  For example, if 
$\ab_2$ is the first node in the dominating set that belongs to a subgraph, we can determine it from the type-31 equations \Ref{step31}, finding
\be
\om_{\ab_2} = \eps_1 \vth^b_{s\ab} +\eps_3\vth_{s\ab_{2}}^{b_3} -\xg_{\ab_{2} \ab}^{b_3 b} +\pi,
\ee
where $\eps_1$ and $\eps_3$ are the signs of the two subgraphs.

\subsection*{Comments}

Given the vast richness of (3-link-connected) graphs, the algorithm we provided can only be indicative: it can be applied straightforwardly to the examples considered below, but the reader should keep in mind that it may need some adapting for specific graphs. If it needs to be adapted, we hope to have at least conveyed the main logic of our technique, which consists of solving the equations in a given order, and collecting step by step the required restrictions on the boundary data. As we will discuss in a moment, these conditions can be identified in fairly general terms as global orientation or angle-matching conditions. If the algorithm applies straightforwardly, the way we presented it has also the more ambitious goal of providing directly the general solution to the critical point equations. In particular, it is then possible to choose the dominating set so that all group elements can be determined using only the smallest cycles. The more complicated equations are then only restriction on the boundary data, automatically satisfied requiring angle-matching conditions everywhere. 

If a graph is not modular, we believe it is always possible to pick a dominating set such that there are no lonely neighbours. In this case, all complex angles are determined by 3-cycles, and the solutions at the end of the algorithm take the generic form 
\begin{subequations}\label{hsols1}\begin{align}
& h_1 = \Id, \qquad h_{a} = \pm\exp \Big(\f i2( \eps\hth_{a1}^{b} - \xg_{a1}^{b})\vec n_{a1}\cdot\vec\s\Big), \\\label{hsols1b}
& h_{a_2} = \pm h_{\ab}\exp \Big(\f i2( \eps\hth_{a_2\ab}^{b_2} - \xg_{a_2\ab}^{b_2})\vec n_{a_2\ab}\cdot\vec\s\Big), \\
& h_{a_3} = \pm h_{\ab} h_{\ab_2}\exp \Big(\f i2( \eps\hth_{a_3\ab_2}^{b_3} - \xg_{a_3\ab_2}^{b_3})\vec n_{a_3\ab_2}\cdot\vec\s\Big),
\end{align}\end{subequations}
and so on, with the choice of dominating set explicitly appearing in the sequence of holonomies. 
For angle-matched data, we can take the spike gauge, then all $\xg$'s vanish, and the critical holonomies are all written in terms of the dihedral angles alone,
with $\hth$ determined by \Ref{defhth}, taking either the Euclidean values $\hth=\th$, or the Lorentzian values 
$\hth=i\th+\Pi$ with conventions \Ref{boostconv} and $\Pi=0$ for the co-chronal and $\pi$ for the anti-chronal cases.

If a graph is 2-modular, there will be a lonely neighbour no matter which set we choose: this coincides with the seed used as first node of the subgraph. 
Suppose this happens at step 2, we will have
\be\label{hsols2}
h_{\ab_2} = \pm h_{\ab}\exp \Big(\f i2\big(  \eps (\vth^b_{s\ab} +\eta\vth_{s\ab_{2}}^{b_3}+\pi) -\xg_{\ab_{2} \ab}^{b_3 b}\big)\vec n_{a_2\ab}\cdot\vec\s\Big), 
\ee
replacing \Ref{hsols1b}. If the boundary data satisfy edge-independence conditions at the face $\ab_2 b_3$, the $\vth$ angles reduce to $\hth$'s and the argument in round brackets equals the definition \Ref{3=4+4} of the 4d dihedral angle for a face not belonging to any 3-cycle,
\be
\hth_{a_2\ab}(\phi,\eta) = \hth^b_{s\ab} +\eta\hth_{s\ab_{2}}^{b_3}+\pi,
\ee
with the same conventions as above for Euclidean and co-chronal and anti-chronal Lorentzian values.
If a graph is $n$-modular, there will be $n-1$ holonomies of the form \Ref{hsols2}, one for each seed entering a subgraph.

The advantage of the gauge-fixed approach used here is that the directions of the individual group elements at the critical point are easily determined by the normals themselves. The vectorial equations determining the complex angles of the four-screws can be projected in scalar components that introduce the edge-dependent 4d dihedral and twist angles described in Section~\ref{SecSCL}: these quantities appear naturally from the saddle point analysis. For 3-cycles, we get directly the spherical cosine laws in terms of boundary data. For higher cycles, we recover the spherical cosine laws only on-shell of equations belonging to 3-cycles.
There is therefore a minor conceptual difference between the geometric formulas, and the interpretation of the critical point equations.

\subsection{Geometric reconstruction}\label{sec:geogen}

The procedure described above permits to find all critical group elements in a certain gauge, and to classify the conditions that must be satisfied by the boundary data. What we see is that each step always introduces the same type of restrictions at the nodes and links involved, namely orientation conditions \Ref{vector} to have one Euclidean solution, and edge-independence of twist and 4d dihedral angles to have a second Euclidean or two Lorentzian solutions, like \Ref{ei1} and \Ref{ei4}, or \Ref{ei2} and \Ref{ei3}. Both types of edge-independences are captured by 
angle-matching conditions, as explained in Section~\ref{SecSCL}. A single step may not force as many angle-matching conditions as the possible maximal valence of a given face. 
But at the end of the algorithm, one typically finds full 2d angle-matchings for all faces dual to the links connecting all nodes to the nodes in the chosen dominating set (in more botanical words, all links corresponding to the trunk and its branches), plus partial 2d angle-matchings at the faces dual to the remaining links. The exact nature of the partial matching can be determined from \Ref{eureka}, but it is not important,
because as we already pointed out when describing the 4-simplex analysis, the critical behaviour of the amplitude does not depend on the dominant path chosen. A moment of reflection observing the angles entering \Ref{eureka} and the multiplicity of possible dominant paths shows that full angle-matching conditions must be satisfied at all faces, since we must be able to solve the equations in whichever order we want. 

For non-modular graphs, the various bracing equations impose a single global sign freedom, and boundary data of the same sector throughout the graph. Therefore, we have at most two solutions, which occur for globally angle-matched data, Euclidean or Lorentzian. For an $n$-modular graph on the other hand, the signs of each subgraph are left free, as are the sectors of the boundary data. As a consequence, we can have up to $2^n$ critical points, for 
data that are angle-matched throughout. Furthermore, $n-1$ of the seeds are necessarily lonely neighbours of some steps. Their complex angles are therefore determined via 4-cycles or higher cycles. The free signs then enter the reconstructed 4d dihedral angle at the bridging link as $\eta_s$ in \Ref{hsols2}. The angle-matched data can be globally Euclidean or Lorentzian, but also mixed, Euclidean in one subgraph and Lorentzian in another subgraph. In the latter case the  bridging 4d angle is neither Euclidean nor Lorentzian.
Fewer critical points occur for data which are angle-matched in some subgraphs but vector geometries in other. The $\eta$ signs are then fixed at the bridges to a vector subgraph.

What emerges from this analysis, and more explicitly from the examples given below, is that 
the vector geometry solution is recovered as the difference between 4d dihedral and twist angles at each link, which is pure gauge, and vanishes in the twisted spike gauge. 
Additional critical points occur generically for angle-matched boundary data, and are thus described by conformal twisted geometries and not just Regge geometries.
In the Lorentzian case, the boundary data determine also which faces of the conformal twisted geometry are thin or thick wedges. This fact, which plays a key role to have a precise geometric reconstruction of the 4-simplex, holds for general graphs.
We use the word generically here, because our analysis does not provide a theorem: there could be special configurations that allow to solve all critical point equations without being angle-matched, or graphs for which not all angle matching conditions arise. The examples and cases treated makes us believe that this is not the case, but we cannot exclude it.

A non-trivial aspect of the geometric reconstruction is whether the critical 3d data admit a 4d embedding. This question is well-posed for the 3d Regge geometries. Still, it can be extended to conformal twisted geometries, once understood in terms of the value of deficit angles defined as sums of the 4d dihedral angles. The answer depends on the graph. Only the 4-simplex graph is known to admit a unique flat embedding for any 3d Regge geometry. In general, a case-by-case analysis is needed to see whether the 4d dihedral angles obtained from the analysis are consistent with a curved or flat embedding.\footnote{An important example of this fact has come to our attention after the publication of this paper, and motivates the addition of this comment in the present v3 on the ArXiv. The example concerns a graph corresponding to a $\{27j\}$ symbol, obtained as the boundary graph of the triangulation called $\D_3$ in \cite{Dona:2020tvv}. In this case, there are only triangular faces, thus only 3d Regge geometries in the critical set. These could have a priori both flat and curved 4d embeddings in terms of three 4-simplices sharing a common face, however the conditions on the holonomies coming from the 5-cycles connecting three triangular subgraphs can be seen to select only those Regge geometries that are 4d flat. The restriction of the critical behaviour to flat-only configurations was missed in the initial numerical analysis presented in \cite{Dona:2020tvv}, but corrected in the v3 once a more efficient numerical code became available. This improvement was motivated by the long discussion and recent results \cite{Engle:2020ffj} on the flatness problem in spin foam models.}

\subsection{Action at the critical points}

Even if the individual critical group elements depends on the choice of  dominating set, the products 
$h_a^{-1}h_b$ entering the action do not. 
Using \Ref{hsols1}, and the fact that orientation or angle-matching conditions have to be satisfied everywhere in the graph, the
evaluation of the on-shell action is identical to the procedure detailed in Section~\Ref{sec:Scrit}. 
For the vector geometries, the result is always \Ref{Svector}.
For angle-matched data, after recursive applications of \Ref{eq:halfangle} or \Ref{eq:halfcompohyp} depending on the sector, we find the same formal expressions \Ref{SER} and \Ref{SLR}, but with the Regge action $S_{\rm R}$ replaced by the general formula
\be\label{Sgen}
S_{\G}(j_{ab},\phi_{ab}^c) := \sum_{(ab)} j_{ab} \th_{ab}(\phi).
\ee
This action is well-defined for any cellular decomposition, and not just a triangulation, and with data which are conformal twisted geometries, and not just Regge geometries.

In the case of modular graphs, one group element per bridge between subgraphs has a critical value like \Ref{hsols2}, with an additional sign freedom.
If we denote ${\cal I}_i$ the sets of links bridging between various subgraphs, with $i$ from 1 to $n-1$, the composition of holonomies introduces the $\eta$-dependent angles for all bridging links. The resulting action this time is 
\be\label{Smod}
S^{\eta_i}_{\G}(j_{ab},\phi_{ab}^c) := \sum_{(ab)\notin \cup{\cal I}_i} j_{ab} \th_{ab}(\phi) + \sum_{i=1}^{n-1}\sum_{(ab)\in {\cal I}_i} j_{ab} \th_{ab}(\phi,\eta_i).
\ee
where $\th_{ab}(\phi,\eta_i)$ is defined by \Ref{3=4+4}.
This formula holds for globally angle-matched configurations, of any sector. For mixed configurations which are angle-matched in one subgraph and  vector geometries in another subgraph, the resulting on-shell action is a mixture of \Ref{Smod} and \Ref{Svector}.
For mixed Euclidean/Lorentzian data, $\g$ appears in front of all links inside the Lorentzian subgraphs. For the bridging links between subgraphs of different sectors, $\g$ appears only in front of the Lorentzian `half-angle' with the auxiliary  hyperplane.

\section{General asymptotic formulae}\label{SectionLO}
Based on the previous analysis, we can now present the asymptotic formulas valid for generic graphs, with all nodes of valence 4 or higher. Assuming that the Hessian determinant does not vanish at the critical points, we have the following leading order behaviours.
\begin{itemize}
\item For vector geometries data:
\be\label{genLOv}
A_\G =\f{(2\pi)^{3(N-1)} 2^{2L+N-1} \J}{\l^{3(N-1)}}  e^{i\Psi}  \f{\Om^{\scr (+)}}{\sqrt{\det{(-H^{\scr (+)})}}} + O(\l^{-3N+2}),
\ee
with $\Psi$ given by the on-shell action 
 \Ref{Svector}.

\item For non-degenerate Euclidean angle-matched data, on non-modular graphs:
\begin{equation}\label{genLOe}
A_\G 
=\f{(2\pi)^{3(N-1)} 2^{2L+N-1} \J }{\l^{3(N-1)}}  
e^{i\Psi}
\bigg( \f{\Om^{\scr (+)} e^{i\lambda S_{\G} } }{\sqrt{\det{(-H^{\scr (+)})}}} 
+ \f{\Om^{\scr (-)} e^{-i\lambda S_{\G} } } {\sqrt{\det{(-H^{\scr (-)})}}} 
 \bigg) + O(\l^{-3N+2}),
 \end{equation} 
with $S_{\G}$ given by \Ref{Sgen} with Euclidean data. Euclidean 3d Regge geometries, and flat 4d polytopes, are subcases with the Regge action appearing.
 \Ref{SER}
\item For non-degenerate Lorentzian angle-matched data, on non-modular graphs:
\begin{equation}\label{genLO}
A_\G=\f{(2\pi)^{3(N-1)} 2^{2L+N-1} \J }{\l^{3(N-1)}}   
 (-1)^\chi e^{i\Psi}
\bigg( \f{\Om^{\scr (+)} e^{i\lambda \g S_{\G} } }{\sqrt{\det{(-H^{\scr (+)})}}} 
+ \f{\Om^{\scr (-)} e^{-i\lambda \g S_{\G} } } {\sqrt{\det{(-H^{\scr (-)})}}} 
 \bigg) + O(\l^{-3N+2}),
\end{equation}
with $S_{\G}$ given by \Ref{Sgen} with Euclidean data, and
 $\chi=\l \sum_{(ab)} j_{ab} \Pi_{ab}$ depends on the number of  anti-chronal pairs of tetrahedra (thin wedges). Lorentzian 3d Regge geometries, and flat 4d polytopes, are subcases with the Regge action appearing.
\end{itemize}
From the general formula \Ref{eq:asymp} or \Ref{genLO}, we see that the fall-off exponent 
of the leading order in $\l$ increases linearly with the number of nodes of the vertex graph, namely with the valence of the vertex. The 4-simplex amplitude is the dominant one among those admitting a 4d Regge interpretation.
The key to the general asymptotic formula \Ref{genLO} is the fact that edge-independent 4d dihedral angles are defined for all conformal twisted geometries, therefore one can write the action \Ref{Sgen} for all multi-critical boundary data, and not just for Regge geometries. For boundary data describing flat-embeddable 3d Regge geometries, the action is a direct generalization of the Regge action from the 4-simplex to a flat 4-polytope. For 3d Regge geometries that are not flat-embeddable, the action could be interpreted as a boundary term for a curved polytope.
We refer the reader to \cite{Dona:2017dvf} for a more detailed discussion of this action, 
 including its relation to area-angle Regge calculus \cite{DittrichSpeziale}, and to a generalized Regge calculus based on 4d polytopes.

One difference with respect to the 4-simplex analysis is that we have not expressed the angle-matched asymptotics using cosines, because we have no evidence that $H^{\scr (+)}=\overline{H^{\scr (-)}}$ for general graphs. The Hessian matrix can be computed analytically as shown in \cite{BarrettLorAsymp,Dona:2019dkf}. For its determinant one will in general have to be content with numerical evaluations, which we did not attempt here.
If one is not interested in the overall phase, the calculation can be simplified using a reduced Hessian like in \cite{Kaminski:2019dld}. It would of course be useful to extend the analytic proof of non-degeneracy of the 4-simplex Hessian presented in \cite{Kaminski:2019dld}, but this would require further work. 

For $n$-modular graphs, there is also the possibility of mixed boundary data, and up to $n$ distinct critical points. We have thus a richer classification.
If the boundary data are globally vector geometries, we have one critical point and the same asymptotics \Ref{genLOv}. If one subgraph satisfies the  angle-matching conditions, Euclidean or Lorentzian, and the rest is vector geometries, we have two critical points, labelled by a sign. In this case the asymptotics is given by \Ref{genLOe}, with $\Psi$ including all links but $S_{\G}$ including only the links with angle-matched data. If two subgraphs are angle-matched, and the rest vector geometries, we have four distinct critical points labelled by two signs, and so on.
When all boundary data are non-degenerate angle-matched, we have $2^n$ critical points. We label them  by signs $\eps$ and $\eta_i$, $i=1$ to $n-1$, and the leading order asymptotic formula for globally Euclidean data is
\begin{align}
A_\G
= \f{(2\pi)^{3(N-1)} 2^{2L+N-1} \J}{\l^{3(N-1)}} 
e^{i\Psi} \sum_{\eps=\pm} \sum_{\eta_i=\pm}
 \f{\Om^{\scr (\eps;\eta_i)}}{\sqrt{\det{(-H^{\scr (\eps;\eta_i)})}}} e^{i\eps \lambda S_{\G}^{\eta_i}}.
\end{align}
The formula for Lorentzian angle-matched data is the same, with $\gamma$ in front of $S_{\G}$, Lorentzian dihedral angles, and the additional $(-1)^\chi$ phase. Similarly for mixed data, with the form of the action discussed below \Ref{Smod}.

We have restricted attention to graphs with nodes of valence 4 or higher. This choice, together with the exclusion of degenerate configurations with coplanar normals, allows us to interpret the 3d boundary data as a collection of polyhedra, and classify them according to whether they represent or not the boundary of a curved, or flat, polytope. Our technique can also be applied to graphs with 3-valent nodes. In this case, the data satisfying closure are necessarily co-planar and represent the edges of triangles. Their scalar products define  2d dihedral angles, and one can use the same construction presented here, with \Ref{defhth} being spherical cosine laws from 2d to 3d dihedral angles. 
The classification of boundary data and determination of critical points will follow a different geometric interpretation, based on 2-dimensional polygonal boundaries, and 3d bulk geometries. Furthermore, the edge-independent conditions are trivially satisfied by the coplanarity of the normal vectors. Therefore all closed boundary data admit two critical points (or more in the modular case), be them Euclidean or Lorentzian. With these differences in mind, we expect our the asymptotic formulas to hold.
On the other hand, for nodes of valence 4 or higher with coplanar normals which are not vector geometries, a different technique is needed.

\subsection{Orientation dependence of the global phase}
In physical applications of the coherent amplitudes and its asymptotics, it is important to keep in mind the relative arbitrariness of the global phase.
As we have discussed above, it depends on the conventions chosen for the irrep matrices and for the $\SU(2)$ coherent states. As well as on the way the parity map to avoid signs in the closure conditions is implemented. But even once all these conventions are uniquely specified, the global phase depends on the orientation of the normals. This is to be contrasted with the norm of the amplitude, which is independent of the orientation, and depends thus only on the geometry reconstructed from areas and normals. In the saddle point analysis, the critical values $h_a^{\scr (c)}$ and $z_{ab}^{\scr (c)}$
depend on the orientation of the normals, and so does the on-shell action \Ref{eq:actioncp}. But the \emph{difference} between the action at two critical points is orientation-invariant.

To prove this, consider boundary data allowing multiple critical point solutions. If we rotate the normals with $R_a\in\SO(3)$ node by node as in \Ref{gauge}, the on-shell action transforms as follows,
\begin{equation}
S(h_a^{\scr (c)},z_{ab}^{\scr (c)};\z_{ab}) \to S(h_a^{\scr (c)} r_a,z_{ab}^{\scr (c)}; r_a\z_{ab}) - i \sum_{(ab)} j_{ab} (\chi_{ab}-\chi_{ba})
\equiv S(h_a^{\scr (c)},z_{ab}^{\scr (c)};\z_{ab}) - i \sum_{(ab)} j_{ab} (\chi_{ab}-\chi_{ba}).
\end{equation}
where for clarity we made explicit the parametric dependence of the action on the boundary spinors $\z_{ab}$.
From this it immediately follows that 
\be
S(h_a^{\scr (c_1)},z_{ab}^{\scr (c_1)}; \z_{ab}) - S(h_a^{\scr (c_2)} ,z_{ab}^{\scr (c_2)}; \z_{ab}) 
\equiv S(h_a^{\scr (c_1)} r_a,z_{ab}^{\scr (c_1)}; r_a\z_{ab}) - S(h_a^{\scr (c_2)} r_a,z_{ab}^{\scr (c_2)}; r_a\z_{ab}). 
\ee

We conclude that in the asymptotic formulas presented above, the global phase $\Psi$ is orientation-dependent, but the relative phase $S_{\G}$ is not.
This is the crucial difference between vector geometries, which admit a single critical point, and conformal twisted geometries, which admit two or more: only for the latter the oscillatory behaviour of the asymptotic formula contains an orientation-invariant information.

\section{Examples} 
\label{sec:examples}

\subsection{Complete graph $K_6$: appearance of conformal twisted geometries}
\label{sec:example1}

\begin{figure}[H]
\centering
\includegraphics[width=4cm]{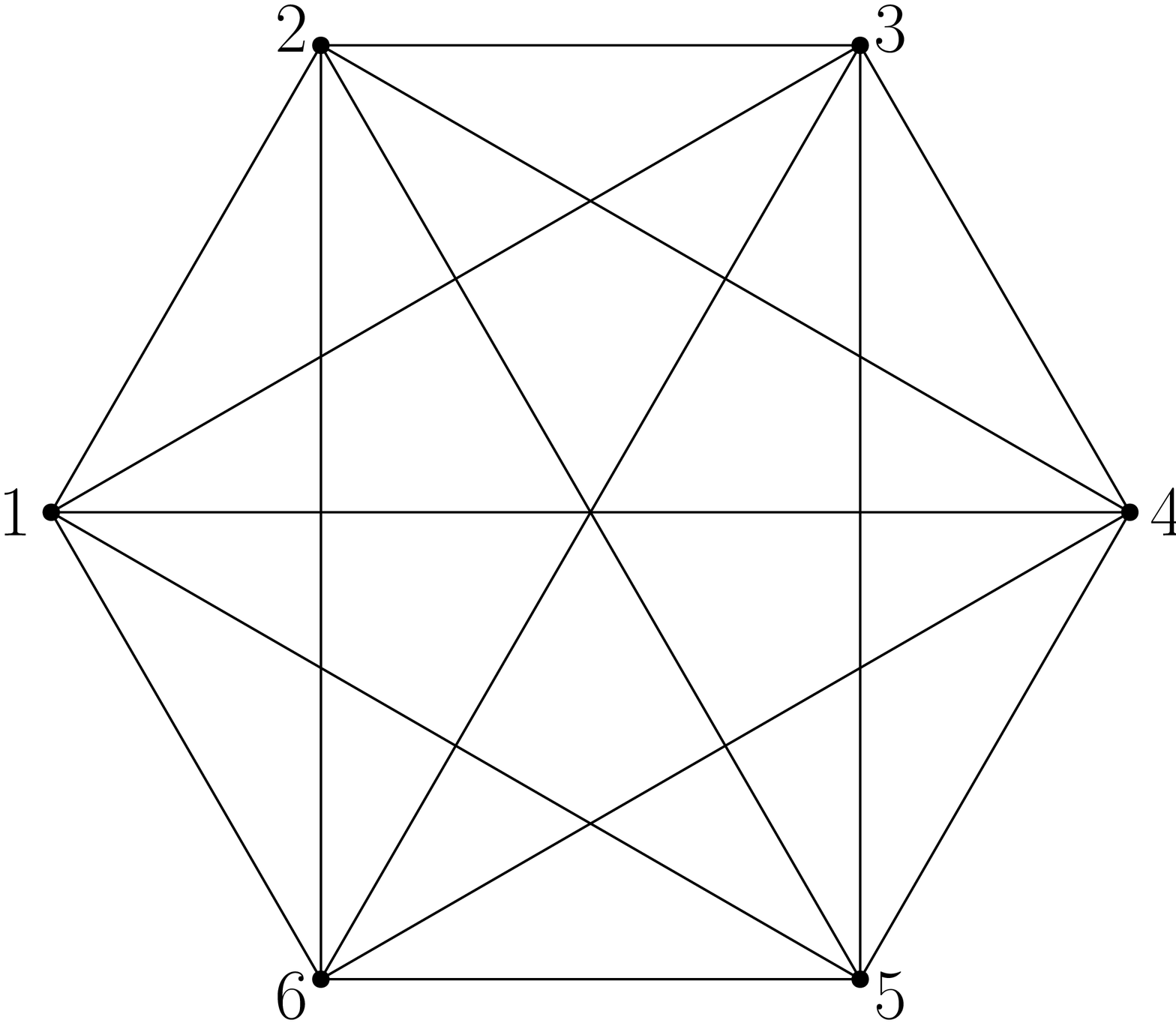}\hspace{2cm}\includegraphics[width=3.6cm]{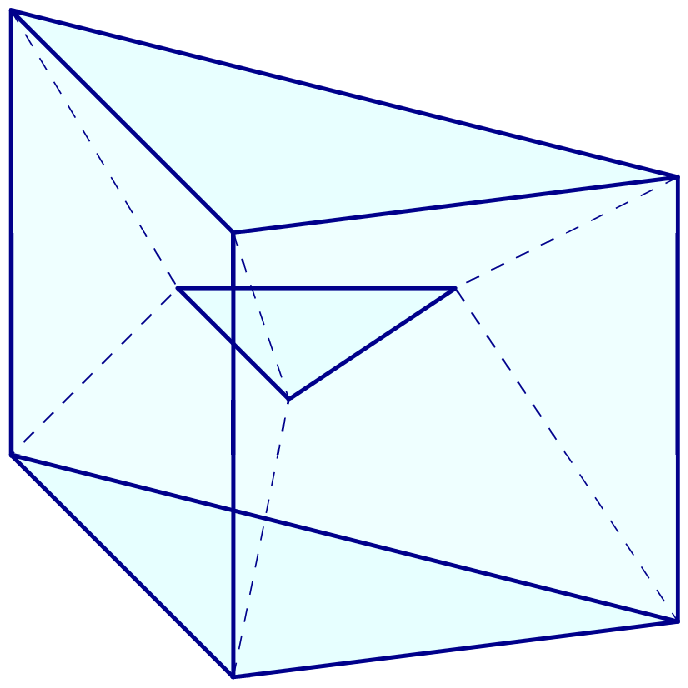}
\caption{\small{\emph{Complete graph with six nodes, combinatorially dual to the boundary of a 3-3 duo-prism. When the boundary data describe a geometric 3-3 duo-prism, the associated vertex asymptotics has two distinct critical points. It can be either Euclidean, or Lorentzian one with all faces space-like. The signature is determined by the boundary data, and can be read off the spherical cosine laws. However, two distinct critical points are also present for more general boundary geometries.
These consist of 3d Regge geometries that cannot be flat embedded as a 3-3 duo-prism, and of angle-matched twisted geometries that are not 3d Regge geometries. When the boundary data are vector geometries, there is a single critical point.
 }}
\label{fig:K6}
}
\end{figure}

Our first case study beyond the graph $K_5$ of the 4-simplex  is $K_6$, the complete graph with $6$ (five-valent) nodes, see Fig~\ref{fig:K6}. Combinatorially, it is dual to the boundary of a flat convex 4d polytope known as 3-3 duo-prism.
We look first at the closure conditions, and describe the geometric interpretation of the boundary data that solve them. Because the nodes are all 5-valent, boundary data satisfying the closure conditions describe a collection of six polyhedra with 5 faces each, all sharing one face with their neighbour following the connectivity of the graph. 
Generically, the polyhedra with five faces are triangular prisms, but (measure-zero) subsets of the data  describe square-based pyramids. See \cite{IoPoly} for more details. These are the (closed) twisted geometry data for this graph. 
Since each face can be at most a tetragon, the conformal twisted geometry subset requires four edge-independence conditions on the twist angles $\xg$ per link.
The subset of 3d Regge data has 18 degrees of freedom, corresponding to the number of edges of a generic shape-matched configuration.
On the other hand, the 3-3 duo-prism has 14 degrees of freedom, as can be easily counted using Minkowski theorem, see Appendix~\ref{AppTacchino}. This means that only a codimension-4 subset of the 3d Regge data can be interpreted as the boundary geometry of a flat convex 3-3 duo-prism. 
For a 4d interpretation of the remaining 3d Regge data, one could look for another polytope with the same boundary graph, convex or concave, and more degrees of freedom than the 3-3 duo-prims; or consider a more general interpretation in terms of a curved polytope, exploring for instance a splitting of the 3-3 duo-prism in 4-simplices. 
 
With this geometric picture in mind, we now look at  the \magg equations for the group elements.
We  require only the first step of the algorithm, since all nodes are first neighbours. 
We pick 1 as the root node for our algorithm, and partially fix the gauge requiring $\vec n_{a1} = -\vec n_{1 a}$.
The critical point equations involving the node $1$, namely  \Ref{eq:direction} with $b=1$ and $a=2$ to 6, fix the directions of the group elements along $\vec{n}_{a1}$, according to \Ref{4screw}. The other critical point equations, with $a,b\neq 1$, lead to \Ref{Eqstep1}, and are solved by \Ref{sol1}.
The plus sign solution requires the orientation conditions \Ref{vector} for all links. The minus sign requires the edge-independence conditions \Ref{ei1} for all links $1a$.
Each link belongs to four 3-cycles, therefore we have four edge-independence conditions per link $1a$. By the analysis of Section~\ref{SecSCL}, we satisfy them with four angle-matching conditions at the faces $1a$, plus additional angle-matchings at the faces $ab$. However, independence of the result from the choice of starting node implies that one needs four angle-matching conditions on every link. 
The faces have a unique area given by the spin and  are either triangles or squares by the closure conditions. 
In the first case, the shapes match. In the second case, we have the freedom of one conformal transformation. 
Therefore four angle-matching conditions are always enough to restrict the boundary data to be conformal twisted geometries.
The two critical points belong to the Euclidean or Lorentzian sector depending on the boundary data, as determined by the domain of the spherical cosine law formula. All $\om_a$'s must be either Euclidean or Lorentzian, and their sign is fixed up to the global freedom.
The boundary data determine also the number of co-chronal and anti-chronal configurations.

The asymptotic formulae for the vertex amplitude associated with this graph can be read from the general ones presented in Section~\ref{SectionLO}, specializing to $L=15$ and $N=6$. For instance for Lorentzian angle-matched data, 
\be
A_\G=\f{2^{50}\pi^{15} \J }{\l^{15}}  
 (-1)^\chi e^{i\Psi}
\bigg( \f{\Om^{\scr (+)} e^{i\lambda \g S_{\G} } }{\sqrt{\det{-H^{\scr (+)}}}} 
+ \f{\Om^{\scr (-)} e^{-i\lambda \g S_{\G} } } {\sqrt{\det{-H^{\scr (-)}}}} 
 \bigg) + O(\l^{-16}).
\ee

An identical analysis can be applied to $K_N$, the complete graph with $N$ nodes. This is in general not dual to the boundary of a 4d polytope, therefore one has to be careful with the geometrical interpretation of the variables.

\subsection{Hypercubic graph: presence of a second neighbour}
\label{sec:example2}
\begin{figure}[H]
\centering
\includegraphics[width=4cm]{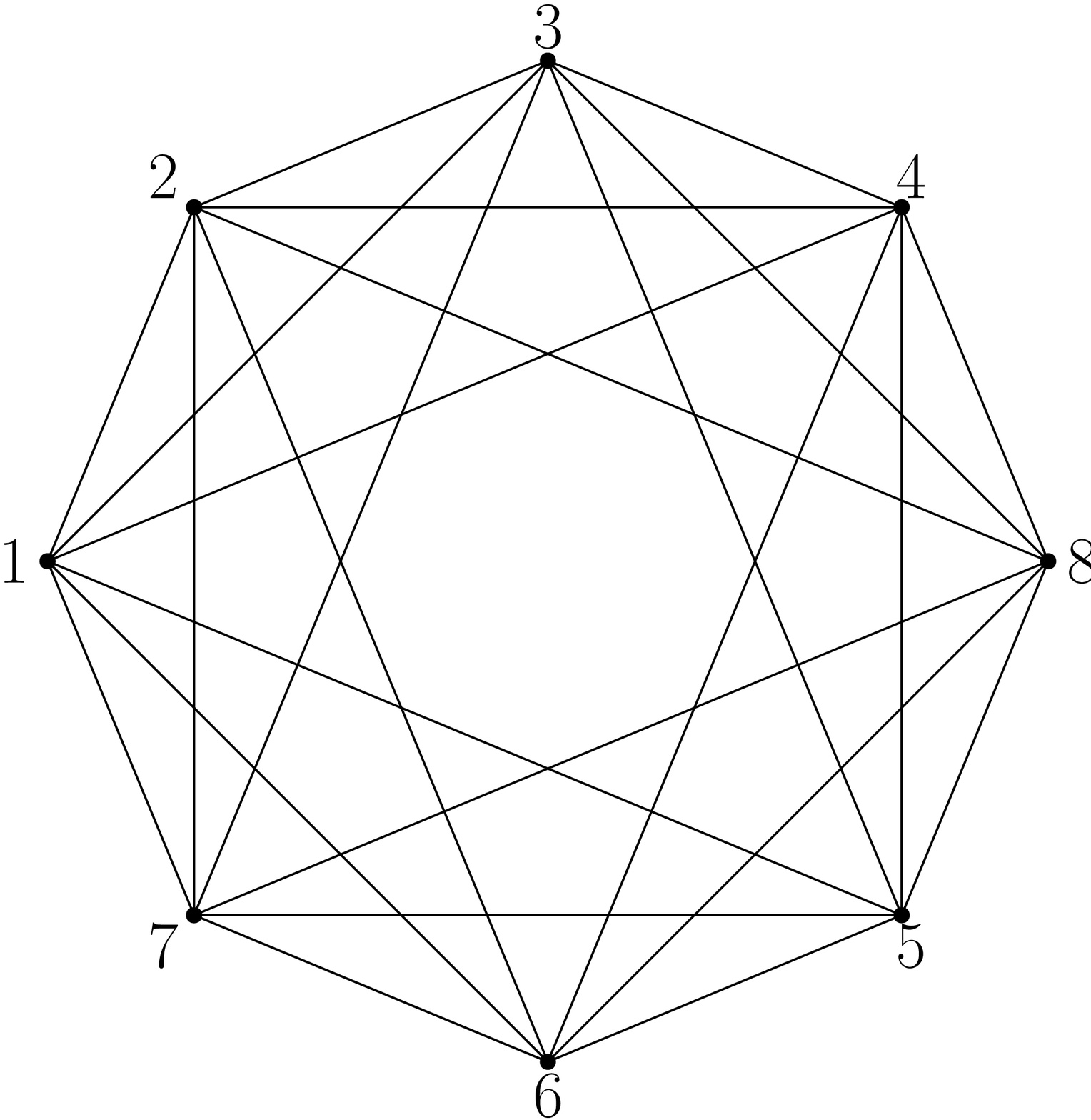}\hspace{2cm}\includegraphics[width=3.6cm]{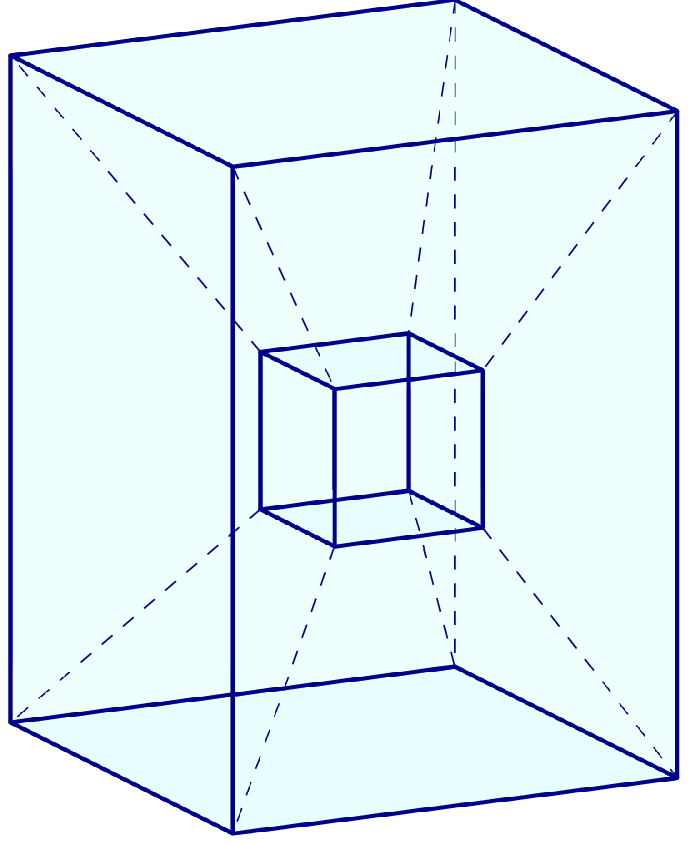}
\caption{\small{\emph{Hypercubic graph with eight nodes: this is combinatorially dual to the boundary of a hypercuboid. When the boundary data describe a geometric hypercuboid, or any other polytope with the same boundary graph, the associated vertex asymptotics has two distinct critical points. There are both Euclidean and Lorentzian sectors, distinguished by the spherical cosine laws. Two distinct critical points are present for more general boundary data that are not the boundary of a hypercuboid. These are 3d Regge geometries that cannot be flat embedded in the hypercuboid, and conformal twisted geometries that are not 3d Regge geometries. When the boundary data are vector geometries, there is  a single critical point.
} }
\label{fig:hypercube}
}
\end{figure}
Our next example of general analysis is the hypercubic graph, see Fig.~\ref{fig:hypercube}. The name refers to the combinatorics of the graph, which coincides with the dual of the boundary of an hypercube, or tesseract.
The closure conditions require boundary data that describe a collection of six-faced polyhedra. In the generic case, they can be either cuboids, or pentagonal wedges, with 12 edges. In special cases, they describe polyhedra with lesser edges. There are five polyhedral classes with codimensions from one to three. For instance, the pinched cuboid with two triangles and four tetragons has codimension one, or the double pyramid with all triangular faces has codimension three. Any 4d flat polytopes with boundary geometries described in terms of these 6-faced polyhedra and the connectivity of the graph, will turn out to provide data admitting two solutions of the critical point equations. While we are not aware of any explicit classification, we know of at least two examples: the hypercuboid, and the pinched hypercuboid with five cuboids and three pinched cuboids.\footnote{
We thank Jonathan Bowers for providing us with this example, and invite the reader interested in polytopes to look at his  webpage at \texttt{http://www.polytope.net/hedrondude/regulars.htm}} 

\medskip
We now run our algorithm and prove that these geometries, as well as more general angle-matched twisted geometries, admit solutions to the critical point equations. We choose 1 as the root node of the algorithm, with $h_1=\Id$. The first step of the algorithm determines all group elements $h_a$ with $a=2$ to $7$,
with directions $\vec n_{a1}$ and angles \Ref{sol1}.
These group elements must furthermore all be Euclidean or Lorentzian, otherwise the sine equations \Ref{DDR} are violated, and there is a single sign freedom $\eps_1$. The plus sign is valid if the normals satisfy the orientation condition at every link involving the nodes 1 to 7, whereas the minus sign requires the edge-independence conditions \Ref{ei1}.
Each link $1a$ belongs to four 3-cycles. Hence these are four edge-independence conditions at the faces $1a$. Since the faces can be at most pentagons, this is not yet enough to always fix angle-matching of the faces $1a$.

It remains to determine the group element of the second neighbour, $a_2=8$. For this we move to step 2 of the algorithm, and pick a new seed among the first neighbours, say $\ab=2$. 
There is one type-12 equation, which imposes $\tilde{H}_{a_{2}}:=H_{\ab}^{-1} H_{a_{2}}$ to be a complex rotation in the direction $\vec{n}_{a_{2}\ab}$,  as in \Ref{4screw2}.
There are no type-22 equations, thus we look at the type-21 to determine $\om_{a_2}$. There are five type-21 equations, $H_{a_{2}}\vec{n}_{a_{2}b}=-H_{b}\vec{n}_{ba_{2}}$, where $b=3$ to $7$. They involve the 4-cycles $1\ab a_2 b = 128b$. 
We rewrite them multiplying by $H_{\ab}^{-1}$ on both sides, 
so to have all directions already known. We obtain \Ref{3H} and their scalar projections \Ref{capitani3}.
The first cosine equations determine the missing complex angle as
\be\label{hc2}
\om_{a_2}=\eps_{a_2}^b \hth^{b}_{a_2 \ab} - \xg^b_{a_2 \ab}, \qquad \eps_{a_2}^b=\pm 1.
\ee
By the usual argument, $\eps_{a_2}^b=1$ is a valid solution if the normals satisfy the orientation equations \Ref{vector}
at the links $a_2 \ab$, $a_2 b$ and $b \ab$, and the minus sign requires five edge-independence conditions \Ref{ei3}, with $\eps_{a_2}^b=\eps_{a_2}$ for all $b$.
Plugging \Ref{hc2} in \Ref{DeRossi3}, we find that the group element at $a_2$ must be in the same sector (Euclidean or Lorentzian) of the group elements determined in the first step, and that
\be\label{epscubo}
\eps_{a_2}=\eps_1.
\ee
There can be at most two solutions. 

It remains to deal with the last set of equations, \Ref{Totti3b}, which are only restrictions on the boundary data. 
In these equations, the
angles $\vth^{b}_{s1}$ and $\vth^{a_2}_{s\ab}$ refer to the hyperplane $s=(\ab 1,b a_2)$ and are defined by
\begin{equation}\label{mocktheta}
\cos\vth_{s1}^{b}:=
\frac{\cos\phi_{a_{2}1}^{b}+\cos\phi_{b\ab}^{1}\cos\varphi^{s}_{b1}}
{\sin\phi_{b\ab}^{1}\sin\varphi^{s}_{b1}},
\qquad
\cos \vth_{s\ab}^{a_2}:=
\frac{\cos\phi_{\ab b}^{a_2}+\cos\phi_{a_{2}1}^{\ab} \cos\varphi^{s}_{b1}}
{\sin\phi_{a_{2}1}^{\ab}\sin\varphi^{s}_{b1}},
\end{equation}
where
\begin{equation}
\label{varphis}
\cos\varphi^{s}_{b1} := \cos(\eps_1\hth_{b1}^d - \xg_{b1}^d +\xg_{b1}^{ a_{2} \ab})\sin\phi_{b \ab}^{1}\sin\phi_{a_{2}1}^{b}-\cos\phi_{b \ab}^{1}\cos\phi_{a_{2}1}^{b}.
\end{equation}
These angles are a `mismatched' version of \Ref{faisola} and \Ref{4daux}, from which they differ because of the presence of the $\xg$'s in \Ref{varphi}.\footnote{The off-shell version of \Ref{Totti3b}, prior to using \Ref{sol1}, 
has $\omega_{b}$ instead of $\eps_1\hth_{b1}^d - \xg_{b1}^d$ as argument of the first cosine defining $\varphi$, and 
$$
\arcsin\left(\f{\sin (|\om_{b}+\xg^{\ab a_2}_{b 1}|) \sin\phi^b_{a_2 1}}{\sin\phi^s_{a_2 1}}\right)
$$
instead of $\xg^{a_2 b}_{\ab 1} - \eps_1 \vth^{b}_{s1}$ in the left-hand side.
As an off-shell equation, \Ref{Totti3b} is thus a relation between $\om_{\ab}$ and $\om_b$.} 
We argue that  \Ref{Totti3b} can be satisfied only if 
\be\label{fifth}
\xg^{a_2 \ab}_{b 1}=\xg^d_{b1}, \qquad \xg^{a_2 b}_{\ab 1}=\xg^c_{\ab 1}.
\ee
This can be proved by a perturbative expansion of \Ref{Totti3b} around \Ref{fifth}, which can be written as
\begin{align}\label{119}
& \cos(\eps_1\hth^c_{\ab 1} - \eps_1 \hth^{b}_{s1} ) + \cos\hth^{a_2}_{s\ab}
\\\nn&\quad + A(\phi^1_{b\ab}, \phi^b_{a_21},\th^{c}_{\ab 1},\th^{d}_{b1})\d\xg^{a_2 b}_{\ab 1} 
+ B(\phi^1_{b\ab}, \phi^b_{a_21},\th^{c}_{\ab 1},\th^{d}_{b1}, \phi^{a_2}_{\ab b},\phi^{\ab}_{a_21})\d\xg^{a_2 \ab}_{b 1} +\ldots =0.
\end{align}
The functions $A$ and $B$ can be computed using
\begin{align}
& \cos\varphi^s_{\ab_2 1} = \cos\phi^s_{\ab_2 1} - \sin\hth^c_{b1} \sin\phi^1_{b\ab}\sin\phi^b_{\ab_21} \, \d\xi^{a_2\ab}_{b1} +\ldots, 
\end{align}
but their explicit expression is not needed here: it suffices to observe that they depend on different angles at $a_2$, which allows us to vary them arbitrarily. Therefore the only way to satisfy the equality is when the parameters of the expansion vanish, namely when \Ref{fifth} hold.
We are then left with the first line of \Ref{119}, which implies \Ref{ei5}. Since the $\hth$'s are all known, this simply fixes 
the sign $\eps_{s}^{a_2}$ in terms of the boundary data and $\eps_1$, as described in the algorithm, provided the right-hand side is edge-independent.
This gives five edge-independence conditions as we vary the 4-cycle and its associated hyperplane $s$.

We have at this point determined all group elements, and exposed the restrictions on the boundary data necessary to solve all \magg equations.
Let us collect them, to summarize the result of the analysis. To have the vector geometry solution, we need the orientation conditions at every face. These are automatically satisfied by our choice of gauge for the links in the dominating path, but restrict non-trivially the data for the other links. To have two solutions, we need 
the $\hth$'s and $\xg$'s 
to satisfy five edge-independent conditions at the faces $1a$ and $\ab a_2$, involving both 3-cycles and 4-cycles.\footnote{Even more than 5, precisely 9, for the link $1\ab$ in the dominating path, but these are redundant matchings among virtual edges which are not in the boundary, as discussed in Section~\ref{SecSCL}. }
But since we could have chosen any other dominant set, the five edge-independence conditions must hold at every face. By the closure conditions the faces can be at most pentagons, hence five edge-independent conditions imply matching of all 2d angles. We conclude that angle-matched, conformal twisted geometries admit two distinct critical points.
Notice also that if the data at every node except 8 satisfy the angle-matching conditions, but the data at 8 do not, the sign restriction \Ref{epscubo} imposes a unique solution: only data that are globally angle-matched have two critical points. 

The asymptotic formulae for this graph can be obtain from the general ones of Section~\ref{SectionLO}, adapting to $L=24$ and $N=8$. For instance for Lorentzian angle-matched data,
\be
A_\G=\f{2^{76}\pi^{21} \J }{\l^{21}}  
e^{i\Psi} (-1)^\chi \bigg( \f{\Om^{\scr (+)} e^{i\lambda \g S_{\G} } }{\sqrt{\det{-H^{\scr (+)}}}} 
+ \f{\Om^{\scr (-)} e^{-i\lambda \g S_{\G} } } {\sqrt{\det{-H^{\scr (-)}}}} 
 \bigg) + O(\l^{-22}).
\ee

\begin{figure}[H]
\centering
\includegraphics[width=5cm]{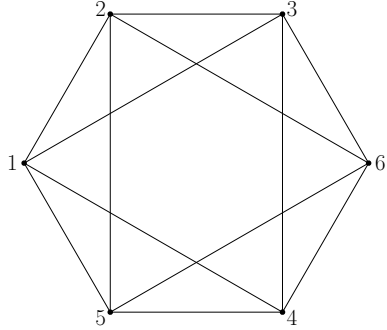}
\caption{\small{\emph{Example of a graph that is not dual to the boundary of a 4d polytope. Two distinct critical points occur for conformal twisted geometries and 3d Regge data, but none with a  straightforward 4d interpretation.}
\label{fig:K6b} }
}
\end{figure}
A simpler graph with a second neighbour can be obtained starting from $K_6$ and removing three links, see the right-panel of Fig.~\ref{fig:K6b}. The amplitude for this vertex graph has been recently considered \cite{CarloNew} in order to study aspects of the spin foam calculation of black hole tunnelling \cite{Christodoulou:2016vny}. The  saddle point analysis proceeds exactly as for the hypercubic example above, with two steps of the algorithm required. One finds a critical point for vector geometries and two for angle-matched geometries, Euclidean or Lorentzian. The latter coincide with 3d Regge geometries, since the nodes are all 4-valent. 
The difference with this graph is that it is not dual to any 4d polytope. Therefore there is no subset of the 3d Regge data with a straightforward 4d interpretation.

For the reader interested in more details, we sketch the algorithm for this graph. We take $1$ and 2 as dominating set. The \magg equations connected to 1 determine the directions of the group elements on the first neighbours 2 to 5. The equations among the first neighbours give \Ref{Eqstep1}, which fix the four-screw angles to a unique solution if the orientation conditions are satisfied for all links except those with node 6, or two  if two edge-independent conditions are satisfied at the faces $1a$. For the second step, we use the \magg equation connecting 2 and 6 to fix the direction of $h_6$, and we are left with three type-21 equations connecting 6 to the nodes 3 to 5. These three determine $\om_6$ through \Ref{Totti3},  provided the orientation conditions are satisfied at the links with node 6, or three edge-independence conditions are satisfied at the face 2-6. By \Ref{DeRossi3} the data must be in the same sector, and with a unique sign freedom.
A third edge-independent condition at $1a$ comes from \Ref{Totti3b} and following the same analysis explained in the case of the hypercube graph.
Together with the those arising at step 1, these conditions are equivalent to angle-matching conditions for the faces $1a$ and $26$. 
By independence of the choice of dominating set, the angle-matching conditions must be satisfied at every face, to have two critical points.

\subsection{The refined hypercubic graph: multiple critical point solutions}
\label{sec:example3}

\begin{figure}[H]
\centering
\includegraphics[width=5cm]{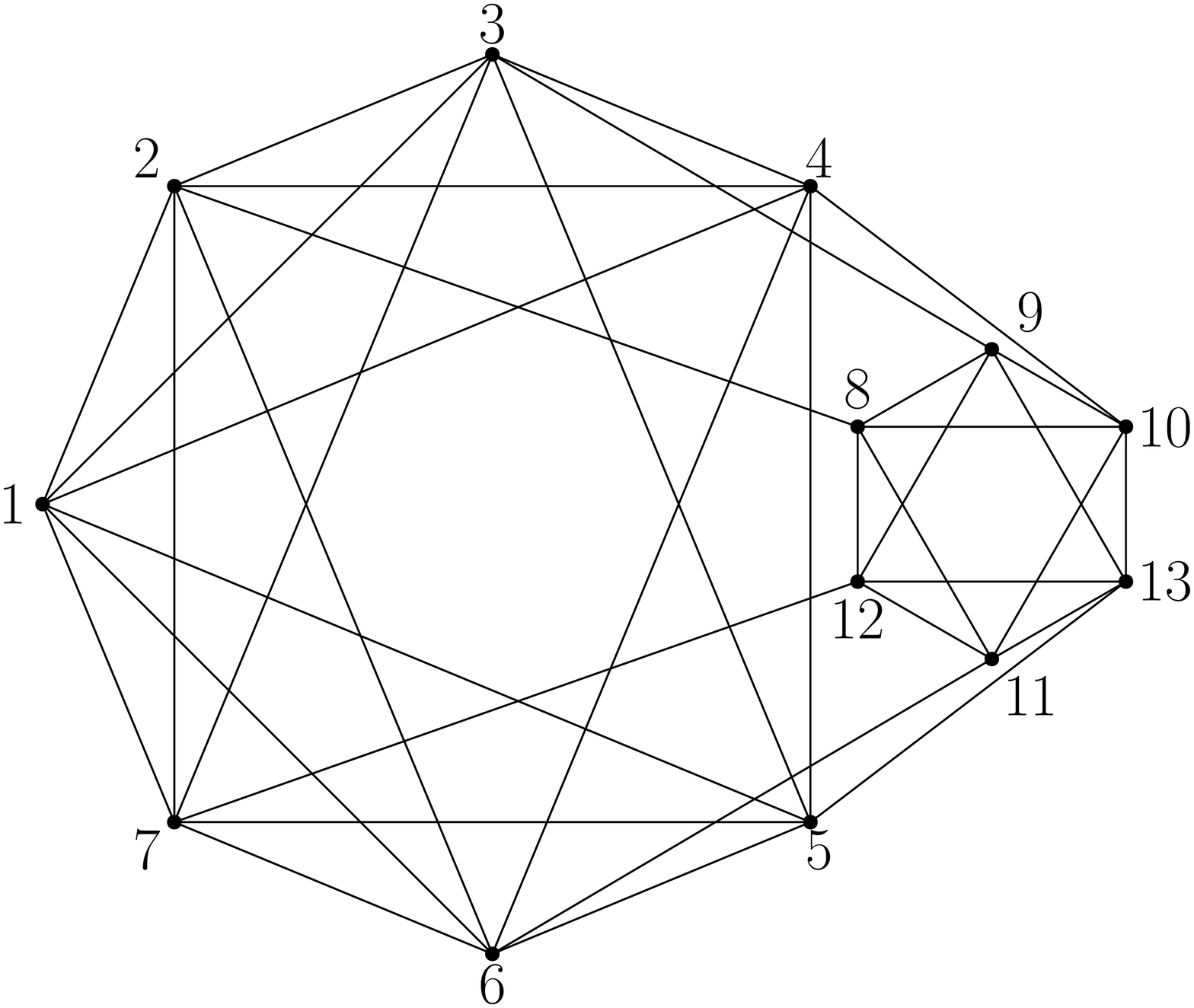}\hspace{2cm}\includegraphics[width=3.6cm]{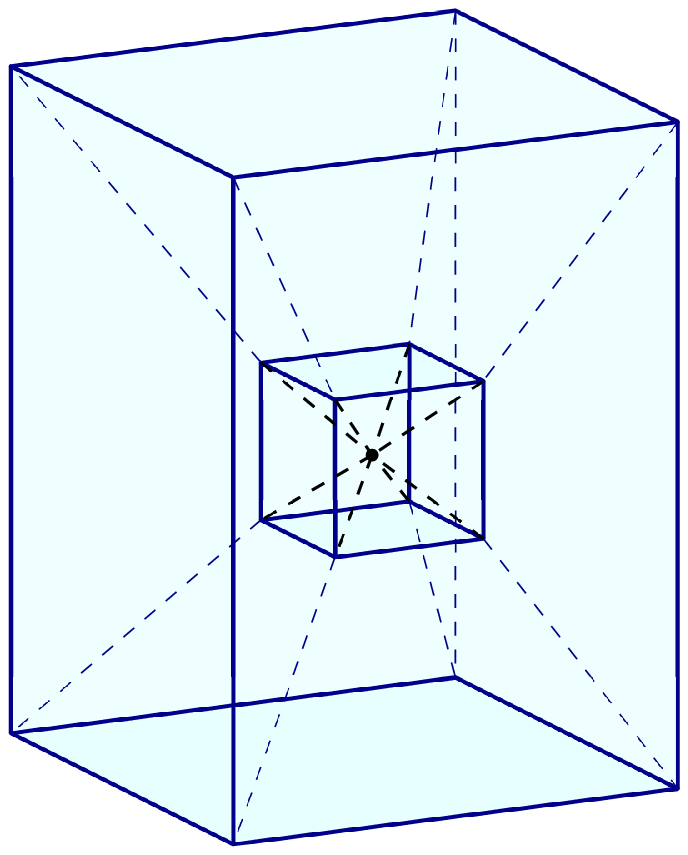}
\caption{\small{\emph{Example of a 2-modular graph, with subgraphs $1-7$ and $8-13$. The dominating set has 4 nodes, e.g. $(1,2,8,9)$, and the algorithm requires 4 steps. The boundary data describing a geometric 4d polytope have four critical points, and corresponding to convex and a concave versions of the polytope, which cannot be distinguished from the 3d data. Four distinct points are present for more general boundary data, corresponding to angle-matched twisted geometries, be them Euclidean or Lorentzian in both subgraphs, or mixed. Two distinct critical points occur for data that are angle-matched in one subgraph and vector geometries in the other subgraph.
}}
\label{fig:santa}
}
\end{figure}

\begin{figure}[H]
\centering
\includegraphics[width=5cm]{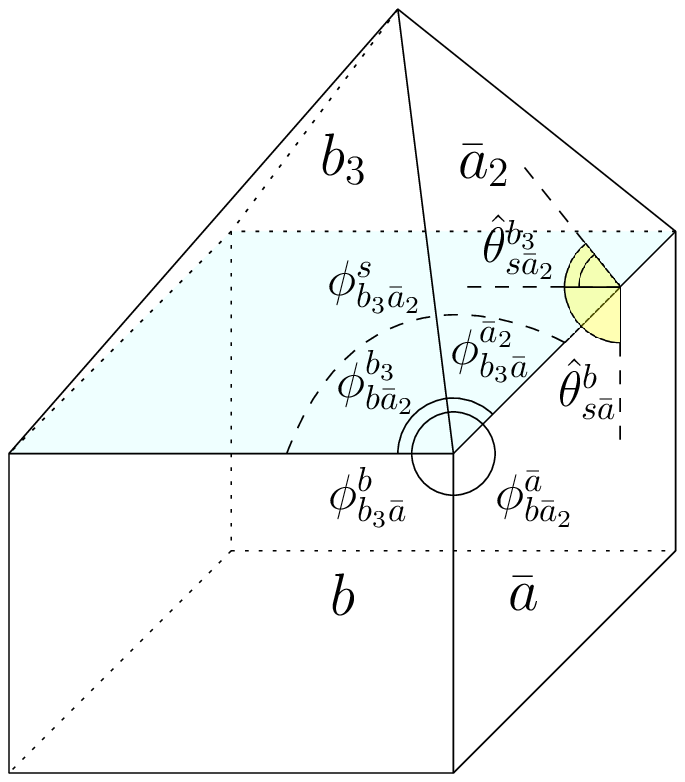} \hspace{2cm}
\includegraphics[width=5cm]{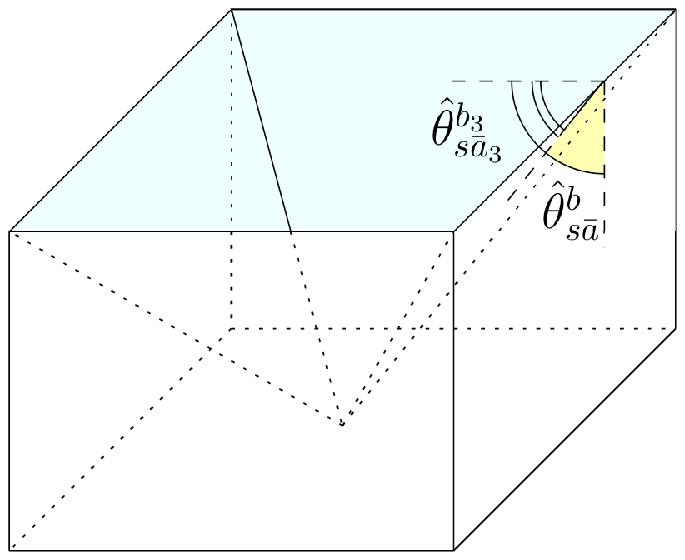}
\caption{\small{\emph{A convex and a concave 3d polyhedron with the same boundary data. This is a 3d version of the refined hypercube of Fig.~\ref{fig:santa}, with squares instead of cubes and triangles instead of pyramids. The convex and concave configuration are shown to differ by the bulk  dihedral angle between one side of the wall, say $\ab$, and one side of the roof, say $\ab_2$. This angle is in one case the sum, and in the other case the difference, of the bulk angles with the auxiliary plane $s$, the `attic' of the house. We use $s$ for speculum, since that plane (or hyperplane in the 4d case) acts as a mirror for the convex and concave configurations. Notice that the convex/concave ambiguity exist for any conformal twisted geometry as boundary data, and not just for 3d Regge geometries: angle-matched configurations are in fact enough to identify a unique (hyper)plane $s$. For more general boundary data on the other hand, $s$ depends on the 4-cycle chosen to define it, e.g. $s=(\ab\ab_2,bb_3)$. 
The labels for the faces have been chosen to provide a visual reference to the critical point analysis performed later. With respect to the numbering of Fig.~\ref{fig:santa}, we can take for instance $s=(28,39)$.
}}
\label{casetta}
}
\end{figure}

Our last case study is a refined hypercube graph considered in \cite{Bahr:2018vvq}, see Fig.~\ref{fig:santa}. It is more than three-link connected and therefore the Lorentzian amplitude well-defined.
It is the boundary graph of a 4d polytope obtained subdividing a boundary cuboid of the hypercube into six pyramids. This polytope can be either convex or concave, depending on the way we embed the six pyramids. It is important to notice that both convex and concave polytopes induce the same 3d boundary geometry, see Fig.~\ref{casetta} for a picture in one less dimension.\footnote{A similar situation occurs for 3-valent graphs, easier to visualize thanks to their 3d interpretation. For the triangular prism graph $Y_3$, corresponding to the reducible $9j$-symbol vertex amplitude of BF theory, the critical group elements describe either a convex or a concave gluing of two tetrahedra into a triangular bi-pyramid, both of which have the same boundary data \cite{Gozzini:2019kui}.} 

The closure condition is solved by 6-faced polyhedra for the 6-valent nodes of the first subgraph, and 5-faced polyhedra for the 5-valent nodes of the second subgraph. The faces can have maximal valence five in the first subgraph, and four in the second subgraph.

The graph is 2-modular: it can be subdivided into two subgraphs, given by the nodes 1 to 7 and 8 to 13 respectively, such that the nodes of each subgraphs are connected to at most one node of the other subgraph. Looking at this graph, one obvious geometric question arises. The spherical cosine laws \Ref{defhth} require three nodes belonging to a 3-cycle.  In the examples considered so far,  each pair of nodes belonged to at least one 3-cycle, so it was possible to define the 4d dihedral angles in this way.  But consider now the links connecting the two subgraphs of Fig~\ref{fig:santa}, e.g. the one from 2 to 8. How can we define the 4d dihedral angle for this pair of polyhedra? The minimal cycle they belong to is a 4-cycle, and we cannot apply \Ref{defhth}. We have to use the 4-cycle construction of Fig.~\ref{fig:genSCL}, with $abcd=2839$ for instance.
The useful observation here is that although the boundary data don't define the 4d dihedral angle between $2$ and 8, they define two auxiliary 4d dihedral angles between either 2 or 8, and the hyperplane identified by one of the 4-cycles. The sum and difference of the auxiliary angles defines a local convex or concave embedding of the polyhedra sharing the (edge dual to the) 4-cycle. It is precisely this combination that will appear at the end of the critical point analysis.
When the angle-matching conditions are satisfied, the hyperplane is independent of the choice of 4-cycle containing 2 and 8, and one obtains a unique notion of 4d dihedral angle between these two polyhedra.

We choose 1 as the root node of the algorithm, with $h_1=\Id$. The first step of the algorithm determines all group elements $h_a$ with $a=2$ to $7$, with directions $\vec n_{a1}$ and angles \Ref{sol1}. The sine equations \Ref{DDR} force all group elements in this step to be all Euclidean or Lorentzian, and allow a single sign freedom $\eps_1$. If the normals at every link involving the nodes 1 to 7 satisfy the orientation condition the plus sign is valid solution. The minus sign requires four angle-matching conditions \Ref{ei1}, since each link of type $1a$ belongs to four 3-cycles. 

For the second step we choose $\ab=2$. There is only one second neighbour, $a_2=8$. We partially fix the gauge at $a_2$ requiring $\vec{n}_{\ab a_2}=-\vec{n}_{a_2 \ab}$. The only type-12 equation \Ref{type12} fixes  $\tilde{H}_{a_2}:= H_{\ab}^{-1}H_{a_2}$ to be a four-screw with direction $\vec{n}_{a_2 \ab}$. There are no type-21 nor type-22 equations: $a_2$ is a lonely neighbour at this step, and to determine the angle $\om_{a_2}$  of $\tilde{H}_{a_2}$ we need to move to the next steps of the algorithm. There are no restrictions on the boundary data from step 2.

In the third and fourth step we pick $\bar a_2=a_2=8$ as third seed, and $\bar a_3=9$ as fourth. The subgraph made of the third neighbours $a_3=9, \ldots, 12$ and the fourth neighbour $a_4=13$ coincides with the one of Fig.~\ref{fig:K6b}, and the procedure is identical to starting over from the first step, with the only difference that $H_{\bar a_2}\neq\Id$. The result will be to fix all group elements of the subgraph in terms of $H_{\bar a_2}$. 
At step 3, we fix the gauge to anti-align the normals at the nodes $a_3$ to $\bar a_2$, and write the four type-23 and four type-33 equations in terms of $\tl H_{a_3}:=H^{-1}_{\bar a_2}H_{a_3}$:
\begin{align}\label{HH3}
\tilde{H}_{a_{3}}\vec{n}_{a_{3}\bar a_{2}}&=-\vec{n}_{\bar a_{2}a_{3}}=\vec{n}_{a_{3}\bar a_{2}},
\qquad \tilde{H}_{a_{3}}\vec{n}_{a_{3}b_{3}}=-\tilde{H}_{b_{3}}\vec{n}_{b_{3}a_{3}}. 
\end{align}
The equations are solved taking projections as for the type-22 equations of step 2 of the algorithm \Ref{Capitani2}. They determine all group elements $\tilde{h}_{a_3}$ with $a_3=9, \ldots, 12$, as four-screws with directions $\vec n_{a_{3}\bar a_{2}}$ and angles
\be
\label{step3cubo}
\om_{a_{3}} = \eps_{3}\hth^{b_3}_{a_3 \ab_2} - \xg^{b_3}_{a_3 \ab_2},
\ee
with a new sign freedom $\eps_3=\pm 1$. 
For step 4, we fix the gauge at $a_4=13$ and anti-align its normal with $\bar a_3$, and write the one type-34 and three type-43 equations in terms of 
$\tilde{H}_{a_{4}}:=H^{-1}_{\bar a_3}H_{a_4}$:
\be\label{HH4}
\tilde{H}_{a_{4}}\vec{n}_{a_{4}\bar a_{3}} = -\vec{n}_{\bar a_{3}a_{4}}=\vec{n}_{a_{4}\bar a_{3}}, \qquad
\tilde{H}_{a_{3}}\vec{n}_{a_{3}a_{4}}=-\tilde{H}_{a_{4}}\vec{n}_{a_{4}a_{3}}.
\ee
These equations are solved taking projections as for the type-21 equations of step 2 of the algorithm \Ref{capitani3}. The group element $\tilde{h}_{a_4}$ is determined in terms of its direction $ \vec{n}_{a_4 \ab_3}$ and angle 
\be\om_{a_4}=\eps_3 \hth^{b_3}_{a_4 \ab_3} - \xg^{b_3}_{a_4 \ab_3},\ee
with the same sign $\eps_3$ of the third step.

The conditions we find on the boundary data in this subgraph are the usual ones: the group elements must be all Euclidean or Lorentzian, and there is a single sign freedom $\eps_3$. 
The solutions with $\eps_3=1$ are valid if the data entering the subgraph 8-13 are vector geometries. This means requiring \Ref{vector} for all normals on links of the subgraph, and not for normals $\vec n_{a_3a}$ and $\vec n_{a_4a}$ leaving the subgraph, such as $\vec n_{93}$.
To prove this we follow the same argument that led to \Ref{varphi} for the 4-simplex. We first complete the gauge fixing to the twisted spike for the  subgraph, to have the vector data in the configuration $\vec n_{ab}=-\vec n_{ba}$ for $a,b$ from 8 to 13. This can be achieved rotating the normals at the nodes 9 to 13, and in this gauge we have $\hth^{b_3}_{a_3 \ab_2}=\xg^{b_3}_{a_3 \ab_2}$. Since $\xg^{b_3}_{a_3 \ab_2}$ is affected only by rotations at $a_3$\footnote{It is affected also by rotations at $\ab_2$, but if we perform one of these, we need to compensate it with an equal rotation at $a_3$, leaving $\xg^{b_3}_{a_3 \ab_2}$ invariant, otherwise \Ref{HH3} does not hold anymore and we cannot write the solutions in this form.} and  $\hth^{b_3}_{a_3 \ab_2}$ is invariant, it follows that leaving this gauge we have $\hth^{b_3}_{a_3 \ab_2}-\xg^{b_3}_{a_3 \ab_2}=\nu_{a_3}$ for some parameter depending only on the rotation at $a_3$. Then $\om_{a_3}\equiv\nu_{a_3}$ so defined is the solution  for a general vector geometry. The vector geometry solution for $\om_4$ follows from a similar argument.
The second solution $\eps_3=-1$ requires that the data of the subgraphs are all Euclidean or Lorentzian, and edge-independence conditions. 
From the type-33 equations, we have  \be
\label{eq:am2}
\hth^{b_3}_{a_3 \ab_2}  = \hth_{a_3 \ab_2}, \qquad \xg^{b_3}_{a_3 \ab_2}  = \xg_{a_3 \ab_2}.
\ee
These are two edge-independence conditions at the faces $a_3\ab_2$.
From the first type-43 cosine equations we find the solution for $\om_{a_4}$ and the conditions
\be
\qquad \hth^{b_3}_{a_4 \ab_3}=\hth_{a_4 \ab_3},\qquad  \xg^{b_3}_{a_4 \ab_3}= \xg_{a_4 \ab_3}
\ee
These are three edge-independence conditions at the faces $a_4\ab_3$.
The second type-43 cosine equations are like \Ref{Totti3b}. They involve the 4-cycles $\ab_2 b_3 a_4 \ab_3$ and become identities provided the folowing edge-independence conditions are satisfied,
\be
\hth^{c_3}_{\ab_3\ab_2} = \hth^{b_3}_{s_4\ab_2} + \eps_3\eps_{s_4}^{a_4}\hth^{a_4}_{s_4\ab_3} +\pi, \qquad \eps_{s_4}^{a_4}=\pm1, \qquad 
\qquad \xg^{a_4\ab_3}_{b_3\ab_2}=\xg_{b_3\ab_2}. 
\ee
These requirements complement the edge-independence conditions \Ref{eq:am2} at the faces $\ab_2 b_3$, bringing the total to three.

Up to this point, the sectors of the partial solutions found in the first step and the partial solutions for the subgraph are independent, and we have two free signs $\eps_1$ and $\eps_3$. This is the consequence of the lack of \bracing equations of type-32 and type-21. 
To complete steps 3 and 4, there are five \magg equations to solve, four of type-31 and one of type-41. There is only one variable left to be determined, $\om_{\ab_2}$. The type-31 can all be written in terms of $\tilde{H}_{\ab_2}$ and $\tilde{H}_{b_{3}}$:
\begin{equation}
\label{mistero}
\tilde{H}_{\ab_{2}}\tilde{H}_{b_{3}}\vec{n}_{b_{3}b}=-H_{\ab}^{-1}H_{b}\vec{n}_{bb_{3}}.
\end{equation} 
We project them on $\vec{n}_{\ab b}$, $\vec{n}_{\ab \ab_2}$ and $\vec{n}_{\ab b}\times\vec{n}_{\ab \ab_{2}}$, and after some lengthy algebra on-shell of the other \magg equations, we find
\begin{subequations}\label{capitani5}\begin{align}
\cos(\om_{\ab_{2}}+\xg_{\ab_{2} \ab}^{b_3 b} - \eps_3\vth_{s\ab_{2}}^{b_3} -\pi)&=\cos \vth_{s\ab }^{b},
\label{totti5}\\\label{totti5bis}
\cos(\eps_1\hth^{c}_{\ab 1}-\xg^c_{\ab 1}+\xg_{\ab 1}^{\ab_2 b} - \eps_1 \vth_{s' 1}^{b} -\pi)&=\cos \vth_{s'\ab }^{s},\\
\sin\varphi^s_{ b_3 \ab_2}\sin(\om_{\ab_{2}}+\xg_{\ab_{2} \ab}^{b_3 b} - \eps_3\vth_{s\ab_{2}}^{b_3}-\pi)&=
\sin\phi^b_{\ab_2 \ab}\sin(\eps_1 \vth_{b\ab}^{s}), \label{derossi5} 
\end{align}
\end{subequations} 
The first equations determine the missing complex angle as
\be\label{138}
\om_{\ab_2} = \eps^b_{\ab_2}\vth^b_{s\ab} +\eps_3\vth_{s\ab_{2}}^{b_3} -\xg_{\ab_{2} \ab}^{b_3 b} +\pi.
\ee
In these equations, the angles $\vth$ are a mismatched version of the auxiliary 4d dihedral angles \Ref{4daux} between respectively $\ab$ and $\ab_2$ and the hyperplane $s=(\ab_2 \ab,b b_3)$. They are the analogue of \Ref{mocktheta} for the hypercube, and are defined by
\begin{equation}\label{mocktheta2}
\cos\vth_{s\ab_2}^{b_3}:=
\frac{\cos\phi_{b\ab_{2}}^{b_3}+\cos\phi_{b_3\ab}^{\ab_2}\cos\varphi^{s}_{ b_3 \ab_2} }
{\sin\phi_{b_3\ab}^{\ab_2}\sin\varphi^{s}_{ b_3 \ab_2}},
\qquad
\cos \vth_{s\ab}^{b}:=
\frac{\cos\phi_{b_3\ab }^{b}+\cos\phi_{b\ab_{2}}^{\ab} \cos\varphi^{s}_{b_3\ab_2}}
{\sin\phi_{b\ab_{2}}^{\ab}\sin\varphi^{s}_{b_3\ab_2}},
\end{equation}
where 
\begin{equation}
\label{varphis2}
\cos\varphi^{s}_{b_3\ab_2} := \cos(\eps_3\hth_{b_3\ab_2}^{d_3} - \xg_{b_3\ab_2}^{d_3} +\xg^{b\ab}_{b_3 \ab_2})
\sin\phi_{b_3 \ab}^{\ab_2}\sin\phi_{b \ab_{2}}^{b_3}-\cos\phi_{b_3 \ab}^{\ab_2}\cos\phi_{b \ab_{2}}^{b_3}.
\end{equation}
Plugging this solution in \Ref{derossi5}, we find
\begin{equation}
\sin\varphi^s_{b_3\ab_2}\sin(\eps^b_{\ab_2}\vth^b_{s\ab})=
\sin\phi^b_{\ab_2 \ab}\sin(\eps_1 \vth_{\ab b}^{s}), \label{DDR2casetta} 
\end{equation}
where
\be\label{vthcasetta2}
\cos \vth_{b\ab}^{s} = \f{\cos\varphi^s_{b_3 \ab_2} + \cos\phi^{\ab}_{b \ab_2} \cos\phi^b_{ b_3 \ab} }{\sin \phi^{\ab}_{b \ab_2} \sin\phi^b_{ b_3 \ab} }
\ee
Notice that the same mismatched $\varphi$ appears in the three $\vth$ angles. One can explicitly check using spherical sine laws that $\vth^{b_3}_{s\ab_2}$ belongs to the same sector of the critical data in the small subgraph, and that $\vth^{b}_{s\ab}$ and $\vth^{s}_{b\ab}$ also belong to the same sector. Then \Ref{DDR2casetta} is just fixing the signs $\eps^b_{\ab_2} = \eps_1$, and the solution reduces to
\be\label{quasi}
\om_{\ab_2} = \eps_1 \vth^b_{s\ab} +\eps_3\vth_{s\ab_{2}}^{b_3} -\xg_{\ab_{2} \ab}^{b_3 b} +\pi.
\ee
The positive signs $\eps_1=\eps_3=1$ correspond to the vector geometry solution, and is valid for all data satisfying the orientation conditions \Ref{vector} also on the links connecting the two subgraphs $b b_3$, i.e. $39$, in addition to the links of the two subgraphs as we discussed in the step 1 and 3 of the algorithm. 
To prove this explicitly, we need to show that in the twisted spike gauge $\xg_{\ab_{2} \ab}^{b_3 b}=\vth^b_{s\ab} + \vth_{s\ab_{2}}^{b_3} +\pi$. This could be done studying trigonometric identities, but we refrain from doing so since it would be very cumbersome, and we know already that $\om_{\ab_2}=0$ is the vector geometry solution in the twisted spike gauge, from the general solution given in Section~\ref{secSpinToVec}.

To have additional solutions, the right-hand side has to be independent of the cycle, namely of the pair $bb_3$. One may think at first sight that this requires $\xg_{\ab_{2} \ab}^{b_3 b} = \xg_{\ab_{2} \ab} $ and also edge-independence of the individual angles 
$\vth^b_{s\ab}$ and $\vth_{s\ab_{2}}^{b_3}$. This is however not the case, thanks to the identity \Ref{matchingconvex}. This shows that it is possible to have the auxiliary angles cycle-dependent, while their sum or difference is cycle-independent. What is required is $\vth\equiv\hth$, for the identity to hold, namely the edge-independence
$\xg^{b\ab}_{b_3\ab_2 }=\xg_{b_3\ab_2}$, which guarantees $\varphi^s_{b_3\ab_2}=\phi^s_{b_3\ab_2}$ .
These conditions are sufficient, but we don't have a proof that they are also necessary. Inspection of the equations and our experience with explicit configurations strongly suggest that if they can be avoided, it will likely be due to numerical coincidences of special configurations, and not to the existence of a different class of data.

In the second cosine equations \Ref{totti5bis} there are no new variables to determine.  
These equations are only a restriction on the boundary data. The new angles appearing here are 4d angles involving the auxiliary hyperplane $s'=(\ab 1, bb_3)$ associated with the 5-cycle $(1\ab\ab_2 b_3b)$. These are defined as follows,
\begin{equation}\label{mocktheta3}
\cos\vth_{s'1}^{b}:=
\frac{\cos\phi_{b_3 1}^{b}+\cos\phi_{\ab b}^1\cos\varphi^{s'}_{ b1} }
{\sin\phi_{\ab b}^1\sin\varphi^{s'}_{ b1}},
\qquad
\cos \vth_{s'\ab}^{s}:=
\frac{\cos\varphi_{b_3\ab_2 }^{s}+\cos\phi_{\ab_{2}1}^{\ab} \cos\varphi^{s'}_{b1}}
{\sin\phi_{\ab_{2}1}^{\ab}\sin\varphi^{s'}_{b1}},
\end{equation}
where
\begin{equation}
\cos\varphi^{s'}_{b1} := \cos(\eps_1\hth_{b1}^{e} - \xg_{b1}^{e} +\xg^{b_3\ab}_{b1})
\sin\phi^1_{b \ab}\sin\phi_{b_3 1}^{b}-\cos\phi_{b\ab}^{1}\cos\phi_{b_3 1}^{b}.
\end{equation}
The equations  \Ref{totti5bis} can be treated as \Ref{Totti3b} in the analysis of the hypercubic graph: they have no solutions in general, unless the following edge independence conditions are satisfied

\be
\hth_{\ab 1}= \hth^c_{\ab 1} = \hth^b_{s'1} + \eps_1\eps_{s'}^{s}\hth^{s}_{s'\ab} +\pi, \qquad \xg^{b_3\ab}_{b1} =\xg_{b1}, \qquad \xg^{\ab_2 b}_{\ab 1}=\xg_{\ab 1},
\ee 
and fixes the value of the sign $\eps_{s'}^{s}$ since $\hth^c_{\ab 1}$ is known.
In other words, having already identified the 4d dihedral angle of $1\ab$ through a 3-cycle, the equation associated to the 5-cycles with $1\ab$ can only be satisfied if there are enough edge-independence conditions so that it becomes an identity.

We are left with only one \magg equation, the type-41
\be
\tilde{H}_{\ab_{2}}\tilde{H}_{\ab_{3}}\tilde{H}_{a_{4}}\vec{n}_{a_{4}b}=-H_{\ab}^{-1}H_{b}\vec{n}_{ba_{4}}.
\end{equation}
Scalar projections of this vectorial equation involve 5-cycles and 6-cycles. Their expression is similar to \Ref{capitani5}, with angles $\vth$ associated with the new auxiliary hyperplanes. We refrain from writing them here. In these equations everything is already determined, and these equations only provide the remaining edge-independence conditions on the faces $\ab_2 \ab$,  $\ab_3 \ab_2$,  $a_4 \ab_3$, $\ab 1$,$b1$.  

We have identified all critical holonomies, and solve all \magg equations. Let us now collect the restrictions on the boundary data that are necessary. To have a vector geometry solution, we need the orientation conditions at every face. These are automatically satisfied by our choice of gauge for the links in the dominating path, but restrict non-trivially the data for the other links. To have more solutions, we need 
the $\hth$'s and $\xg$'s 
to satisfy five edge-independent conditions at the faces $1a$ and $\ab \ab_2$, involving cycles of length from three to six, and four edge-independent conditions at the faces $b_3 \ab_2$ and $\ab_3 a_4$, involving both 3-cycles and 4-cycles. 
Since we could have chosen any other dominant set, five edge-independence conditions must hold at every face in the first subgraph and at each face bridging between the two subgraphs, while four edge-independence conditions must hold at every face in the second subgraph. By the closure conditions, the faces in the first subgraph can be at most pentagons and the faces in the second subgraph can be at most squares. Hence we need edge-independence conditions at every link of the graph, and by \Ref{at}, 2d angle-matching conditions of every face.

The main novelty of this graph is that the 
angle-matched solutions are labelled by two independent signs $\eps_1$ and $\eps_3$. We conclude that conformal twisted geometries admit at most \emph{four} distinct solutions. 
If the angles match in a subgraph only, including the bridging faces, we only have one free sign, and two distinct solutions.
This freedom goes hand in hand with the freedom of taking independent sectors of the boundary data. Therefore the global angle-matched geometries need not be globally Euclidean or Lorentzian. We could have mixed ones as well. For these, $\om_{\ab_2}$ is neither a pure rotation nor a pure boost (in the spike gauge and up to $\Pi_{\ab_2}$).  The mixed signature geometries have no possible 4d embedding.
For the global Euclidean or Lorentzian geometries with a flat 4d embedding possible, the relative sign $\eta =\eps_1\eps_3$ describe a local convex or concave embedding as exemplified by Fig.~\ref{casetta}.

The asymptotic formulae for this graph can be obtain from the general ones of Section~\ref{SectionLO}, adapting to $L=36$ and $N=13$. For instance for Lorentzian angle-matched data,
\begin{align}
A_\G =  &=
\f{2^{120} \pi^{36} \J }{\l^{36}} 
e^{i\Psi} 
\bigg( \f{\Om^{\scr (+;+)} e^{i\lambda S_{\G}^{\scr + } }}{\sqrt{\det{-H^{\scr (+;+)}}}} 
+ \f{\Om^{\scr (-;+)} e^{-i\lambda S_{\G}^{\scr +} } }{\sqrt{\det{-H^{\scr (-;+)}}}} 
+\f{\Om^{\scr (+;-)} e^{i\lambda S_{\G}^{\scr -} }}{\sqrt{\det{-H^{\scr (+;-)}}}}   
+ \f{\Om^{\scr (-;-)} e^{-i\lambda S_{\G}^{\scr -} } }{\sqrt{\det{-H^{\scr (-;-)}}}}  \bigg)  + O(\l^{-37}).
\end{align}

\section{Conclusions}
We have introduced a technique and a step-by-step algorithm that can be used to compute the asymptotic behaviour of the Lorentzian EPRL-KKL amplitude for an arbitrary graph, and with any choice of boundary data. One of the main technical results of this paper is to have obtained convenient explicit scalar projections of the \magg equations. These are given by trigonometric equations, like \Ref{Eqstep1} or \Ref{step31}, and can be used to solve the critical point equations for general graphs. We found that critical points exist only for special types of boundary data, that can be identified on general grounds for any graph. For non-modular graphs, the critical data are vector geometries with one critical point, and  angle-matched data, defining conformal twisted geometries, with two critical points. The angle-matched data can be either Euclidean or Lorentzian, in the sense determined by the spherical cosine laws, and contain 3d Regge geometries as a subset. We also explained that not all 3d Regge geometries are the boundary of flat polytopes, and referred to \cite{Dona:2017dvf} for a discussion of the required conditions, and identification of the flat polytope subset.

The characterization of the (non-degenerate) critical configurations as conformal twisted geometries is based on two assumptions: the equivalence of global angle-matching and edge-independence conditions \Ref{at}, and the inexistence of numerical coincidences as discussed at the end of Section~\ref{sec:geogen}.
As a consequence, we do not have a theorem that conformal twisted geometries are always the most general class of data admitting more than one critical point for an arbitrary graph. It would be great if this result could be tightened up, but that may likely require a graph by graph analysis.

The fact that the critical data describe conformal twisted geometries and not just Regge geometries, is after all quite natural from the point of view of the saddle point approximation:
the key equations \Ref{maggica} are scale invariant. Therefore it is to be expected that any restriction on the boundary data that may arise from these equations will only involve angles, and thus not be as strong as selecting Regge geometries.
What emerges from our analysis is that these restrictions are actually the strongest ones that can arise from scale-invariant equations:
all 2d angles have to match, making all 4d dihedral and twist angles edge-independent. No further restrictions appear rather than closure and angle matching, making thus conformal twisted geometries the natural object to discuss asymptotics of unitary $\SL(2,\C)$ invariants. It is a special feature of 4-simplices that conformal twisted geometries coincide with Regge geometries.

\begin{figure}[H]
\centering
\includegraphics[width=9cm]{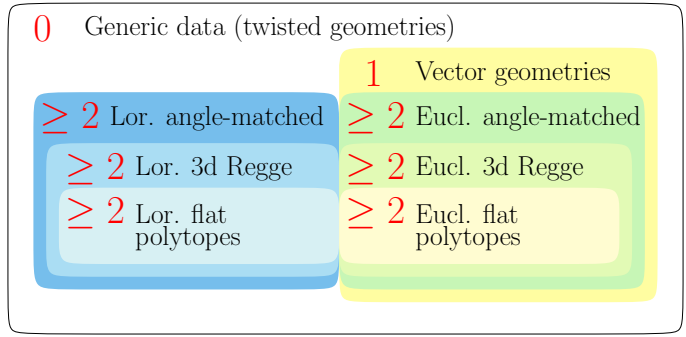}
\caption{\small{\emph{Schematic classification of boundary data, with their geometric interpretation and number of distinct critical points (up to the spin lift). The angle-matched data describe conformal twisted geometries, and only a subset of them 
represents 3d Regge geometries, unless all reconstructed faces are triangular. Furthermore, not all 3d Regge geometries are flatly embeddable. For the flat embeddable ones, the boundary data can be put in corresponds with the boundary geometries of flat 4d polytopes.
For modular graphs, mixed configurations are also possible and more than two critical points.}}
\label{FigSchemone}
}
\end{figure}

The generic relation between classes of boundary data and critical behaviour of the amplitude is schematically represented in Fig.~\ref{FigSchemone}.
The diagram is necessarily qualitative, and questions like the actual location of the boundaries of the different subsets are left to a more precise analysis once a specific graph is picked.
The diagram is very similar to the one derived for $\SU(2)$ BF theory \cite{Dona:2017dvf}, with the only addition of the Lorentzian configurations, which have no critical point and thus an exponential fall-off in $\SU(2)$ BF theory.
The vertex asymptotics shares many features between $\SU(2)$ BF theory, and the Euclidean and Lorentzian EPRL-KKL models. The boundary data are the same in the three cases, 
and the asymptotic formulas are also very similar: This was known for the 4-simplex graph \cite{BarrettSU2}, and a comparison of the results presented here and in \cite{Dona:2017dvf} shows that it is true for generic graphs. 

Our results help assess the viability of the generalized EPRL-KKL vertex.
On the one hand, the fact that (non-vector-geometry) critical behaviour occurs for the EPRL-KKL model on an arbitrary graph, with a well-defined geometric interpretation and rotation-invariant action, is a positive result: not only we have a notion of transition amplitudes for all spin networks, but these transition amplitudes can be interpreted geometrically in the large spin limit. Furthermore, 3d Regge data, flat embeddable or not, are always part of the critical behaviour.

On the other hand, the fact that critical configurations occur for a larger class than Regge geometries may be undesirable, and hinder the spin foam dynamics with many vertices and its continuum limit, adding to the non-trivial open questions that the presence of vector geometries and cosines of Regge actions already pose.
From this perspective, let us recall that the rationale to define the EPRL-KKL model is the imposition of the quantum simplicity constraints as restrictions of the irreps appearing in the state sum.
This is based on a direct discretization of the continuum covariant constraints \cite{Barrett:1997gw,BaezBarrett} which is complete only for the 4-simplex (see also Appendix of \cite{EPR}). In particular, it is only on the 4-simplex that a third set of simplicity constraints,\footnote{After the diagonal and off-diagonal, implemented respectively with $\r_{ab}=\g k_{ab}$ and $k_{ab}=j_{ab}$.}  namely the discretization of the continuum constraints on non-adjacent faces, become redundant and can be dropped. This prompts the question of whether the amplitude for non-simplicial vertices should be modified to include this third set. It is not an easy question to address, because the third set  depends a priori not just on fluxes but also on holonomies. Some preliminary considerations in this direction have appeared in \cite{Bahr:2017ajs,Bahr:2018vvq}. 
It is possible that a proper discretization of the third set, usually referred to as `volume simplicity constraints', will break the scale invariance of the \magg equations, and lead to new types of constraints explicitly introducing lengths.

This said about the generalized vertex amplitude, modifications of the model have also been discussed for the 4-simplex amplitude, in order to eliminate the vector geometries and/or the cosine of the Regge action in favour of a single exponential, see e.g. \cite{Livine:2002rh,Engle:2011un, Engle:2015zqa,Han:2018fmu,BiancaHalnuovo}. The analysis discussed here also offers a new perspective on this program. As we have pointed out, the Lorentzian data can be characterized in terms of inequalities on the scalar products of the normals. Or, for the 4-simplex, a negative squared 4-volume computed from the spins. Imposing such inequalities has the effect of removing all Euclidean critical points, including the vector geometries. As for the presence of a cosine instead of a single exponential of the Regge action, namely the fact that angle-matched critical points come in pairs, our technique traces this to the evenness of the \magg equations, in particular their scalar projections \Ref{Eqstep1}: the cosine there is ultimately the reason why we have a cosine in the asymptotic formulae (provided the Hessians are complex conjugated). Having solutions coming in pairs appears thus to be unavoidably related to the nature of the irreps used in the model, and our analysis does not suggest any obvious simple modifications to change this fact, on top of those already considered in the literature.
We believe that our techniques and algorithm can be of use also for the modified models. 

Let us conclude with two possible extensions of the analysis presented here. We considered only irreps with the lowest SU(2) weights, which are the ones that enter the EPRL-KKL model. The general case with non-minimal SU(2) spins is also interesting, at least from a mathematical perspective. It has more complicated critical point equations, and a richer geometric interpretation.
In the simplest case of a product of generalized Clebsch-Gordan coefficients, with no graph structure, the critical point equations define a more elaborated map between the 3d vectors appearing in the boundary data and bivectors than the one appearing here: $\g$-simple bivectors instead of bivectors with vanishing magnetic part \cite{Pierre}. For a tensor invariant associated to a graph, it should be possible to combine the technique presented here with the ones of \cite{Pierre} to obtain asymptotics of all unitary $\SL(2,\C)$ invariants on graphs. 
Secondly, it would be also interesting to extend our technique to more general settings with time-like or null faces, see e.g. \cite{Barrett:1997gw,Conrady:2010vx,Kaminski:2017eew}. These involve the other little groups of $\SL(2,\C)$ and typically use their associated representation basis instead of the canonical basis, and other coherent states such as those for $\SU(1,1)$.

We hope to have convinced the reader that the beautiful link between $\SL(2,\C)$ Clebsch-Gordan coefficients and Minkowski geometry goes far beyond the simplicial case.

\subsection*{Acknowledgments}
The work of P.D. is partially supported by the grant 2018-190485 (5881) of  the Foundational Questions Institute and the Fetzer Franklin Fund. 

\begin{appendix}
\setcounter{equation}{0}
\renewcommand{\theequation}{\Alph{section}.\arabic{equation}}

\section{Vectorial representation of $\SL(2,\C)$}
\label{app:pauli}

In this Appendix we show how to derive  \Ref{3dHirrep}, namely the explicit form of the 3d, non-unitary irrep of $\SL(2,\C)$ in terms of the fundamental one. We start from the polar decomposition $h=b u$. We parametrize 
\begin{equation}
\label{eq:fundamental}
u= \exp(i \frac{\theta}{2} \vec{n}\cdot\vec{\sigma})=\mathds{1}\cos\frac{\theta}{2}+i\vec{n}\cdot\vec{\sigma}\sin\frac{\theta}{2} \in \SU(2),
\end{equation}
and compute its adjoint action on $\vec{\s}$ using the properties of Pauli matrices,
\begin{equation}
g^{-1}\vec{\sigma}g
=\cos\theta \,\vec{\sigma}-\sin\theta\,\vec{n}\times\vec{\sigma}+\left(1-\cos\theta\right)\left(\vec{n}\cdot\vec{\sigma}\right)\,\vec{n} 
= R_{\vec n}(\th) \vec \sigma. 
\label{rotrep}
\end{equation}
where $U:=R_{\vec n}(\th)$ is the vectorial representation of $u$, a rotation of angle $\th$ and axes $\vec{n}$. 
This equation is actually valid for $\th\in\C$. Therefore, for a pure boost 
\begin{equation}
\label{eq:boost}
b= \exp(\frac{\beta}{2} \vec{m}\cdot\vec{\sigma})=\mathds{1}\cosh\frac{\beta}{2}+\vec{m}\cdot\vec{\sigma}\sinh\frac{\beta}{2},
\end{equation}
we immediately derive 
\begin{equation}
b^{-1}\vec{\sigma}b
=\cosh\beta \,\vec{\sigma}+i\sinh\beta\,\vec{m}\times\vec{\sigma}+\left(1-\cosh\beta\right)\left(\vec{m}\cdot\vec{\sigma}\right)\,\vec{m} =B_{\vec m}(\b) \vec\sigma. 
\label{rotrep2}
\end{equation}
Combining these two we obtain the expression  
\begin{equation}
h^{-1}\vec{\sigma}h
=B_{\vec m}(\b) R_{\vec n}(\th) \vec\sigma \equiv H \vec{\s},
\end{equation}
with $H$ given by \Ref{3dHirrep}.
This 3d, non-unitary irrep of $\SL(2,\C)$ is usually labelled ${\bf (1,0)}$ in terms of the self-dual and antiself-dual spins $j_\pm$, however we avoided this notation in the main text to avoid confusion with a unitary irrep $(\r,k)$ of fixed $\r$ and $k$.

\section{Orientation conditions imply Euclidean geometry}
\label{app:vector}
In this Appendix we prove that boundary data satisfying the orientation conditions \Ref{vector} satisfy the Euclidean spherical cosine laws.
Consider the quantity
\begin{equation}
\label{def:C}
C:=\frac{\vec{n}_{1b}\times\vec{n}_{1a}}{\norm{\vec{n}_{1b}\times\vec{n}_{1a}}}\cdot\frac{R_{a}\vec{n}_{a1}\times R_{a}\vec{n}_{ab}}{\norm{\vec{n}_{a1}\times\vec{n}_{ab}}}.
\end{equation}
By the Cauchy-Schwarz inequality and the invariance of the scalar product under $R_{a}\in SO(3)$,
\begin{equation*}
\left|C\right|\leq\frac{\norm{\vec{n}_{1b}\times\vec{n}_{1a}}}{\norm{\vec{n}_{1b}\times\vec{n}_{1a}}}\frac{\norm{R_{a}\vec{n}_{a1}\times R_{a}\vec{n}_{ab}}}{\norm{ \vec{n}_{a1}\times\vec{n}_{ab}} }=1.
\end{equation*}
On the other hand, we can expand the numerator of \Ref{def:C} as follows,
\begin{align*}
C=&\frac{\left(\vec{n}_{1b}\cdot R_{a}\vec{n}_{a1}\right)\left(\vec{n}_{1a}\cdot R_{a}\vec{n}_{ab}\right)-\left(\vec{n}_{1b}\cdot R_{a}\vec{n}_{ab}\right)\left(\vec{n}_{1a}\cdot R_{a}\vec{n}_{a1}\right)}{\norm{\vec{n}_{1b}\times\vec{n}_{1a}} \norm{\vec{n}_{a1}\times\vec{n}_{ab}}}=\\
&\frac{\left(\vec{n}_{1b}\cdot R_{a}\vec{n}_{a1}\right)\left(R_{a}^{-1}\vec{n}_{1a}\cdot\vec{n}_{ab}\right)-\left(R_{b}^{-1}\vec{n}_{1b}\cdot R_{b}^{-1}R_{a}\vec{n}_{ab}\right)\left(R_{a}^{-1}\vec{n}_{1a}\cdot\vec{n}_{a1}\right)}{\norm{\vec{n}_{1b}\times\vec{n}_{1a}} \norm{\vec{n}_{a1}\times\vec{n}_{ab}}}.
\end{align*}
If the data satisfy \Ref{vector} we can remove all rotations and obtain
\begin{align*}
C=\frac{\left(\vec{n}_{1b}\cdot\vec{n}_{1a}\right)\left(\vec{n}_{a1}\cdot\vec{n}_{ab}\right)+\left(\vec{n}_{b1}\cdot\vec{n}_{ba}\right)\left(\vec{n}_{a1}\cdot\vec{n}_{a1}\right)}{\left\Vert \vec{n}_{1b}\times\vec{n}_{1a}\right\Vert \left\Vert \vec{n}_{a1}\times\vec{n}_{ab}\right\Vert }=\frac{\vec{n}_{b1}\cdot\vec{n}_{ba}+\left(\vec{n}_{1b}\cdot\vec{n}_{1a}\right)\left(\vec{n}_{a1}\cdot\vec{n}_{ab}\right)}{\left\Vert \vec{n}_{1b}\times\vec{n}_{1a}\right\Vert \left\Vert \vec{n}_{a1}\times\vec{n}_{ab}\right\Vert }=\cos\hat{\theta}_{1a}^{b}.
\end{align*}
The bound derived above therefore implies that the 3d angles of vector geometries satisfy the Euclidean condition $|\cos\hth_{1a}^{b}|\leq 1$
for the spherical cosine laws.

\section{Evaluating the on-shell action}\label{app:razzetto}
In this Appendix we give more details on the evaluation of the action at the critical points presented in Section~\ref{sec:Scrit}.
First, we recall some basic properties of SU(2) coherent states:
\begin{align}\label{eq:eigen}
&\vec{n}_{ab} \cdot \vec{\s} \, \ket{\z_{ab}}=-\ket{\z_{ab}}, \qquad \vec{n}_{ba} \cdot \vec{\s} \, |\z_{ba}]=|\z_{ba}],
\end{align}
and for non-proportional spinors, 
\begin{align} \label{eq:matrixelement}
& [\z_{ba}|\vec{\sigma}\ket{\z_{ab}} = \f{[\z_{ba}\ket{\z_{ab}}}{1-\vec{n}_{ab}\cdot\vec{n}_{ba}}\left( 
-\vec{n}_{ab}+\vec{n}_{ba}-i \vec{n}_{ab} \times \vec{n}_{ba}\right).
\end{align}
Then, composing four-screws using \Ref{eq:fundamental} with a complex angle and properties of the Pauli matrices, we arrive at \Ref{hrocket}.
To proceed, it is convenient to choose the spike gauge, which is available for both Euclidean and Lorentzian angle-matched data.

In the spike gauge the vectors $\vec{n}_{ab}\times\vec{n}_{ba}$ and $\vec{n}_{1a}\times\vec{n}_{1b}$ are anti-aligned,
\be
\f{\vec{n}_{a1}\times\vec{n}_{b1}}{\norm{\vec{n}_{a1}\times\vec{n}_{b1}}} = 
- \f{ \vec{n}_{ab}\times\vec{n}_{ba}} { \norm{\vec{n}_{ab}\times\vec{n}_{ba}} }.
\ee
We can use this property to eliminate  some  scalar products from \Ref{hrocket}, obtaining
\begin{align}
\label{eq:step2razzetto}
& [{\z_{ba}}|h_{b}^{-1}h_{a}\ket{\z_{ab}}= 
{[\z_{ba}\ket{\z_{ab}}}\bigg(\cos{\frac{\om_{a}}{2}}\cos{\frac{\om_{b}}{2}}-\sin{\frac{\om_{a}}{2}}\sin{\frac{\om_{b}}{2}}\vec{n}_{a1}\cdot\vec{n}_{b1} \\
&\ - i \f{ \vec{n}_{ab}-\vec{n}_{ba} }{1-\vec{n}_{ab}\cdot\vec{n}_{ba}} \cdot
\big(\cos{\frac{\om_{a}}{2}} \sin{\frac{\om_{b}}{2}} \,\vec{n}_{a1}-\sin{\frac{\om_{a}}{2}}\cos{\frac{\om_{b}}{2}}\,\vec{n}_{b1}\big)
- \sin{\frac{\om_{a}}{2}}\sin{\frac{\om_{b}}{2}}\,
\frac{\vec{n}_{a1}\times\vec{n}_{b1}\cdot\vec{n}_{ab}\times\vec{n}_{ba}}{1-\vec{n}_{ab}\cdot\vec{n}_{ba}} \bigg). \nn
\end{align}
\begin{figure}[H]
\centering
\includegraphics[width=5cm]{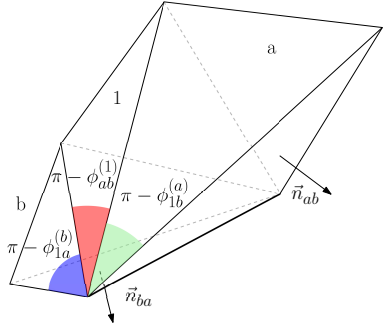}
\caption{\small{\emph{In the spike gauge the tetrahedra $1$, $a$, and $b$ share an edge vector that is orthogonal to both $\vec{n}_{ab}$ and $\vec{n}_{ba}$. The angle between these two vectors is given in terms of the external 3d dihedral angles  by
$\pi - (\pi-\phi^{a}_{b1})- (\pi-\phi^{1}_{ab})- (\pi-\phi^{b}_{a1})=-2\pi + \phi^{a}_{b1}+\phi^{1}_{ab}+\phi^{b}_{a1}=\phi^{a}_{b1}+\phi^{1}_{ab}+\phi^{b}_{a1}$. This and analogue observations lead to \Ref{staring}.}}
\label{fig:nabnba}
}
\end{figure}
With the help of Fig.~\ref{fig:nabnba}, we deduce 
\begin{align}\label{staring}
& \vec{n}_{ab}\cdot\vec{n}_{ba}=\cos\left(\phi^{1}_{ab} + \phi^{b}_{a1} + \phi^{a}_{b1} \right),
\\\nn
& \vec{n}_{b1}\cdot\left(-\vec{n}_{ab}+\vec{n}_{ba}\right)=
-\cos\left(      \phi_{ab}^{1}
                +\phi_{b1}^{a}\right)+
 \cos\left(      \phi_{a1}^{b}\right)=
2\sin\left(\frac{\phi_{ab}^{1}+\phi_{b1}^{a}+\phi_{a1}^{b}}{2}\right)
 \sin\left(\frac{\phi_{ab}^{1}+\phi_{b1}^{a}-\phi_{a1}^{b}}{2}\right),
\\\nn
& \vec{n}_{a1}\cdot\left(-\vec{n}_{ab}+\vec{n}_{ba}\right)=-\cos\left(\phi_{b1}^{a}\right)+\cos\left(\phi_{ab}^{1}+\phi_{a1}^{b}\right)
=-2\sin\left(\frac{\phi_{ab}^{1}+\phi_{b1}^{a}+\phi_{a1}^{b}}{2}\right)
   \sin\left(\frac{\phi_{ab}^{1}-\phi_{b1}^{a}+\phi_{a1}^{b}}{2}\right).
\end{align}
The last ingredient we need is the scalar product 
\begin{equation}
[\z_{ba}\ket{\z_{ab}} = \sqrt{\frac{1-\vec{n}_{ab}\cdot\vec{n}_{ba}}{2}} e^{i \arg [\z_{ba}\ket{\z_{ab}} }
= \sin\left(\frac{\phi^{1}_{ab} + \phi^{b}_{a1} + \phi^{a}_{b1}}{2} \right)e^{i  \arg [\z_{ba}\ket{\z_{ab}} }.
\end{equation}
Plugging these relations in \Ref{eq:step2razzetto} we obtain \Ref{hrocketspike}.

\section{Spherical and hyperbolic trigonometry}
\label{app:spherical}

In evaluating the action at the critical points, we used formulas of spherical and hyperbolic trigonometry which relate 3d and 4d dihedral angles. 
We report them here for completeness, and to fix our conventions which use external angles as opposed to the internal angles more commonly found in the literature.

\begin{figure}[H]
\centering
\includegraphics[width=5cm]{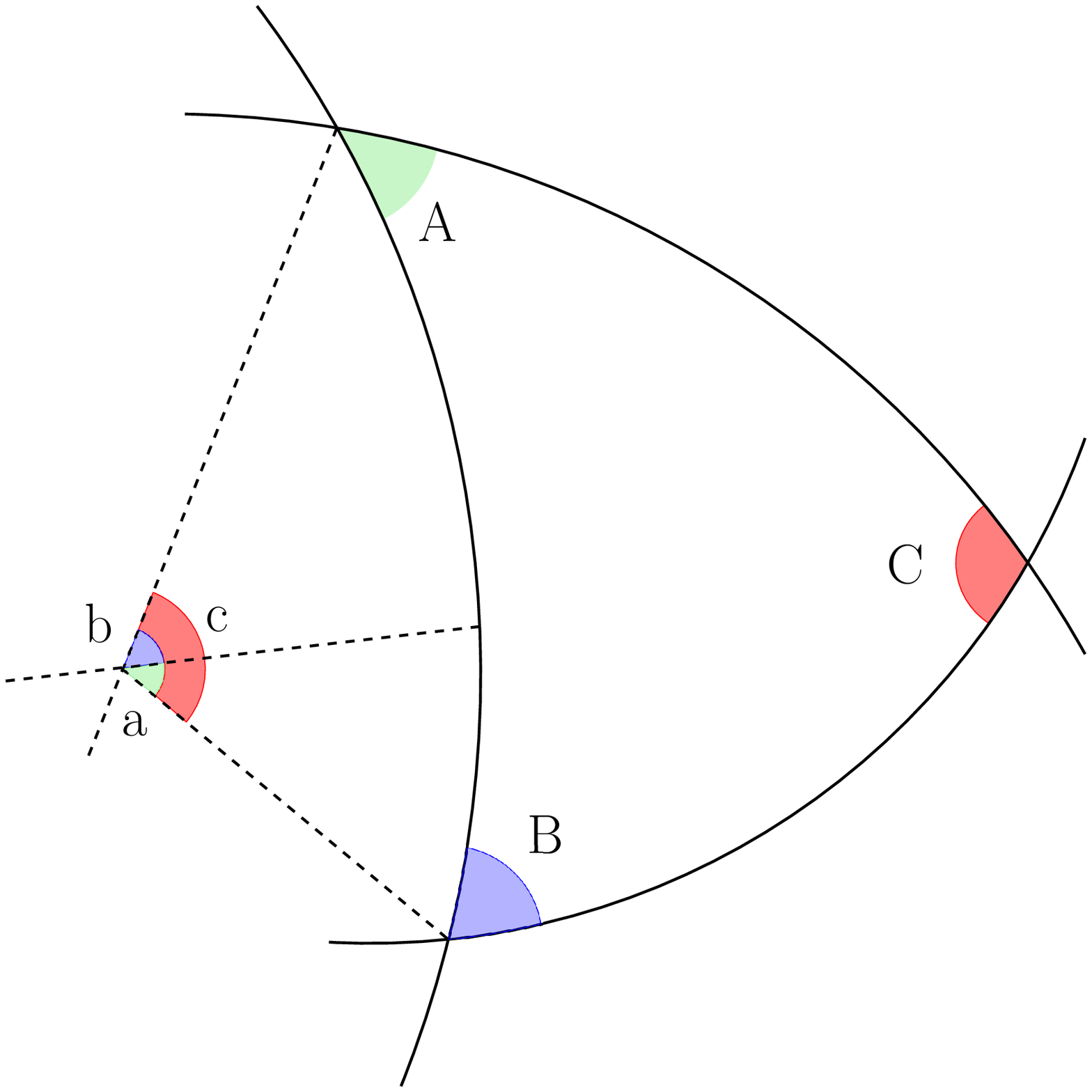} \hspace{2cm}
\includegraphics[width=5cm]{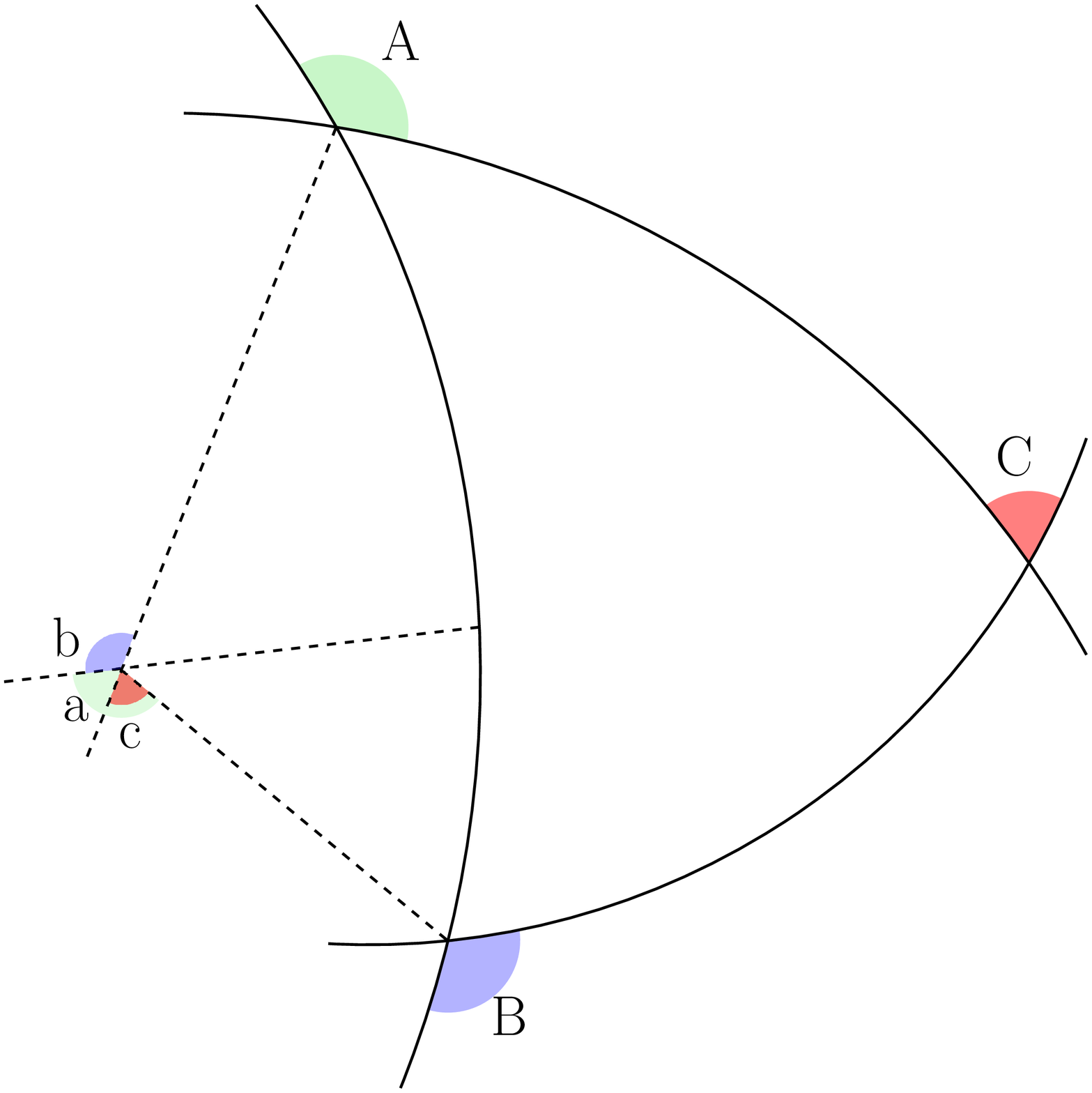}
\caption{\small{\emph{Spherical triangle, and spherical cosine laws.} Left panel: \emph{The more common case of internal dihedral angles}.
Right panel: \emph{The use of external dihedral angles.}}
\label{fig:sphericaltriangle} 
}
\end{figure}
Consider a spherical triangle with  spherical (external) angles $A$, $B$, $C$, and area $2\pi-A-B-C$. With reference to Fig.~\ref{fig:sphericaltriangle}, right panel, we have the following identities
\begin{equation}
\cos A=\frac{\cos a +\cos b\cos c}{\sin c\sin b}.
\end{equation} 
The identifies for the internal dihedral angles, depicted in the left panel, have a relative minus sign.
From these identities and trigonometric identities we derive the half-angle equations
\begin{equation}
\label{eq:halfangle}
\cos \frac{A}{2} =\sqrt{\frac{\cos(s-c)\cos(s-b)}{\sin c\sin b}} \qquad \sin\frac{A}{2}=\sqrt{\left|\frac{\cos(s)\cos(s-a)}{\sin c\sin b}\right|},
\end{equation} 
where $s= \frac{a+b+c}{2}$. And from these we derive the composition formulas
\begin{subequations}\label{eq:halfcompo}
\begin{align}
\cos \frac{C}{2} &= \cos \frac A 2 \cos \frac B 2 \sin(s)  +\sin \frac A 2 \sin \frac B2 \sin(s-c),\\
\sin \frac{C}{2} &= \sin \frac B 2 \cos \frac A 2 \sin(s-a)+\sin \frac A 2 \cos \frac B2 \sin(s-b),
\end{align} 
\end{subequations}
used to obtain \Ref{eq:euclrazz}.
Similiar relations hold also in the case of a hyperbolic triangle substituting the trigonometric functions with hyperbolic ones.
\begin{subequations}\label{eq:halfcompohyp}
\begin{align}
\cosh \frac{C}{2} &= \cosh \frac A 2 \cosh \frac B 2 \sin(s)  +\sinh \frac A 2 \sinh \frac B2 \sin(s-c),\\
\sinh \frac{C}{2} &= \sinh \frac B 2 \cosh \frac A 2 \sin(s-a)+\sinh \frac A 2 \cosh \frac B2 \sin(s-b),
\end{align} 
\end{subequations}
used to obtain \Ref{eq:lorerazz}.

\section{Degrees of freedom of polytopes}
\label{AppTacchino}

Minkowski theorem states that a set of $N$ vectors in $\R^n$ identifies a unique convex polytope, up to rotations. The norms of the vectors determine the $n-1$-volumes of the facets (the $n-1$ dimensional cells), and the scalar products the $n$-dimensional dihedral angles. This information determines the whole geometry of the polytope, including its adjacency matrix, and the number and geometry of faces, edges, etc. It follows that the number of degrees of freedom of a convex polytope, up to rotations in the flat ambient space, is $n(N-1)-n(n-1)/2$. The $n$-simplex has $n+1$ facets, hence $n(n+1)/2$ degrees of freedom.
This coincides with the number of edges, and it is the reason why Regge calculus uses $n$-simplices and edge lengths. For more general polytopes in the dominant adjacency classes (i.e. all vertices $n$-valent\footnote{Special  vector configurations have the same number of facets but fewer edges, hence higher-valence vertices, and correspond to polytopes with subdominant adjacency classes.}), the number of edges can be estimated using the Euler characteristic and the relation $nV=2E$.
In 2d, we have $2N-3$ degrees of freedom, which are always less than the edges, except for the triangle. Polygons are thus generically not edge-rigid. In 3d, we have $3N-6$ degrees of freedom, which always coincides with the number of edges of the dominant classes. Polyhedra are thus generically edge-rigid. But not always: there exist special configurations where the Jacobian between closed vectors and lengths is degenerate, e.g. a regular parallelepiped. In 4d, we have $4N-10$ degrees of freedom. This is typically \emph{less} than the number of edges, except for the 4-simplex. 4d polytopes are thus generically edge-constrained, a fact that can be understood in terms of geometric embeddings. The edge lengths define a 3d Regge tessellation of $S^3$, and not every such geometry can be flat embedded in $\R^4$. It should be possible to understand the difference between the degrees of freedom of a 3d Regge geometry and a flat 4d polytope in terms of flat-embedding conditions, but we are not aware of an explicit construction.

For a 4-simplex, not only the edge lengths determine uniquely its geometry. It is also the case that their number matches the number of triangles. Therefore it is possible to locally invert from lengths to areas. The reconstruction of a 4-simplex from its ten areas is something heavily used in simplicial spin foam models. 
In the main text we have used this fact to argue that in the asymptotic analysis of the vertex amplitude, the 4-simplex is `already there' once the spins are correctly chosen; and the role of the critical point equations is to verify the compatibility of the 3d normals with the given the 4-simplex. It allows us also to distinguish a Euclidean from a Lorentzian 4-simplex using the spins. This is established looking at the sign of the squared 4-volume, which can be written in terms of the areas by writing it first in terms of the lengths via the Caley determinant formula, and then inverting the lengths for the areas.

This inversion is however not always guaranteed, as there may exist special configurations at which the Jacobian determinant vanishes. An example mentioned in the literature is the Tuckey configuration \cite{Barrett:1994nn}: two different sets of lengths both giving all areas  equal. 
To study this example in more details, we wrote a code on Mathematica that computes all areas, 3d volumes and the 4d volume of a 4-simplex, strarting from the edge lengths and using the Heron formula and its generalization with Caley determinants.
Requiring that the areas are all equal, we actually found \emph{three} sets of  compatible edge lengths.

The first one is the equilateral 4-simplex: all lengths $\ell$ are equal, all the three-dimensional volumes are equal to $V=1/(6 \sqrt{2}) \ell^3$, and the 4d  volume is $V_{\s}=\sqrt{5}/96 \ell^4$. 
The second one is Tuckey's example: nine  lengths are equal to $\ell$, while the last one is $\sqrt{3} \ell$. Two of the three-dimensional volumes are $1/(6 \sqrt{2}) \ell^3$ while the remaining three are 0. The squared 4d volume is negative, $V_{\s}=i/(32 \sqrt{3})\ell^4$.
The last configuration has eight lengths equal to $\ell$ and two equal to $\sqrt{3} \ell$. Four tetrahedra are degenerate with vanishing volume, and the last one has negative square volume,  $V=i/\sqrt{8}$. The squared 4d volume is also negative, $V_{\s}=i \sqrt{3}/32 \ell^4$.
Therefore, Tuckey's configuration is not so relevant after all: having vanishing 3d volumes, it is part of the degenerate data usually disregarded. There exist however other configurations with singular area-length Jacobian, but corresponding to non-degenerate geometries with both 3d and 4d volumes non-vanishing.  Specific examples using two different values of the areas were pointed out to us by Bianca Dittrich. For these configurations, there are two or more  (up to rotations) configurations of normals compatible with the areas, which allow to distinguish the geometry and reconstruct a unique set of edge lengths.

\end{appendix}

\providecommand{\href}[2]{#2}\begingroup\raggedright\endgroup


\begin{thebibliography}{10}

\bibitem{EPRL}
J.~Engle, E.~Livine, R.~Pereira and C.~Rovelli, {\it {LQG vertex with finite
  Immirzi parameter}},  Nucl.Phys. {\bf B799} (2008) 136--149
  [\href{http://arXiv.org/abs/0711.0146}{{\tt 0711.0146}}].

\bibitem{FK}
L.~Freidel and K.~Krasnov, {\it {A New Spin Foam Model for 4d Gravity}},
  Class.Quant.Grav. {\bf 25} (2008) 125018
  [\href{http://arXiv.org/abs/0708.1595}{{\tt 0708.1595}}].

\bibitem{BarrettLorAsymp}
J.~W. Barrett, R.~Dowdall, W.~J. Fairbairn, F.~Hellmann and R.~Pereira, {\it
  {Lorentzian spin foam amplitudes: Graphical calculus and asymptotics}},
  Class.Quant.Grav. {\bf 27} (2010) 165009
  [\href{http://arXiv.org/abs/0907.2440}{{\tt 0907.2440}}].

\bibitem{Barrett:1998gs}
J.~W. Barrett and R.~M. Williams, {\it {The Asymptotics of an amplitude for the
  four simplex}},  Adv. Theor. Math. Phys. {\bf 3} (1999) 209--215
  [\href{http://arXiv.org/abs/gr-qc/9809032}{{\tt gr-qc/9809032}}].

\bibitem{Barrett:2002ur}
J.~W. Barrett and C.~M. Steele, {\it {Asymptotics of relativistic spin
  networks}},  Class. Quant. Grav. {\bf 20} (2003) 1341--1362
  [\href{http://arXiv.org/abs/gr-qc/0209023}{{\tt gr-qc/0209023}}].

\bibitem{Freidel:2002mj}
L.~Freidel and D.~Louapre, {\it {Asymptotics of 6j and 10j symbols}},  Class.
  Quant. Grav. {\bf 20} (2003) 1267--1294
  [\href{http://arXiv.org/abs/hep-th/0209134}{{\tt hep-th/0209134}}].

\bibitem{BarrettEPRasymp}
J.~W. Barrett, R.~Dowdall, W.~J. Fairbairn, H.~Gomes and F.~Hellmann, {\it
  {Asymptotic analysis of the EPRL four-simplex amplitude}},  J.Math.Phys. {\bf
  50} (2009) 112504 [\href{http://arXiv.org/abs/0902.1170}{{\tt 0902.1170}}].

\bibitem{BarrettSU2}
J.~W. Barrett, W.~J. Fairbairn and F.~Hellmann, {\it {Quantum gravity
  asymptotics from the SU(2) 15j symbol}},  Int. J. Mod. Phys. {\bf A25} (2010)
  2897--2916 [\href{http://arXiv.org/abs/0912.4907}{{\tt 0912.4907}}].

\bibitem{Bianchi:2011hp}
E.~Bianchi and Y.~Ding, {\it {Lorentzian spinfoam propagator}},  Phys. Rev. D
  {\bf 86} (2012) 104040 [\href{http://arXiv.org/abs/1109.6538}{{\tt
  1109.6538}}].

\bibitem{HellmannFlatness}
F.~Hellmann and W.~Kaminski, {\it {Holonomy spin foam models: Asymptotic
  geometry of the partition function}},  JHEP {\bf 1310} (2013) 165
  [\href{http://arXiv.org/abs/1307.1679}{{\tt 1307.1679}}].

\bibitem{HanZhangLor}
M.~Han and M.~Zhang, {\it {Asymptotics of Spinfoam Amplitude on Simplicial
  Manifold: Lorentzian Theory}},  Class. Quant. Grav. {\bf 30} (2013) 165012
  [\href{http://arXiv.org/abs/1109.0499}{{\tt 1109.0499}}].

\bibitem{Kaminski:2017eew}
W.~Kaminski, M.~Kisielowski and H.~Sahlmann, {\it {Asymptotic analysis of the
  EPRL model with timelike tetrahedra}},
  \href{http://arXiv.org/abs/1705.02862}{{\tt 1705.02862}}.

\bibitem{Riello:2013bzw}
A.~Riello, {\it {Self-energy of the Lorentzian Engle-Pereira-Rovelli-Livine and
  Freidel-Krasnov model of quantum gravity}},  Phys. Rev. D {\bf 88} (2013),
  no.~2 024011 [\href{http://arXiv.org/abs/1302.1781}{{\tt 1302.1781}}].

\bibitem{Engle:2011un}
J.~Engle, {\it {Proposed proper Engle-Pereira-Rovelli-Livine vertex
  amplitude}},  Phys. Rev. {\bf D87} (2013), no.~8 084048
  [\href{http://arXiv.org/abs/1111.2865}{{\tt 1111.2865}}].

\bibitem{Engle:2015zqa}
J.~Engle, I.~Vilensky and A.~Zipfel, {\it {Lorentzian proper vertex amplitude:
  Asymptotics}},  Phys. Rev. D {\bf 94} (2016), no.~6 064025
  [\href{http://arXiv.org/abs/1505.06683}{{\tt 1505.06683}}].

\bibitem{Dona:2019dkf}
P.~Dona, M.~Fanizza, G.~Sarno and S.~Speziale, {\it {Numerical study of the
  Lorentzian Engle-Pereira-Rovelli-Livine spin foam amplitude}},  Phys. Rev. D
  {\bf 100} (2019), no.~10 106003 [\href{http://arXiv.org/abs/1903.12624}{{\tt
  1903.12624}}].

\bibitem{Han:2020fil}
M.~Han, Z.~Huang, H.~Liu and D.~Qu, {\it {Numerical computations of
  next-to-leading order corrections in spinfoam large-$j$ asymptotics}},
  \href{http://arXiv.org/abs/2007.01998}{{\tt 2007.01998}}.

\bibitem{KKL}
W.~Kaminski, M.~Kisielowski and J.~Lewandowski, {\it {Spin-Foams for All Loop
  Quantum Gravity}},  Class.Quant.Grav. {\bf 27} (2010) 095006
  [\href{http://arXiv.org/abs/0909.0939}{{\tt 0909.0939}}].

\bibitem{Ding:2010fw}
Y.~Ding, M.~Han and C.~Rovelli, {\it {Generalized Spinfoams}},  Phys. Rev. {\bf
  D83} (2011) 124020 [\href{http://arXiv.org/abs/1011.2149}{{\tt 1011.2149}}].

\bibitem{Bianchi:2010zs}
E.~Bianchi, C.~Rovelli and F.~Vidotto, {\it {Towards Spinfoam Cosmology}},
  Phys. Rev. D {\bf 82} (2010) 084035
  [\href{http://arXiv.org/abs/1003.3483}{{\tt 1003.3483}}].

\bibitem{Christodoulou:2016vny}
M.~Christodoulou, C.~Rovelli, S.~Speziale and I.~Vilensky, {\it {Planck star
  tunneling time: An astrophysically relevant observable from background-free
  quantum gravity}},  Phys. Rev. {\bf D94} (2016), no.~8 084035
  [\href{http://arXiv.org/abs/1605.05268}{{\tt 1605.05268}}].

\bibitem{Bahr:2016hwc}
B.~Bahr and S.~Steinhaus, {\it {Numerical evidence for a phase transition in 4d
  spin foam quantum gravity}},  Phys. Rev. Lett. {\bf 117} (2016), no.~14
  141302 [\href{http://arXiv.org/abs/1605.07649}{{\tt 1605.07649}}].

\bibitem{Bahr:2015gxa}
B.~Bahr and S.~Steinhaus, {\it {Investigation of the Spinfoam Path integral
  with Quantum Cuboid Intertwiners}},  Phys. Rev. D {\bf 93} (2016), no.~10
  104029 [\href{http://arXiv.org/abs/1508.07961}{{\tt 1508.07961}}].

\bibitem{Bahr:2018gwf}
B.~Bahr, G.~Rabuffo and S.~Steinhaus, {\it {Renormalization of symmetry
  restricted spin foam models with curvature in the asymptotic regime}},  Phys.
  Rev. D {\bf 98} (2018), no.~10 106026
  [\href{http://arXiv.org/abs/1804.00023}{{\tt 1804.00023}}].

\bibitem{Sarno:2018ses}
G.~Sarno, S.~Speziale and G.~V. Stagno, {\it {2-vertex Lorentzian Spin Foam
  Amplitudes for Dipole Transitions}},  Gen. Rel. Grav. {\bf 50} (2018), no.~4
  43 [\href{http://arXiv.org/abs/1801.03771}{{\tt 1801.03771}}].

\bibitem{Barrett:1997gw}
J.~W. Barrett and L.~Crane, {\it {Relativistic spin networks and quantum
  gravity}},  J. Math. Phys. {\bf 39} (1998) 3296--3302
  [\href{http://arXiv.org/abs/gr-qc/9709028}{{\tt gr-qc/9709028}}].

\bibitem{Immirzi96NPPS}
G.~Immirzi, {\it {Quantum gravity and Regge calculus}},  Nucl.Phys.Proc.Suppl.
  {\bf 57} (1997) 65--72 [\href{http://arXiv.org/abs/gr-qc/9701052}{{\tt
  gr-qc/9701052}}].

\bibitem{DittrichRyan}
B.~Dittrich and J.~P. Ryan, {\it {Phase space descriptions for simplicial 4d
  geometries}},  Class.Quant.Grav. {\bf 28} (2011) 065006
  [\href{http://arXiv.org/abs/0807.2806}{{\tt 0807.2806}}].

\bibitem{twigeo}
L.~Freidel and S.~Speziale, {\it {Twisted geometries: A geometric
  parametrisation of SU(2) phase space}},  Phys. Rev. {\bf D82} (2010) 084040
  [\href{http://arXiv.org/abs/1001.2748}{{\tt 1001.2748}}].

\bibitem{IoWolfgang}
S.~Speziale and W.~M. Wieland, {\it {The twistorial structure of loop-gravity
  transition amplitudes}},  Phys. Rev. {\bf D86} (2012) 124023
  [\href{http://arXiv.org/abs/1207.6348}{{\tt 1207.6348}}].

\bibitem{Freidel:2013fia}
L.~Freidel and J.~Hnybida, {\it {A Discrete and Coherent Basis of
  Intertwiners}},  Class. Quant. Grav. {\bf 31} (2014) 015019
  [\href{http://arXiv.org/abs/1305.3326}{{\tt 1305.3326}}].

\bibitem{IoFabio}
F.~Anz{\`a} and S.~Speziale, {\it {A note on the secondary simplicity
  constraints in loop quantum gravity}},  Class. Quant. Grav. {\bf 32} (2015),
  no.~19 195015 [\href{http://arXiv.org/abs/1409.0836}{{\tt 1409.0836}}].

\bibitem{Bahr:2018vvq}
B.~Bahr, {\it {Non-convex 4d polytopes in Spin Foam Models}},
  \href{http://arXiv.org/abs/1812.10314}{{\tt 1812.10314}}.

\bibitem{LS}
E.~R. Livine and S.~Speziale, {\it {A New spinfoam vertex for quantum
  gravity}},  Phys.Rev. {\bf D76} (2007) 084028
  [\href{http://arXiv.org/abs/0705.0674}{{\tt 0705.0674}}].

\bibitem{Dona:2017dvf}
P.~Dona, M.~Fanizza, G.~Sarno and S.~Speziale, {\it {SU(2) graph invariants,
  Regge actions and polytopes}},  Class. Quant. Grav. {\bf 35} (2018), no.~4
  045011 [\href{http://arXiv.org/abs/1708.01727}{{\tt 1708.01727}}].

\bibitem{BahrSteinhaus15}
B.~Bahr and S.~Steinhaus, {\it {Investigation of the Spinfoam Path integral
  with Quantum Cuboid Intertwiners}},  Phys. Rev. {\bf D93} (2016), no.~10
  104029 [\href{http://arXiv.org/abs/1508.07961}{{\tt 1508.07961}}].

\bibitem{Boosting}
S.~Speziale, {\it {Boosting Wigner's nj-symbols}},  J. Math. Phys. {\bf 58}
  (2017), no.~3 032501 [\href{http://arXiv.org/abs/1609.01632}{{\tt
  1609.01632}}].

\bibitem{EPR}
J.~Engle, R.~Pereira and C.~Rovelli, {\it {The Loop-quantum-gravity
  vertex-amplitude}},  Phys.Rev.Lett. {\bf 99} (2007) 161301
  [\href{http://arXiv.org/abs/0705.2388}{{\tt 0705.2388}}].

\bibitem{LS2}
E.~R. Livine and S.~Speziale, {\it {Consistently Solving the Simplicity
  Constraints for Spinfoam Quantum Gravity}},  Europhys.Lett. {\bf 81} (2008)
  50004 [\href{http://arXiv.org/abs/0708.1915}{{\tt 0708.1915}}].

\bibitem{Baez:2001fh}
J.~C. Baez and J.~W. Barrett, {\it {Integrability for relativistic spin
  networks}},  Class. Quant. Grav. {\bf 18} (2001) 4683--4700
  [\href{http://arXiv.org/abs/gr-qc/0101107}{{\tt gr-qc/0101107}}].

\bibitem{Kaminski:2010qb}
W.~Kaminski, {\it {All 3-edge-connected relativistic BC and EPRL spin-networks
  are integrable}},  \href{http://arXiv.org/abs/1010.5384}{{\tt 1010.5384}}.

\bibitem{Dona:2018nev}
P.~Dona and G.~Sarno, {\it {Numerical methods for EPRL spin foam transition
  amplitudes and Lorentzian recoupling theory}},  Gen. Rel. Grav. {\bf 50}
  (2018) 127 [\href{http://arXiv.org/abs/1807.03066}{{\tt 1807.03066}}].

\bibitem{Perelomov}
A.~M. Perelomov, {\em {Generalized coherent states and their applications}}.
\newblock Springer, 1986.

\bibitem{Ruhl}
W.~Ruhl, {\em {The Lorentz Group and Harmonic Analysis}}.
\newblock W. A. Benjamin, 1970.

\bibitem{Bianchi:2006uf}
E.~Bianchi, L.~Modesto, C.~Rovelli and S.~Speziale, {\it {Graviton propagator
  in loop quantum gravity}},  Class.Quant.Grav. {\bf 23} (2006) 6989--7028
  [\href{http://arXiv.org/abs/gr-qc/0604044}{{\tt gr-qc/0604044}}].

\bibitem{IoPoly}
E.~Bianchi, P.~Dona and S.~Speziale, {\it {Polyhedra in loop quantum gravity}},
   Phys. Rev. {\bf D83} (2011) 044035
  [\href{http://arXiv.org/abs/1009.3402}{{\tt 1009.3402}}].

\bibitem{twigeo2}
L.~Freidel and S.~Speziale, {\it {From twistors to twisted geometries}},  Phys.
  Rev. {\bf D82} (2010) 084041 [\href{http://arXiv.org/abs/1006.0199}{{\tt
  1006.0199}}].

\bibitem{WielandTwistors}
W.~M. Wieland, {\it {Twistorial phase space for complex Ashtekar variables}},
  Class.Quant.Grav. {\bf 29} (2012) 045007
  [\href{http://arXiv.org/abs/1107.5002}{{\tt 1107.5002}}].

\bibitem{Rennert:2016rfp}
J.~Rennert, {\it {Timelike twisted geometries}},  Phys. Rev. D {\bf 95} (2017),
  no.~2 026002 [\href{http://arXiv.org/abs/1611.00441}{{\tt 1611.00441}}].

\bibitem{IoNull}
S.~Speziale and M.~Zhang, {\it {Null twisted geometries}},  Phys.Rev. {\bf D89}
  (2014) 084070 [\href{http://arXiv.org/abs/1311.3279}{{\tt 1311.3279}}].

\bibitem{Conrady:2010vx}
F.~Conrady, {\it {Spin foams with timelike surfaces}},  Class. Quant. Grav.
  {\bf 27} (2010) 155014 [\href{http://arXiv.org/abs/1003.5652}{{\tt
  1003.5652}}].

\bibitem{Dona:2020tvv}
P.~Dona, F.~Gozzini and G.~Sarno,
{\it {Numerical analysis of spin foam dynamics and the flatness problem,}}
Phys. Rev. D \textbf{102} (2020) no.10, 106003  [\href{http://arXiv.org/abs/2004.12911}{{\tt
  2004.12911}}].
  
\bibitem{Engle:2020ffj}
J.~Engle, W.~Kaminski and J.~Oliveira,
{\it {Addendum to ``EPRL/FK asymptotics and the flatness problem'',}}
Class.Quant.Grav. 38 (2021) 11, 119401 (addendum)
[\href{http://arXiv.org/abs/2012.148221}{{\tt
  2012.14822}}].

\bibitem{DittrichSpeziale}
B.~Dittrich and S.~Speziale, {\it {Area-angle variables for general
  relativity}},  New J.Phys. {\bf 10} (2008) 083006
  [\href{http://arXiv.org/abs/0802.0864}{{\tt 0802.0864}}].

\bibitem{Dittrich:2012rj}
B.~Dittrich and J.~P. Ryan, {\it {On the role of the Barbero-Immirzi parameter
  in discrete quantum gravity}},  Class.Quant.Grav. {\bf 30} (2013) 095015
  [\href{http://arXiv.org/abs/1209.4892}{{\tt 1209.4892}}].


\bibitem{IoMiklos}
M.~Langvik and S.~Speziale, {\it {Twisted geometries, twistors and conformal
  transformations}},  Phys. Rev. {\bf D94} (2016), no.~2 024050
  [\href{http://arXiv.org/abs/1602.01861}{{\tt 1602.01861}}].

\bibitem{Barrett:1993db}
J.~W. Barrett and T.~Foxon, {\it {Semiclassical limits of simplicial quantum
  gravity}},  Class. Quant. Grav. {\bf 11} (1994) 543--556
  [\href{http://arXiv.org/abs/gr-qc/9310016}{{\tt gr-qc/9310016}}].

\bibitem{Kaminski:2019dld}
W.~Kaminski and H.~Sahlmann, {\it {The hessian in spin foam models}},  Annales
  Henri Poincare {\bf 20} (2019), no.~12 3927--3953
  [\href{http://arXiv.org/abs/1906.05258}{{\tt 1906.05258}}].

\bibitem{CarloNew}
F.~D'Ambrosio, M.~Christodoulou, P.~Martin-Dussaud, C.~Rovelli and F.~Soltani,
``The End of a Black Hole's Evaporation -- Part I,''
[arXiv:2009.05016 [gr-qc]]. Part 2 is in preparation.

\bibitem{Gozzini:2019kui}
P.~Dona, F.~Gozzini and G.~Sarno, {\it {Searching for classical geometries in
  spin foam amplitudes: a numerical method}},  Class. Quant. Grav. {\bf 37}
  (2020), no.~9 094002 [\href{http://arXiv.org/abs/1909.07832}{{\tt
  1909.07832}}].

\bibitem{BaezBarrett}
J.~C. Baez and J.~W. Barrett, {\it {The Quantum tetrahedron in three-dimensions
  and four-dimensions}},  Adv.Theor.Math.Phys. {\bf 3} (1999) 815--850
  [\href{http://arXiv.org/abs/gr-qc/9903060}{{\tt gr-qc/9903060}}].

\bibitem{Bahr:2017ajs}
B.~Bahr and V.~Belov, {\it {Volume simplicity constraint in the
  Engle-Livine-Pereira-Rovelli spin foam model}},  Phys. Rev. D {\bf 97}
  (2018), no.~8 086009 [\href{http://arXiv.org/abs/1710.06195}{{\tt
  1710.06195}}].

\bibitem{Livine:2002rh}
E.~R. Livine and D.~Oriti, {\it {Implementing causality in the spin foam
  quantum geometry}},  Nucl. Phys. B {\bf 663} (2003) 231--279
  [\href{http://arXiv.org/abs/gr-qc/0210064}{{\tt gr-qc/0210064}}].

\bibitem{Han:2018fmu}
M.~Han, Z.~Huang and A.~Zipfel, {\it {Emergent four-dimensional linearized
  gravity from a spin foam model}},  Phys. Rev. D {\bf 100} (2019), no.~2
  024060 [\href{http://arXiv.org/abs/1812.02110}{{\tt 1812.02110}}].

\bibitem{BiancaHalnuovo}
S.~K. Asante, B.~Dittrich and H.~M. Haggard, {\it {Effective Spin Foam Models
  for Four-Dimensional Quantum Gravity}},
  \href{http://arXiv.org/abs/2004.07013}{{\tt 2004.07013}}.

\bibitem{Pierre}
P.~Dona, M.~Fanizza, P.~Martin-Dussaud and S.~Speziale, {\it {Asymptotics of
  SL(2,C) coherent invariant tensors}},  in preparation.

\bibitem{Barrett:1994nn}
J.~W. Barrett, {\it {First order Regge calculus}},  Class. Quant. Grav. {\bf
  11} (1994) 2723--2730 [\href{http://arXiv.org/abs/hep-th/9404124}{{\tt
  hep-th/9404124}}].

\end{thebibliography}
\end{document}